\documentclass[10pt, a4paper, dvipsnames]{article}

\usepackage[utf8]{inputenc}
\usepackage[english]{babel}

\usepackage[margin=72pt]{geometry}
\usepackage{ragged2e}
    \justifying
\usepackage[skip=0pt, indent=12pt]{parskip}
\usepackage[toc]{appendix}
\usepackage[numbers, sort&compress]{natbib}
\usepackage[justification=justified, font=small, labelsep=period, margin=24pt]{caption}
\usepackage{fancyhdr}

\usepackage[T1]{fontenc}
\usepackage{dsfont}
\usepackage{siunitx}
\usepackage{dutchcal}
\usepackage{soul}

\usepackage{mathtools}
\usepackage{amssymb}
\usepackage{braket}

\usepackage{graphicx}
    \graphicspath{{./images/}}
\usepackage{xcolor}
\usepackage{tikz}
    \usetikzlibrary{math}
    \usetikzlibrary{calc} 
    \usetikzlibrary{arrows.meta}
    \usetikzlibrary{decorations.pathmorphing}
    \usetikzlibrary{decorations.markings}
    \usetikzlibrary{fadings}
    \usetikzlibrary{shapes.geometric}
\usepackage{pgfmath}

\usepackage{array}
    
\usepackage{multirow}
\usepackage{dcolumn}

\usepackage{enumitem}

\usepackage[hidelinks]{hyperref}
    \hypersetup{
        colorlinks=true,
        linkcolor=blue,
        filecolor=blue,      
        urlcolor=blue,
        citecolor=blue
    }

\definecolor{white}{rgb}{1.0, 1.0, 1.0}
\definecolor{black}{rgb}{0.0, 0.0, 0.0}
\definecolor{red}{rgb}{0.8, 0.2, 0.2}
\definecolor{blue}{rgb}{0.0, 0.3, 0.7}
\definecolor{green}{rgb}{0.2, 0.7, 0.2}
\definecolor{yellow}{rgb}{1.0, 0.9, 0.2}
\definecolor{purple}{rgb}{0.6, 0.0, 0.6}
\definecolor{orange}{rgb}{1.0, 0.6, 0.0}

\renewcommand{\d}{\mathrm{d}} 
\newcommand{\e}{\mathrm{e}} 
\renewcommand{\i}{\mathrm{i}} 

\newcommand{\tr}{\mathrm{Tr}} 

\newcommand{\h}[1]{#1} 
\renewcommand{\b}[1]{\mathbf{#1}} 

\newcommand{\ab}[1]{|#1|} 

\newcommand{\ip}[2]{\braket{#1|#2}} 
\newcommand{\op}[2]{\ket{#1}\!\!\bra{#2}} 
\newcommand{\me}[3]{\bra{#1}\!#2\!\ket{#3}} 

\newcommand{\kket}[1]{\ket{#1}\!\rangle} 
\newcommand{\bbra}[1]{\langle\!\bra{#1}} 

\newcommand{\ipe}[2]{\langle\!\ip{#1}{#2}\!\rangle} 

\begin{document}

\thispagestyle{empty}


\begin{center}
    {\Large{\bf 
        Impact of micromotion and field-axis misalignment on the excitation of Rydberg states of ions in a Paul trap
    }\par}
    \vskip 12pt
    \renewcommand{\thefootnote}{$\dagger$}
    {\normalsize{
        Wilson S. Martins$^{1,}$\footnote{Email: \href{mailto:wilson.santana-martins@uni-tuebingen.de}{wilson.santana-martins@uni-tuebingen.de}},
        Joseph W. P. Wilkinson$^{1}$,
        Markus Hennrich$^{2}$,
        Igor Lesanovsky$^{1,3}$
    }\par}
    \renewcommand{\thefootnote}{\arabic{footnote}}
    \vskip 4pt
    {\small{
        $^{1}${\em
            Institut f{\"u}r Theoretische Physik and Center for Integrated Quantum Science and Technology, \\
            Universit{\"a}t T{\"u}bingen,
            Auf der Morgenstelle 14,
            72076 T{\"u}bingen,
            Germany
        } \\
        $^{2}${\em
           Department of Physics,
           Stockholm University,  SE-106 91 Stockholm, Sweden
        } \\
        $^{3}${\em
            School of Physics and Astronomy and Centre for the Mathematics and Theoretical Physics of Quantum Non-Equilibrium Systems,
            University of Nottingham,
            Nottingham, NG7 2RD,
            United Kingdom
        }
    }\par}
    \vskip 4pt
    {\small{
        (Dated: \today)
    }\par}
\end{center}


\begin{abstract}
    Trapped ions are among the most advanced platforms for quantum simulation and computation.
    Their capabilities can be further augmented by making use of electronically highly excited Rydberg states, which enable the realization of long-ranged electric dipolar interactions.
    Most experimental and theoretical studies so far focus on the excitation of ionic Rydberg states in linear Paul traps, which generate confinement by a combination of static and oscillating electric fields.
    These two fields need to be carefully aligned to minimize so-called micromotion, caused by the time-dependent electric field.
    The purpose of this work is to systematically understand the qualitative impact of micromotion on the Rydberg excitation spectrum, when the symmetry axes of the two electric fields do not coincide.
    Considering this scenario is not only important in the case of possible field misalignment, but becomes inevitable for Rydberg excitations in 2D and 3D ion crystals.
    We develop a minimal model describing a single trapped Rydberg ion, which we solve numerically via Floquet theory and analytically using a perturbative approach.
    We calculate the excitation spectra and analyze in which parameter regimes addressable and energetically isolated Rydberg lines persist, which are an important requirement for conducting coherent manipulations.
\end{abstract}


\tableofcontents

\pagestyle{fancy}
\fancyhf{}
    \fancyhead[L]{\leftmark}
    \fancyhead[R]{\thepage}


\section{Introduction}\label{sec:introduction}

Trapped ions are versatile platforms for quantum technology applications that include quantum computing~\cite{cirac1995, cirac2000, blatt2008, haffner2008, schindler2013, duan2010, bermudez2017, bruzewicz2019}, quantum simulation~\cite{porras2004, friedenauer2008, kim2010, barreiro2011, blatt2012, georgescu2014, monroe2021}, and quantum sensing~\cite{kotler2011, hempel2013, degen2017, campbell2017, gilmore2021}.
In conventional trapped ion quantum simulators, quantum information is encoded in qubits, which are formed from the energetically low-lying electronic states of the individual ions~\cite{kim2009, muller2011, schneider2012}.
The interactions between the ions employed to implement the necessary coherent quantum operations are then realized by electronic state-dependent light shifts in combination with the excitation of collective vibrational modes of the trapped ion crystal~\cite{leibfried2003, haljan2005, behrle2023}.
Recent research efforts (see, e.g., Refs.~\cite{muller2008, schmidtkaler2011, feldker2015, bachor2016, higgins2017a, higgins2017b, higgins2018, higgins2019, mokhberi2019, vogel2019, andrijauskas2021}) have theoretically and experimentally explored the possibility of combining the benefits of trapped ions with Rydberg interactions (see the recent review in Ref.~\cite{mokhberi2020} and references therein).
This is accomplished by laser-exciting the ions to energetically high-lying electronic states, so-called \textit{Rydberg states}, which interact strongly via long-ranged electric dipolar forces~\cite{weber2017}.
This mechanism can be used to implement fast coherent interactions between the trapped Rydberg ions and recently led to the demonstration of entangling gate operations on submicrosecond timescales~\cite{zhang2020}.

\begin{figure}[t]
    \centering
    \includegraphics[scale=0.46]{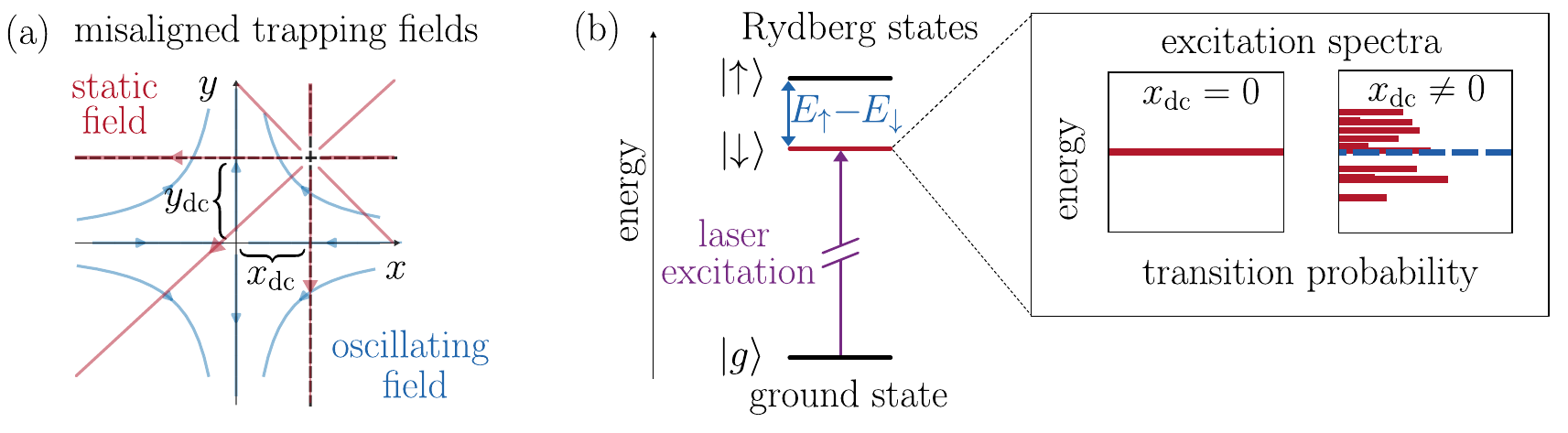}
    \caption{
        \textbf{Electric field configuration of the Paul trap and Rydberg excitation spectrum.}
        (a) Electric field lines in a misaligned linear Paul trap.
        The electric quadrupole potential forming the trap is composed of a static (red) and an oscillating (blue) radio frequency electric field.
        We consider the $1$D case in which these fields are misaligned, i.e., where the symmetry axes of the fields are shifted relative to one another, here by an amount $x_{\mathrm{dc}}$ (see main text for details).
        This gives rise to excess micromotion that affects the Rydberg excitation spectrum.
        (b) Relevant energy level scheme and excitation spectrum of the trapped Rydberg ion, with ground state $\ket{g}$ and Rydberg states $\ket{\downarrow}$ and $\ket{\uparrow}$.
        Micromotion causes coupling between the vibrational degrees of freedom and the nearby Rydberg state $\ket{\uparrow}$, which is separated from $\ket{\downarrow}$ by the energy $E_{\uparrow} - E_{\downarrow}$.
        The impact of the micromotion is enhanced the larger the relative field displacement $x_{\mathrm{dc}}$, which becomes visible in the formation of sidebands in the excitation spectra, as seen in the zoomed-in view.
    }
    \label{fig:model}
\end{figure}

In current trapped Rydberg ion experiments, the ions are confined within a linear Paul trap~\cite{paul1990, wineland1998, major2005}.
These traps rely on the combination of a static and an oscillating radio frequency electric quadrupole field~\cite{brown1991, raizen1992, drewsen2000}, as illustrated in Fig.~\ref{fig:model}a.
Longitudinally (i.e., in the $z$-direction), this creates static harmonic confinement, whilst transversal confinement is generated by a ponderomotive force.
As was demonstrated in Ref.~\cite{cook1985}, the resultant ponderomotive trapping potential is well approximated by an effectively static harmonic potential if the field radio frequency is sufficiently large relative to the harmonic trap frequency.
However, the remaining time-dependent terms may cause transitions between vibrational states and controlling them, such as to avoid resonances, is necessary for stable operation of the trap.
Typically, this residual time-dependence is referred to as \textit{micromotion}.
In linear, i.e., 1D, ion crystals, the impact of micromotion is minimized by aligning the symmetry axes of the static and oscillating electric quadrupole fields.
Nevertheless, deviations from this ideal situation can emerge due to stray electric fields, which are typically offset by micromotion compensation electrodes~\cite{wineland1998, berkeland1998}.
However, when considering 2D and 3D ion crystals within linear Paul traps, such compensation is not possible for all ions.
Consequently, this inherent micromotion may alter the transition rates and spectral properties of the ions located off-axis (i.e., away from the $z$-axis).

In this work, we investigate the impact of micromotion on the Rydberg excitation spectrum of an ion confined within a linear Paul trap.
In particular, we explore how a relative displacement of the symmetry axes of the static and oscillating electric fields (as depicted in Fig.~\ref{fig:model}a) affects the transition energies and laser excitation probabilities of the energetically high-lying Rydberg states of a trapped ion (Fig.~\ref{fig:model}b).
Starting from the full trapped Rydberg ion Hamiltonian, we develop a simplified model consisting of a phonon mode (describing the vibrational motion of the ion) coupled to a spin (modeling a Rydberg $S$- and $P$-state).
This model, which resembles a generalized quantum Rabi model~\cite{braak2011, eckle2017, xie2017}, is numerically solved by Floquet theory~\cite{sambe1973, eckardt2015, novicenko2017, rodriguez2021}.
From the quasienergy spectrum, we compute the relative strength of transitions from the ground state to the Rydberg states due to laser excitation.
This allows us to understand the impacts of the electric field displacement and the oscillating field frequency on the emergence of spectral sidebands induced by the coupling between electronic and vibrational motion.
These numerical studies are complemented by time-independent perturbation theory.
Our results may be relevant for assessing the applicability of Rydberg excitations in 2D and 3D ion crystals for the realization of scalable quantum computers and quantum simulators.


\section{Trapped Rydberg ion Hamiltonian}\label{sec:derivation}

In this section, we outline the derivation of the Hamiltonian for a singly positively charged alkaline earth metal Rydberg ion confined within the electric quadrupole potential of a Paul trap.
Here, the focus is on coupling terms introduced by the relative displacement between the static and oscillating electric fields.
These induce excess micromotion, which we investigate throughout the remainder of this paper.

\subsection{Rydberg ion in a Paul trap}

Throughout this work, we consider an alkaline earth metal Rydberg ion confined in a Paul trap, which is currently what is used in typical trapped Rydberg ion experiments (see the review in Ref.~\cite{mokhberi2020}).
These ions possess a valence electron (charge $-e$, mass $m_{\mathrm{e}}$) that interacts with an ionic core (charge $2e$, mass $m_{\mathrm{c}}$) formed by the remaining core electrons and nucleus~\cite{djerad1991}.
The interaction between these charged particles can be well approximated by an effective central potential.
This incorporates the Coulomb interaction and additionally contains contributions accounting for the screening of the nucleus, the polarizability of the ionic core, and relativistic effects~\cite{gallagher2023}.
Taking into account the interaction of these charges with the electric field of the Paul trap via minimal coupling~\cite{steck2024}, the Hamiltonian describing the dynamics of the trapped Rydberg ion is well approximated by
\begin{equation}\label{eq:lab-frame-hamiltonian}
    H = \frac{\b{p}_{\smash{\mathrm{c}}}^{2}}{2 m_{\smash{\mathrm{c}}}} + \frac{\b{p}_{\mathrm{e}}^{2}}{2 m_{\mathrm{e}}} + V(\ab{\b{r}_{\mathrm{e}} - \b{r}_{\mathrm{c}}}) + 2 e \Phi(\b{r}_{\mathrm{c}}, t) - e \Phi(\b{r}_{\mathrm{e}}, t).
\end{equation}
Here $\b{r}_{\mathrm{c}}$, $\b{p}_{\mathrm{c}}$, $\b{r}_{\mathrm{e}}$, and $\b{p}_{\mathrm{e}}$ are the positions and momenta of the ionic core and valence electron, respectively, and $V(\ab{\b{r}_{\mathrm{e}} - \b{r}_{\mathrm{c}}})$ is the effective central potential approximating the interaction between them~\cite{schmidtkaler2011}.

The electric potential confining the ions within the Paul trap is composed of an oscillating and static electric field~\cite{leibfried2003}.
When the electric fields are aligned, namely, when the symmetry axes (i.e., the $z$-axis, see Fig. \ref{fig:model}a) of the fields coincide, the electric potential near the center of the trap is well approximated by an electric quadrupole potential of the form
\begin{equation}\label{eq:perfect-trap}
    \Phi(\b{r}, t) = A \cos(\nu t) [r_{x}^{2} - r_{y}^{2}] - B [[1 + \epsilon] r_{x}^{2} + [1 - \epsilon] r_{y}^{2} - 2 r_{z}^{2}].
\end{equation}
Here, $A$ and $B$ are the oscillating and static electric field gradients, $\nu$ the oscillating electric field (radio) frequency, and $\epsilon$ the static electric field eccentricity (i.e., a dimensionless quantity that parametrizes the axial trap symmetry).
When $\epsilon = 0$, the radial modes are degenerate.

For the purpose of this work, we consider a Paul trap wherein which the oscillating electric potential is centered along the $z$-axis, but where the static electric potential is shifted in the $xy$-plane (see Fig.~\ref{fig:model}).
Here, the electric quadrupole potential is modified from that in Eq.~\eqref{eq:perfect-trap} and becomes
\begin{equation}
    \tilde{\Phi}(\b{r}, t) = A \cos(\nu t) [r_{x}^{2} - r_{y}^{2}] - B [[1 + \epsilon] [r_{x} - x_{\mathrm{dc}}]^{2} + [1 - \epsilon] [r_{y} - y_{\mathrm{dc}}]^{2} - 2 r_{z}^{2}],
\end{equation}
with $x_{\mathrm{dc}}$ and $y_{\mathrm{dc}}$ the displacements in the $xy$-plane of the $z$-axis of the static field relative to that of the oscillating field.
Using the definition of the electric potential in Eq.~\eqref{eq:perfect-trap}, this can then be recast as
\begin{equation}\label{eq:imperfect-trap}
    \tilde{\Phi}(\b{r}, t) = \Phi(\b{r}, t) + \delta \Phi(\b{r}), \qquad
    \delta \Phi(\b{r}) = 2 B \left[[1 + \epsilon] x_{\mathrm{dc}} r_{x} + [1 - \epsilon] y_{\mathrm{dc}} r_{y}\right],
\end{equation}
where the static electric potential $\delta \Phi(\b{r})$ accounts for contributions due to stray electric fields or, more generally, from the off-axis position or out-of-equilibrium motion of the ion due to being displaced from the trap axis (i.e., $z$-axis), as is the case in 2D and 3D ion crystals.

In a trap with aligned electric fields, i.e. $x_{\mathrm{dc}} = y_{\mathrm{dc}} = 0$ as in Eq.~\eqref{eq:perfect-trap}, the ionic motion separates into time-independent ``slow'' harmonic motion called secular motion and time-dependent ``fast'' driven motion often referred to as micromotion~\cite{major2005}.
However, in a trap where the symmetry axes of the static and oscillating fields are misaligned, as described by Eq.~\eqref{eq:imperfect-trap}, the motion of the ionic core and valence electron acquire additional contributions, which are also termed micromotion~\cite{wineland1998}.
To avoid this ambiguity, we term the micromotion due to the oscillating field frequency \textit{intrinsic} and that caused by the relative field displacement \textit{extrinsic}.

\subsection{Oscillating center of mass frame}

For treating this problem, it is beneficial to introduce the canonical center of mass and relative positions and momenta,
\begin{equation}
    \b{R} = \frac{m_{\mathrm{c}} \b{r}_{\mathrm{c}} + m_{\mathrm{e}} \b{r}_{\mathrm{e}}}{m_{\mathrm{c}} + m_{\mathrm{e}}}, \qquad
    \b{r} = \b{r}_{\mathrm{e}} - \b{r}_{\mathrm{c}}, \qquad
    \b{P} = \b{p}_{\mathrm{c}} + \b{p}_{\mathrm{e}}, \qquad
    \b{p} = \frac{m_{\mathrm{c}} \b{p}_{\mathrm{e}} - m_{\mathrm{e}} \b{p}_{\mathrm{c}}}{m_{\mathrm{c}} + m_{\mathrm{e}}},
\end{equation}
together with the center of mass and relative masses, i.e. the total mass and reduced mass,
\begin{equation}
    M = m_{\mathrm{c}} + m_{\mathrm{e}}, \qquad
    m = \frac{m_{\mathrm{c}} m_{\mathrm{e}}}{m_{\mathrm{c}} + m_{\mathrm{e}}}.
\end{equation}
Expressed in terms of the center of mass and relative coordinates, and exploiting the fact that the masses satisfy $m_{\mathrm{c}} \gg m_{\mathrm{e}}$, the Hamiltonian in Eq.~\eqref{eq:lab-frame-hamiltonian} can then be approximated by~\cite{wilkinson2024b}
\begin{equation}\label{eq:stationary-frame-hamiltonian}
    H \approx \frac{\b{P}^{2}}{2 M} + e \Phi(\b{R}, t) + e \delta \Phi(\b{R}) + \frac{\b{p}^{2}}{2 m} + V(\ab{\b{r}}) - e \Phi(\b{r}, t) - e \delta \Phi(\b{r}) - e \b{r} \cdot \boldsymbol{\nabla} \Phi(\b{R}, t).
\end{equation}

In order to separate the terms describing the ionic center of mass motion into a slow harmonic motion and fast driven motion, we perform a unitary transformation that takes the system into a reference frame that oscillates at the electric field radio frequency $\nu$. This is achieved via the unitary operator
\begin{equation}
    U = \exp \! \bigg( \!\! - \! \frac{e A}{\i \hbar \nu} \sin(\nu t) [R_{x}^{2} - R_{y}^{2}] \bigg).
\end{equation}
Within the oscillating frame, the unitarily transformed Hamiltonian can then be recast explicitly in terms of its external, internal, and coupled degrees of freedom as~\cite{wilkinson2024b}
\begin{equation}\label{eq:unitary-transformation}
    H \mapsto H' = U H U^{\dagger} + \i \hbar \frac{\partial U}{\partial t} U^{\dagger} = H_{\mathrm{ex}} + H_{\mathrm{in}} + H_{\mathrm{co}}.
\end{equation}
The Hamiltonian governing the external motion of the trapped Rydberg ion follows as
\begin{equation}
    H_{\mathrm{ex}} = \frac{1}{2 M} \sum_{u} [P_{u}^{2} + M^{2} \omega_{u}^{2} R_{u}^{2}] - \frac{2 e A}{M \nu} \sin(\nu t) [R_{x} P_{x} - R_{y} P_{y}] - \frac{e^{2} A^{2}}{M \nu^{2}} \cos(2 \nu t) [R_{x}^{2} + R_{y}^{2}] + e \delta \Phi(\b{R}),
\end{equation}
where we have introduced the harmonic frequencies of the oscillations of the ions along the $u$-axis within the (effectively) static ponderomotive potential~\cite{berkeland1998}:
\begin{equation}\label{eq:trap-frequencies}
    \omega_{x} = \sqrt{\frac{2 e^{2} A^{2}}{M^{2} \nu^{2}} - \frac{2 [1 + \epsilon] e B}{M}}, \qquad
    \omega_{y} = \sqrt{\frac{2 e^{2} A^{2}}{M^{2} \nu^{2}} - \frac{2 [1 - \epsilon] e B}{M}}, \qquad
    \omega_{z} = \sqrt{\frac{4 e B}{M}}.
\end{equation}
The Hamiltonian describing the motion of the valence electron (i.e., the internal motion) reads
\begin{equation}\label{eq:electron-trap-interactions}
    H_{\mathrm{in}} = \frac{\b{p}^{2}}{2 m} + V(\ab{\b{r}}) - e A \cos(\nu t) [r_{x}^{2} - r_{y}^{2}] - e B [3 r_{z}^{2} - \b{r}^{2}] + \epsilon e B [r_{x}^{2} - r_{y}^{2}] - e \delta \Phi(\b{r}).
\end{equation}
The remaining Hamiltonian, which accounts for the coupling between the external and internal motion, is then accordingly given by
\begin{equation}\label{eq:electron-ion-trap-interactions}
    H_{\mathrm{co}} = - 2 e A \cos(\nu t) [R_{x} r_{x} - R_{y} r_{y}] + 2 e B [[1 + \epsilon] R_{x} r_{x} + [1 - \epsilon] R_{y} r_{y} - 2 R_{z} r_{z}].
\end{equation}
For a more detailed derivation of these Hamiltonian terms, see Refs.~\cite{muller2008,schmidtkaler2011,wilkinson2024a, wilkinson2024b}.


\section{Spin-phonon coupled model Hamiltonian}\label{sec:model}

With a general expression for the trapped Rydberg ion Hamiltonian derived, we now focus on developing a model Hamiltonian that allows us to understand the general features of the Rydberg excitation spectrum.
To this end, we use a two-level approximation for the electronic states and restrict the vibrational motion to one dimension.

\subsection{Isolated electronic and reduced vibrational subspaces}\label{sec:state-spaces}

In the absence of the electric quadrupole potential of the Paul trap, the bound electronic quantum states of the ion can be completely characterized by the principal $n$, orbital angular momentum $l$, spin $s$, total angular momentum $j$, and total magnetic $m_{j}$ quantum numbers.
Gathering these quantum numbers in the symbol $\b{n} = (n, l, s, j, m_{j})$, we can then identify the states $\ket{\b{n}}$ as the eigenstates of energy $E_{\b{n}}$ of the field-free Hamiltonian~\cite{muller2008}.
These satisfy the Schr{\"o}dinger equation
\begin{equation}\label{eq:electronic-states}
    \bigg[ \frac{\b{p}^{2}}{2 m} + V(\ab{\b{r}}) \bigg] \! \ket{\b{n}} = E_{\b{n}} \! \ket{\b{n}} \!.
\end{equation}
In what follows, we restrict ourselves to a pair of low angular momentum Rydberg states ($n \gg 1$), which we identify as (fictitious) spin degrees of freedom,
\begin{equation}\label{eq:electronic-spin-states}
    \ket{n, 0, 1/2, 1/2, -1/2} \equiv \ket{n S_{\smash{1/2}}(- 1/2)} \equiv \ket{\downarrow} \!, \qquad
    \ket{n, 1, 1/2, 1/2, +1/2} \equiv \ket{n P_{\smash{1/2}}(+ 1/2)} \equiv \ket{\uparrow} \!.
\end{equation}
All operators acting solely on the electronic states can then be expressed through the Pauli matrices,
\begin{equation}
    \sigma_{x} = \op{\uparrow}{\downarrow} + \op{\downarrow}{\uparrow} \!, \qquad
    \sigma_{y} = - \i \! \op{\uparrow}{\downarrow} + \i \! \op{\downarrow}{\uparrow} \!, \qquad
    \sigma_{z} = \op{\uparrow}{\uparrow} - \op{\downarrow}{\downarrow} \!.
\end{equation}
This simplification is motivated by the fact that low angular momentum states are spectrally well isolated from other electronic levels and that they can be excited from the electronic ground state via few-photon transitions.
Moreover, under typical trapping conditions and for an appropriate choice of Rydberg states (see, e.g., Ref.~\cite{mokhberi2020}), the spectral isolation guarantees that these electronic states are only weakly mixed through the terms in Eqs.~\eqref{eq:electron-trap-interactions} and~\eqref{eq:electron-ion-trap-interactions}, which result from the electric fields of the Paul trap. 

The quantum state of the external motion of the ion is expanded in a harmonic oscillator basis $\ket{N_{u}}$ with the quantum numbers $N_{u}$ for $u = x, y, z$, which correspond to occupation numbers of the vibrational (i.e., phonon) modes along each axis~\cite{muller2008}.
These states satisfy the Schr{\"o}dinger equation 
\begin{equation}\label{eq:vibrational-modes}
    \bigg[ \frac{P_{u}^{2}}{2 M} + \frac{M \omega_{u}^{2} R_{u}^{2}}{2} \bigg] \! \ket{N_{u}} = \hbar \omega_{u} N_{u} \! \ket{N_{u}} \!,
\end{equation}
where $\hbar \omega_{u} N_{u}$ is the eigenenergy of the eigenstate (Fock state) $\ket{N_{u}}$.
Introducing the bosonic creation and annihilation operators $a_{u}^{\dagger}$ and $a_{u}$, the center of mass position $R_{u}$ and the momentum $P_{u}$ can be written as
\begin{equation}\label{eq:bosonic-operators}
    R_{u} = \ell_{u} [a_{u}^{\dagger} + a_{u}], \qquad
    P_{u} = \i M \omega_{u} \ell_{u} [a_{u}^{\dagger} - a_{u}],
\end{equation}
with the vibrational lengths
\begin{equation}
    \ell_{u} = \sqrt{\frac{\hbar}{2 M \omega_{u}}}.
\end{equation}

The expression of the full Hamiltonian for the trapped Rydberg ion represented in terms of the field-free electronic and vibrational bases is presented in App.~\ref{app:derivation}.
In principle, this Hamiltonian can be numerically treated within an appropriate finite-dimensional Hilbert space of electronic states and vibrational modes (e.g., the time evolution of the system under the corresponding Schr{\"o}dinger equation can be efficiently solved). 
However, the purpose of our work is to understand the general features that arise from the relative displacement of the static and oscillating electric fields of the Paul trap on the Rydberg excitation spectrum.
Therefore, we resort to a simplified model, which describes the essential features of the trapped Rydberg ion system, but involves only the (fictitious) spin degree of freedom in Eq.~\eqref{eq:electronic-spin-states} coupled to a single phonon degree of freedom.
Specifically, we choose the vibrational motion in the $x$-direction.
The derivation of this model, which turns out to be a generalized quantum Rabi model, is presented in App.~\ref{app:model}.

\subsection{Generalized quantum Rabi model Hamiltonian}\label{sec:generalized-quantum-rabi-model}

According to Eq.~\eqref{eq:unitary-transformation}, the trapped Rydberg ion Hamiltonian can be decomposed into a term that governs the internal (electronic) dynamics, $H_{\mathrm{in}}$, a term that describes the external (vibrational) motion, $H_{\mathrm{ex}}$ and a term $H_{\mathrm{co}}$, which accounts for the coupling between these degrees of freedom.
For the following analysis it is convenient to decompose them further into their Fourier components with respect to the oscillating electric field frequency.
To this end, we introduce dimensionless units, for which frequencies are given in units of the trap frequency $\omega_{x}$.
The dimensionless radio frequency of the oscillating electric field is then $\zeta \equiv \nu / \omega_{x}$ and the dimensionless time is $\tau = \omega_x t$. 
Written in these units, the model Hamiltonian describing the trapped Rydberg ion is given by
\begin{equation}\label{eq:spin-phonon-coupled-hamiltonian}
    H(\tau) = H^{0} + V(\tau), \qquad 
    V(\tau) = \e^{\i \zeta \tau} H^{1} + \e^{\i 2 \zeta \tau} H^{2} + \mathrm{H.c.},
\end{equation}
where the terms associated to the Fourier components are
\begin{equation}
    H^{0} = H_{\mathrm{ex}}^{0} + H_{\smash{\mathrm{in}}}^{0} + H_{\mathrm{co}}^{0}, \qquad
    H^{1} = H_{\mathrm{ex}}^{1} + H_{\mathrm{co}}^{1}, \qquad
    H^{2} = H_{\mathrm{ex}}^{2}.
\end{equation}
The Hamiltonian terms describing the external, internal, and coupled dynamics of the trapped Rydberg ion then follow as (see App.~\ref{app:model} for details)
\begin{equation}\label{eq:spin-phonon-coupled-hamiltonian-terms}
\begin{gathered}
    H_{\mathrm{ex}}^{0} = a^{\dagger} a + \frac{\delta}{4 \gamma^{2}} [a^{\dagger} + a], \qquad
    H_{\smash{\mathrm{in}}}^{0} = \varepsilon \sigma_{z} - \frac{\delta \lambda}{4 \gamma^{2}} \sigma_{x}, \qquad
    H_{\mathrm{co}}^{0} = \frac{\lambda}{4 \gamma^{2}} [a^{\dagger} + a] \sigma_{x}, \\    
    H_{\mathrm{ex}}^{1}  = - \frac{\sqrt{2 \gamma^{2} + 1}}{4 \gamma} [a^{\dagger} + a] [a^{\dagger} - a],
    \qquad
    H_{\mathrm{co}}^{1} = - \frac{\sqrt{2 \gamma^{2} + 1} \zeta \lambda}{4 \gamma} [a^{\dagger} + a] \sigma_{x}, \\
    H_{\mathrm{ex}}^{2} = - \frac{2 \gamma^{2} + 1}{16 \gamma^{2}} [a^{\dagger} + a] [a^{\dagger} + a].
\end{gathered}
\end{equation}
They depend on the dimensionless quantities
\begin{equation}\label{eq:dimensionless-quantities}
    \delta \equiv \frac{x_{\mathrm{dc}}}{\ell_{x}}, \qquad
    \varepsilon \equiv \frac{E_{\uparrow} - E_{\downarrow}}{2 \hbar \omega_{x}}, \qquad
    \lambda \equiv \frac{\ab{\!\me{\uparrow}{r_{x}}{\downarrow}\!}}{\ell_{x}}, \qquad \gamma\equiv\frac{\omega_x}{\omega_z}.
\end{equation}
Here, $\delta$ parametrizes the relative displacement of the electric fields of the Paul trap and $\varepsilon$ is the energy separation of the two considered Rydberg states with energies $E_{\uparrow}$ and $E_{\downarrow}$.
The parameter $\lambda$ yields the magnitude of the transition dipole matrix element between the two Rydberg states and $\gamma$ is the ratio of the trap frequencies in the $x$- and $z$-direction.
To get an idea of the typical values that these parameters assume, we consider as example a strontium $^{88}\mathrm{Sr}^{+}$ ion in Rydberg states with principal quantum number $n = 46$~\cite{mokhberi2020}.
In a Paul trap with axial trap frequency $\omega_{z} = 2 \pi \times 1 \, \unit{\mega\hertz}$, this yields a vibrational length~of $\ell_{x} \approx 10 \, \unit{\nano\meter}$ and we obtain
\begin{equation}
    \varepsilon = 3 \times 10^{4}, \qquad
    \lambda = 4.
\end{equation}
Moreover, in a typical experiment, the ratio of the radial and axial trapping frequency is $\gamma = 2$.
We will use these values throughout the remainder of this paper.
In App.~\ref{app:model}, we analyze how the parameters $\varepsilon$ and $\lambda$ change with $n$ and discuss their impact on the system's physics.

For the relative field displacement $\delta$, we will consider here values up to a maximum of $\delta = 100$, which corresponds to $x_{\mathrm{dc}} \approx 1 \, \unit{\micro\meter}$ (see Fig.~\ref{fig:model}).
An inspection of Eq. (\ref{eq:spin-phonon-coupled-hamiltonian-terms}) shows that the displacement introduces a term linear in $\delta$ in the Hamiltonian $H_{\mathrm{ex}}^{0}$.
As this part of the Hamiltonian provides the static trapping, the equilibrium position of the ion is apparently shifted. 
To simplify our analysis, we undo this shift by introducing the displaced creation and annihilation operators, $b^{\dagger} = a^{\dagger} + \delta / 16$ and $b = a + \delta / 16$.
Then, the model Hamiltonian takes the form (see App.~\ref{app:model})
\begin{equation}\label{eq:spin-phonon-coupled-hamiltonian-terms-transformed}
\begin{gathered}
    H_{\mathrm{ex}}^{0} = b^{\dagger} b, \qquad
    H_{\smash{\mathrm{in}}}^{0} = \varepsilon \sigma_{z} - \frac{9 \delta}{32} \sigma_{x}, \qquad
    H_{\mathrm{co}}^{0} = \frac{1}{4} [b^{\dagger} + b] \sigma_{x}, \\
    H_{\mathrm{ex}}^{1} = - \frac{3}{8} [b^{\dagger} + b] [b^{\dagger} - b] + \frac{3 \delta}{64} [b^{\dagger} - b], \qquad
    H_{\smash{\mathrm{in}}}^{1} = \frac{3 \zeta \delta}{16} \sigma_{x}, \qquad
    H_{\mathrm{co}}^{1} = - \frac{3 \zeta}{2} [b^{\dagger} + b] \sigma_{x}, \\
    H_{\mathrm{ex}}^{2} = - \frac{9}{64} [b^{\dagger} + b] [b^{\dagger} + b]  + \frac{9 \delta}{256} [b^{\dagger} + b].
\end{gathered}
\end{equation}
Here, the terms depending on the relative field displacement $\delta$ can be understood as contributing to the extrinsic micromotion (i.e., due to the relative displacement of the symmetry axes of the electric fields).
Those that depend on the oscillating field frequency $\zeta$, including through the time-dependent exponents $\e^{\i m \zeta \tau}$, and not on $\delta$ can then be interpreted as causing the intrinsic micromotion. 
Note that we have omitted constant energy shifts arising from the introduction of the displaced operators, since these can be eliminated via time-dependent unitary transformation (see App.~\ref{app:model} for details).


\section{Floquet theory and quasienergy operator}

In order to analyze the Rydberg excitation spectrum in the presence of the Paul trap, we employ Floquet theory.
To start, we introduce the necessary concepts, quantities, and notations, before utilizing them to represent the model Hamiltonian within the extended Floquet Hilbert space.
This allows us to obtain the quasienergies and relative transition probabilities, which characterize the impact of the micromotion on the excitation of the trapped Rydberg ion.

\subsection{Periodic motion and Floquet states}

Systems whose time evolution is governed by a time-periodic Hamiltonian, such as in Eq.~\eqref{eq:spin-phonon-coupled-hamiltonian-terms-transformed}, can be described within the framework of Floquet theory~\cite{shirley1965, barone1977, ho1983}. 
The main result of Floquet theory---\textit{Floquet's theorem}---is that any homogeneous system of linear equations with periodic coefficients can be mapped, via a linear basis transformation, to an equivalent system with constant coefficients.

A quantum system described by a time-periodic Hamiltonian $H(\tau) = H(\tau + T)$, with $T = 2 \pi / \zeta$ being the oscillation period, possesses generalized stationary states $\ket{\psi_{\alpha} (\tau)}$ called Floquet states.
They are of the form~\cite{shirley1965},
\begin{equation}\label{Floq_theorem}
    \ket{\psi_{\alpha} (\tau)} = \e^{- \i \epsilon_{\alpha} \tau} \! \ket{u_{\alpha} (\tau)} \!,
\end{equation}
where $\epsilon_{\alpha}$ is the so-called \textit{quasienergy} and $\ket{u_{\alpha} (\tau)} = \ket{u_{\alpha} (\tau + T)}$ the corresponding time-periodic Floquet mode.
The quasienergies and Floquet modes are not defined uniquely.
Introducing an integer index $m$, they can be labelled according to
\begin{equation}\label{eq:quasi_energies_exact}
    \epsilon_{\alpha, m} = \epsilon_{\alpha} + m \zeta, \qquad
    \ket{u_{\alpha, m} (\tau)} = \e^{\i m \zeta \tau} \! \ket{u_{\alpha} (\tau)} \!.
\end{equation}
Entering these into the time-dependent Schr{\"o}dinger equation with Hamiltonian $H(\tau)$ then returns
\begin{equation}\label{quasi_eq}
    \mathcal{H}(\tau) \! \ket{u_{\alpha, m}(\tau)} = \epsilon_{\alpha, m} \! \ket{u_{\alpha, m}(\tau)} \!,
\end{equation}
where $\mathcal{H}(\tau) = H(\tau) - \i \frac{\d}{\d \tau}$ is the instantaneous Hamiltonian (or Floquet operator) and the time $\tau$ is treated as a continuous parameter.
The eigenvalue problem in Eq.~\eqref{quasi_eq} defines a set of time-independent quasienergies that will form the object of our study to understand the effects of the micromotion on the excitation spectrum of the trapped Rydberg ion.

\subsection{Extended Floquet Hilbert space and quasienergy operator}\label{sec:extended-floquet-space}

We now represent the time-dependent Hamiltonian, which is composed of the terms given in Eq.~\eqref{eq:spin-phonon-coupled-hamiltonian-terms-transformed}, in the extended Floquet Hilbert space~\cite{sambe1973, eckardt2015, novicenko2017}.
This is constructed from the tensor product of the Hilbert space of the quantum system and the space of square-integrable $T$-periodic time-dependent functions where the time $\tau$ is treated as a coordinate under periodic boundary conditions.
The extended Floquet Hilbert space is spanned by the states $\kket{\beta, m}$, where $\beta$ labels the basis of the Hilbert space of the quantum system under consideration and $m$ is the so-called Fourier index.
Within this space, the inner product is defined as
\begin{equation}
    \ipe{\beta'\!, m'}{\beta, m} = \frac{1}{T} \int^{T}_{0} \! \d \tau \, \e^{\i [m - m'] \zeta \tau} \! \ip{\beta'}{\beta} \!.
\end{equation}
In the extended Floquet Hilbert space, the instantaneous Hamiltonian in Eq.~\eqref{quasi_eq} is represented by the \textit{quasienergy operator}, whose matrix elements follow as
\begin{equation}\label{eq:floquet-basis-states}
    [\mathcal{H}]^{\beta'\!, m'}_{\beta, m} = \bbra{\beta'\!, m'} \! \mathcal{H} \! \kket{\beta, m} = \me{\beta'}{H^{m'\! - m}}{\beta} + \delta_{\beta, \beta'} \delta_{m, m'} m \zeta,
\end{equation}
with the Fourier transform of the time-periodic Hamiltonian $H(\tau)$ given by
\begin{equation}
    H^{m} = \frac{1}{T} \int^{T}_{0} \! \d \tau \, \e^{- \i m \zeta \tau} H(\tau) = [H^{- m}]^{\dagger}.
\end{equation}

Since our Hamiltonian defined by the terms in Eq.~\eqref{eq:spin-phonon-coupled-hamiltonian-terms-transformed} only contains Fourier components with index $m = 0, \pm 1, \pm 2$ (i.e., with frequencies $0$, $\pm \zeta$, and $\pm 2 \zeta$), the quasienergy operator acquires the block form
\begin{equation}\label{eq:quasi_op}
    \mathcal{H} = \begin{bmatrix}
        \ddots & \vdots & \vdots & \vdots & \vdots & \vdots & \\
        \ldots & H^{0} - 2 \zeta & H^{-1} & H^{-2} & 0 & 0 & \ldots \\
        \ldots & H^{1} & H^{0} - 1 \zeta & H^{-1} & H^{-2} & 0 & \ldots \\
        \ldots & H^{2} & H^{1} &H^{0} + 0 \zeta & H^{-1} & H^{-2} & \ldots \\
        \ldots & 0 & H^{2} & H^{1} & H^{0} + 1 \zeta & H^{-1} & \ldots \\
        \ldots & 0 & 0 & H^{2} & H^{1} & H^{0} + 2 \zeta & \ldots \\
        & \vdots & \vdots & \vdots & \vdots & \vdots & \ddots \\ 
    \end{bmatrix} \!.
\end{equation}
The Hamiltonians appearing in each of these blocks are the ones in Eq.~\eqref{eq:spin-phonon-coupled-hamiltonian}. 
Note that the quasienergy operator is exact only if an infinite number of blocks are considered, which leads to periodically repeating quasienergies $\epsilon_{\alpha, m}$.
In general, the exact diagonalization of the quasienergy operator~\eqref{eq:quasi_op} is not possible and so perturbative methods are often utilized to obtain approximate solutions (see, e.g., Ref.~\cite{eckardt2015}).
In the high-frequency regime, the diagonal blocks are widely separated and the spectrum of the quasienergy operator can be obtained via the rotating wave approximation for first order perturbations and using the Floquet-Magnus expansion for higher orders~\cite{blanes2009, mananga2011, mananga2016, kuwahara2016}.
Similarly, the low-frequency regime approximations may be calculated through adiabatic elimination or, more generally, using perturbation theory~\cite{shavitt1980, kirtman1981, sibert1988}.
Alternatively, one can employ an effective Hamiltonian obtained by truncating the number of blocks in the quasienergy operator and decoupling the Fourier modes~\cite{giovannini2019, vogl2020}.
In this work, the approximated spectrum is obtained numerically by diagonalization of the truncated quasienergy operator and analytically with the aid of time-independent perturbation theory. 

In what follows, we use a finite basis to obtain the truncated matrix representation of the quasienergy operator in Eq.~\eqref{eq:quasi_op}, which we numerically diagonalize.
The electronic degrees of freedom are represented by a fictitious spin, see Eq.~\eqref{eq:electronic-spin-states}, in a two-dimensional Hilbert space.
The infinite-dimensional state space of the vibrational degrees of freedom is truncated by choosing a basis set that includes all phonon number states, see Eq.~\eqref{eq:vibrational-modes}, from the vacuum up to a maximum occupation number $N_{\mathrm{tr}}$.
From the expansion in Eq.~\eqref{eq:floquet-basis-states}, this corresponds to choosing the product basis $\ket{\beta} = \ket{\sigma, N}$, where $\sigma = \ \uparrow, \downarrow$ labels the electronic states and $N$ the vibrational mode.
For the Floquet space (i.e., the space of square-integrable $T$-periodic time-dependent functions), without loss of generality, we center the expansion around the Fourier block with $m = 0$, as shown in Eq.~\eqref{eq:quasi_op}, and include all blocks for $|m| \leq m_\mathrm{tr}$.
When the truncation parameters $m_{\mathrm{tr}}$ and $N_{\mathrm{tr}}$ are chosen sufficiently large, the numerically calculated eigenenergies that are located in the central region of the spectrum are expected to converge to the exact quasienergies.
Note, however, that the strict periodicity of the spectrum, indicated by Eq.~\eqref{eq:quasi_energies_exact}, will be lost in this finite basis set expansion.


\section{Quasienergy spectrum and perturbation theory}
\label{sec:quasienergy-spectrum}

In the following, we compute the quasienergies and relative transition probabilities involved in the laser excitation of Rydberg states by numerically diagonalizing the truncated quasienergy operator.
Explicitly, we solve the eigenvalue problem $\mathcal{H} \! \kket{E} = E \! \kket{E}$ for the matrix in Eq.~\eqref{eq:quasi_op}.
We then compare the numeric results to analytic expressions which are calculated using time-independent perturbation theory.

\begin{figure}[t]
    \centering
    \includegraphics[scale=0.38]{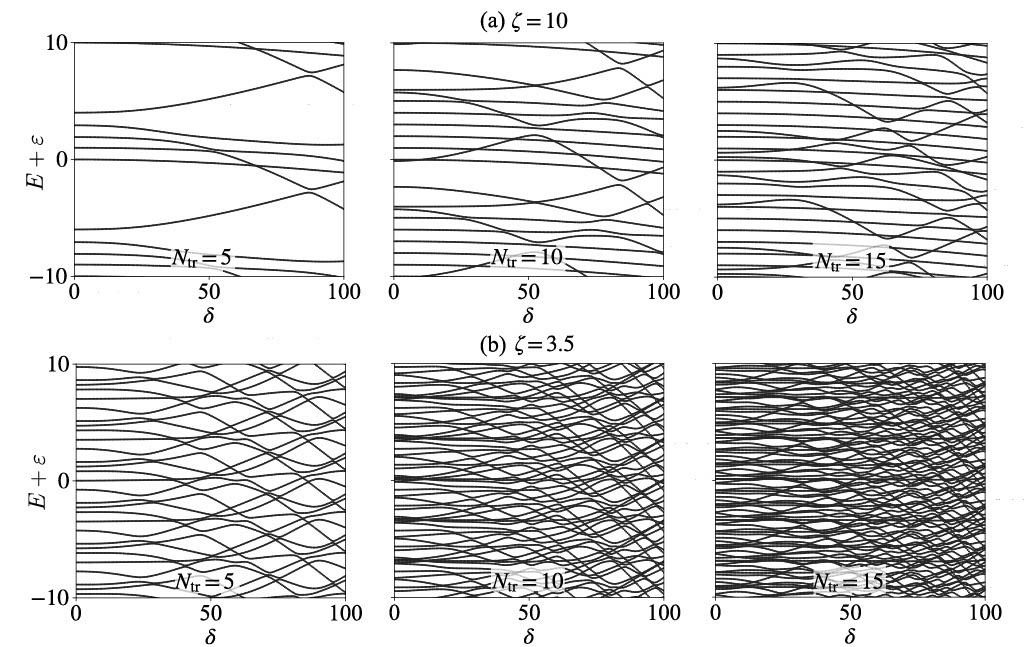}
    \caption{
        \textbf{Numerically computed spectra of the quasienergy operator as a function of the field displacement $\delta$.}
        Quasienergy spectra for two values of the oscillating field radio frequency: (a) $\zeta = 10$ and (b) $\zeta = 3.5$.
        The spectra are computed for three different values of the truncation parameter $N_{\mathrm{tr}}$, which denotes the maximum number of vibrational excitations considered for the numerical diagonalization.
        As $N_{\mathrm{tr}}$ increases, the number of quasienergy eigenstates in the shown energy interval increases.
        For details regarding the parameter values and numerical diagonalization, see Secs.~\ref{sec:generalized-quantum-rabi-model} and~\ref{sec:extended-floquet-space}, respectively.
    } 
    \label{fig:quasienergy-spectrum-delta}
\end{figure}

\begin{figure}[t]
    \centering
    \includegraphics[scale=0.38]{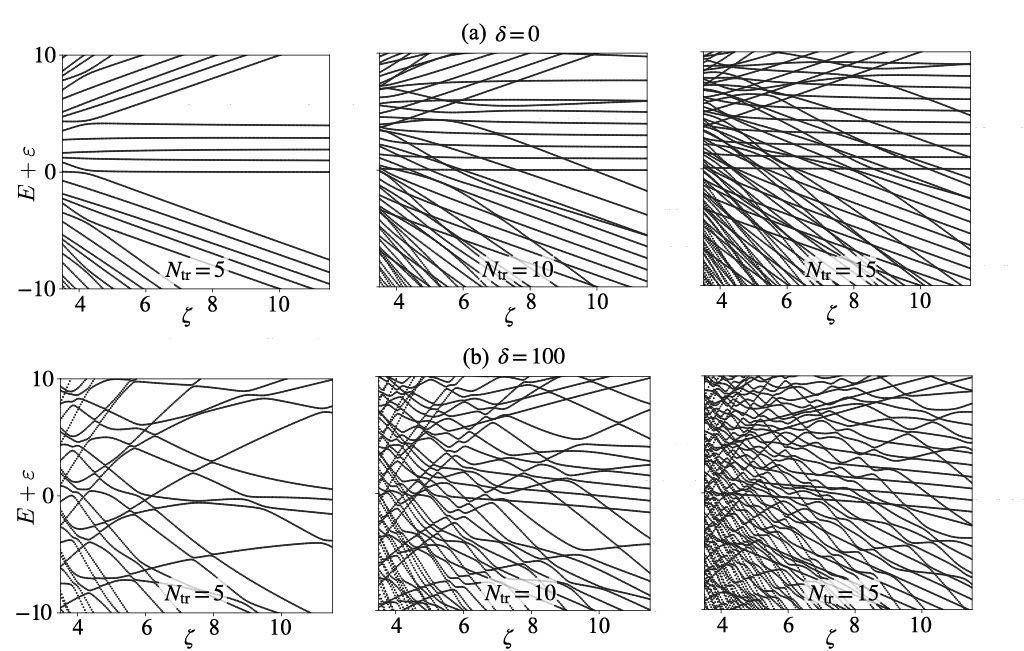}
    \caption{
        \textbf{Numerically computed spectra of the quasienergy operator as a function of the field frequency $\zeta$.}
        Spectra for two relative field displacements: (a) $\delta = 0$ and (b) $\delta = 100$.
        As in Fig.~\ref{fig:quasienergy-spectrum-delta}, spectra are computed by numerically diagonalizing the quasienergy operator.
        We show the results for three different values of the phonon number truncation parameter, $N_{\mathrm{tr}} = 5, 10, 15$.
        For further details, see Fig~\ref{fig:quasienergy-spectrum-delta}.
    }
    \label{fig:quasienergy-spectrum-eta}
\end{figure}

\subsection{Quasienergy level structure}

In Figs.~\ref{fig:quasienergy-spectrum-delta} and~\ref{fig:quasienergy-spectrum-eta}, we plot the quasienergy spectrum as we vary the relative field displacement $\delta$ and the oscillating field radio frequency $\zeta$, respectively.
The spectra are computed by numerically diagonalizing the matrix in Eq.~\eqref{eq:quasi_op}, which is truncated at $m_\mathrm{tr} = 35$.
In these figures, we focus on an energy window located about the bare energy $- \varepsilon$ of the state $\ket{\downarrow}$ so that the observable signatures of the mixing between the electronic and vibrational degrees of freedom are clearly visible.
Generally, the smaller $\zeta$ and larger $\delta$, the stronger this mixing, which manifests in the appearance of avoided crossings.
To obtain faithful quasienergies, it is necessary to increase the maximum occupation number $N_{\mathrm{tr}}$ of the phonon mode under consideration in the basis set expansion, until convergence is established.
As can be seen in Figs.~\ref{fig:quasienergy-spectrum-delta} and~\ref{fig:quasienergy-spectrum-eta}, increasing $N_{\mathrm{tr}}$ leads to a significant increase of states in the displayed energy window.
This follows from the fact that more vibrational states from other Fourier blocks enter the energy window, making it harder to interpret the data.
We therefore pursue in the following section a different route for analyzing the data.

\subsection{Relative transition probability to the Rydberg states}
\label{sec:transition-probability}

\begin{figure}[t]
    \centering
    \includegraphics[scale=0.4]{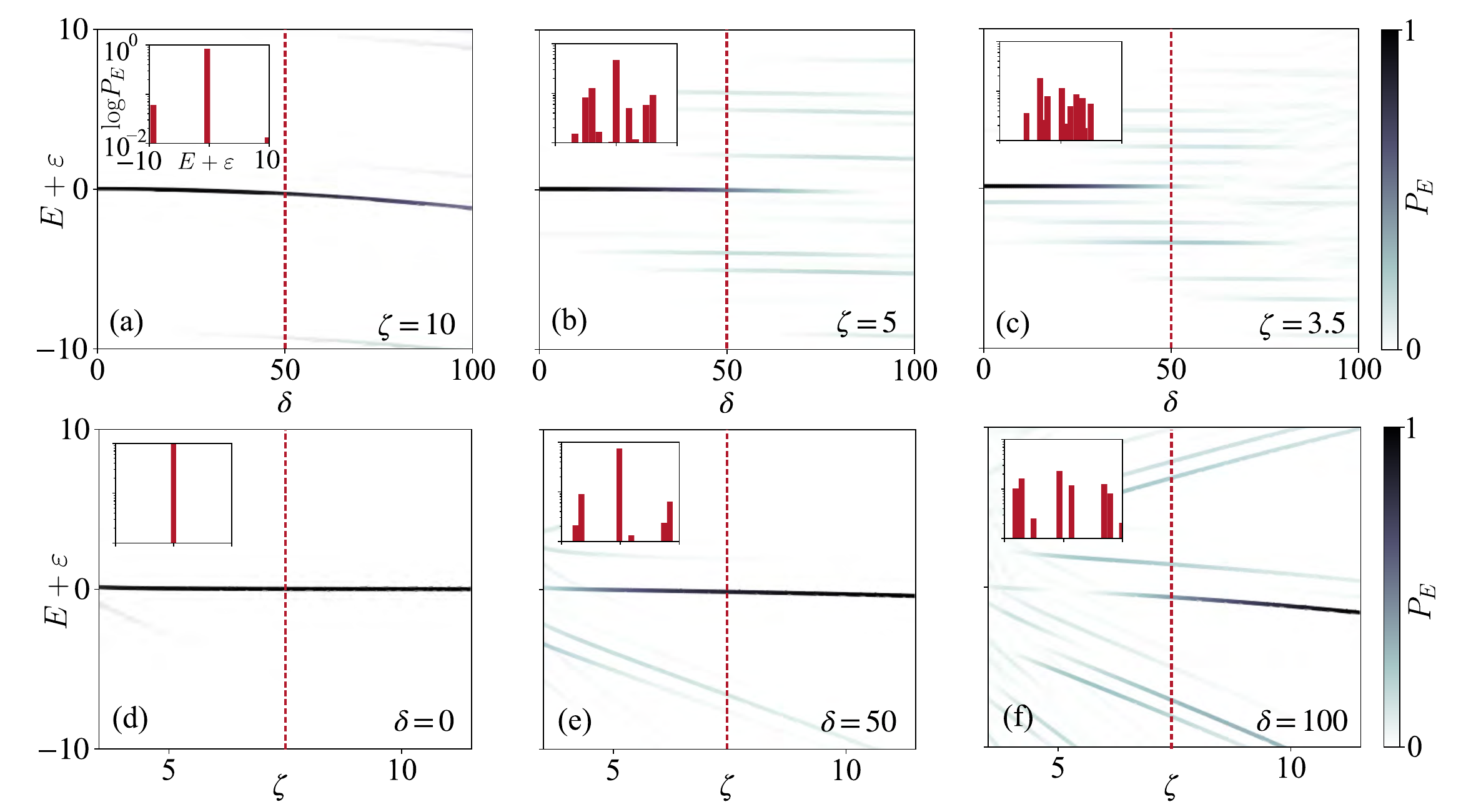}
    \caption{
        \textbf{Relative transition probability from the initial state $\kket{g, 0, 0}$ to the quasienergy eigenstates $\kket{E}$ as a function of the field displacement $\delta$ and field frequency $\zeta$.}
        The data is calculated following Eq.~\eqref{eq:state-transition-probabilities}, which assumes that initially the vibrational state is at zero temperature, $T = 0$, and contains zero phonons.
        For the smaller $\delta$ and larger $\zeta$ values considered, the relative transition probability is concentrated on a single state.
        As $\delta$ increases or $\zeta$ decreases, the coupling between the vibrational and electronic degrees of freedom leads to the emergence of sidebands.
        Eventually, the relative transition probability distributes over many states and no dominant spectral feature persists.
        The insets show a cut along the dashed line.
        The results are obtained using the same parameters as in Figs.~\ref{fig:quasienergy-spectrum-delta} and~\ref{fig:quasienergy-spectrum-eta}.
    } 
    \label{fig:quasienergy-spectrum}
\end{figure}

From an experimental perspective the quasienergy spectrum itself may actually not be very informative.
A pertinent question is rather how the Rydberg excitation spectrum changes under the coupling of the electronic states to the vibrational motion.

To understand this, we investigate which quasienergy eigenstates $\kket{E}$ can be actually excited from a given initial state (see Fig. \ref{fig:model}b).
We begin by analyzing the zero-temperature case ($T = 0$), which provides a baseline for our discussion.
Subsequently, we extend our analysis to finite temperatures ($T > 0$), examining how thermal effects modify the excitation behavior.
For simplicity, we first assume the initial state $\kket{\sigma, N, m} = \kket{g, 0, 0}$, that is, the ion is in the electronic ground state (first index), its vibrational state contains zero phonons (second index), and the Fourier index is zero.
The latter choice is, in principle, arbitrary, since the quasienergy spectrum is periodic.
However, we choose $m = 0$, since this is the Fourier block around which we expand the quasienergy operator in Eq.~\eqref{eq:quasi_op}.
Note that $\kket{g, 0, 0}$ is not strictly a state in which a ground state ion can be prepared, even at zero temperature.
The reason is that the ground state of the vibrational motion is in fact not the Fock state with zero phonons, i.e., the ground state of the Hamiltonian $H_{\mathrm{ex}}^{0}$ in Eq.~\eqref{eq:spin-phonon-coupled-hamiltonian-terms-transformed}.
Rather, due to the influence of the terms $H_{\mathrm{ex}}^{1}$ and $H_{\mathrm{ex}}^{2}$, this zero phonon Fock state is dressed through couplings to nearby Fourier blocks and vibrational states.
In App.~\ref{app:ground-state} we discuss this issue in detail.
However, the qualitative findings remain unchanged compared to the subsequent simplified discussion, which we proceed by calculating the dipole transition matrix element
\begin{equation}\label{eq:fermi-golden-rule}
    \mathcal{T} = \bbra{g, 0, 0} \! \b{d} \! \kket{E}.
\end{equation}
This quantity, which involves the dipole operator $\b{d}$, is only non-zero when transitions can occur from the chosen initial state $\kket{g, 0, 0}$ to the quasienergy eigenstate $\kket{E}$.
Note that we do not need to know the exact form of the operator $\b{d}$, which depends on the details of the employed laser excitation scheme.
For example, Rydberg states of trapped ions are often excited using multi-photon transitions~\cite{mokhberi2020}.
For strontium $^{88}\mathrm{Sr}^{+}$ ions, one utilizes the metastable state $\ket{4 D_{\smash{5/2}}}$ and intermediate state $\ket{6 P_{\smash{3/2}}}$, whilst the ground state is $\ket{g} = \ket{5 S_{\smash{1/2}}}$ (see, e.g., Ref.~\cite{zhang2020}).
However, what is essential for the upcoming analysis is that $\b{d} \kket{g, 0, 0}$ is proportional to the Rydberg $S$-state $\kket{\downarrow,0,0}$.
Therefore, in order to get an indication of which of the quasienergy eigenstates can actually be excited, we essentially only need calculate their inner product with the Rydberg $S$-state and zero phonon state.
We refer to this quantity as the relative transition probability, which at zero temperature takes the form
\begin{equation}\label{eq:state-transition-probabilities}
    P_{E} = \ab{\ipe{\downarrow, 0, 0}{E}}^{2},
\end{equation}
and assumes values between zero and one.

At finite temperature the initial state is no longer described by the pure state $\kket{g, 0, 0}$, but rather by the thermal density operator 
\begin{equation}
    \begin{aligned}
    \label{eq:thermal_state}
    \rho_{\mathrm{th}}  = \sum^{\infty}_{N = 0} \frac{\e^{-N \omega_{x}/T}}{Z} \kket{g, N, 0}\bbra{g, N, 0}, \qquad \text{with} \qquad Z = \sum^{\infty}_{N = 0} \e^{-N\omega_{x}/T},
    \end{aligned}
\end{equation}
in which $T$ is the temperature and $Z$ the partition function.
Here, we set $\hbar = k_{\mathrm{B}} = 1$.
Consequently, the excitations do not originate from a single reference energy, but rather from a distribution of initial states with energy $N\omega_{x}$ with $N \ge 0$.
The relative transition probability for a specific initial state $N$ is then given by
\begin{equation}
    p_{N} (E + N\omega_{x}) = \frac{\e^{-N \omega_{x}/T}}{Z} \ab{\ipe{\downarrow, N, 0}{E}}^{2},
\end{equation}
where we assigned the corresponding Boltzmann weight. Note that the argument of $p_{N}$ is shifted in order to account for the energy shift of the initial state. The relative transition probability is then given by 
\begin{equation}
    P_{E} (T) = \sum^{\infty}_{N = 0} p_{N} (E).
    \label{eq:relative-transition_probability-th}
\end{equation}
In the limit of very low temperatures, $T \approx 0$, we recover Eq.~\eqref{eq:state-transition-probabilities}, since the system is found in the lowest vibrational state, $N = 0$.

\begin{figure}[t]
    \centering
    \includegraphics[scale=0.38]{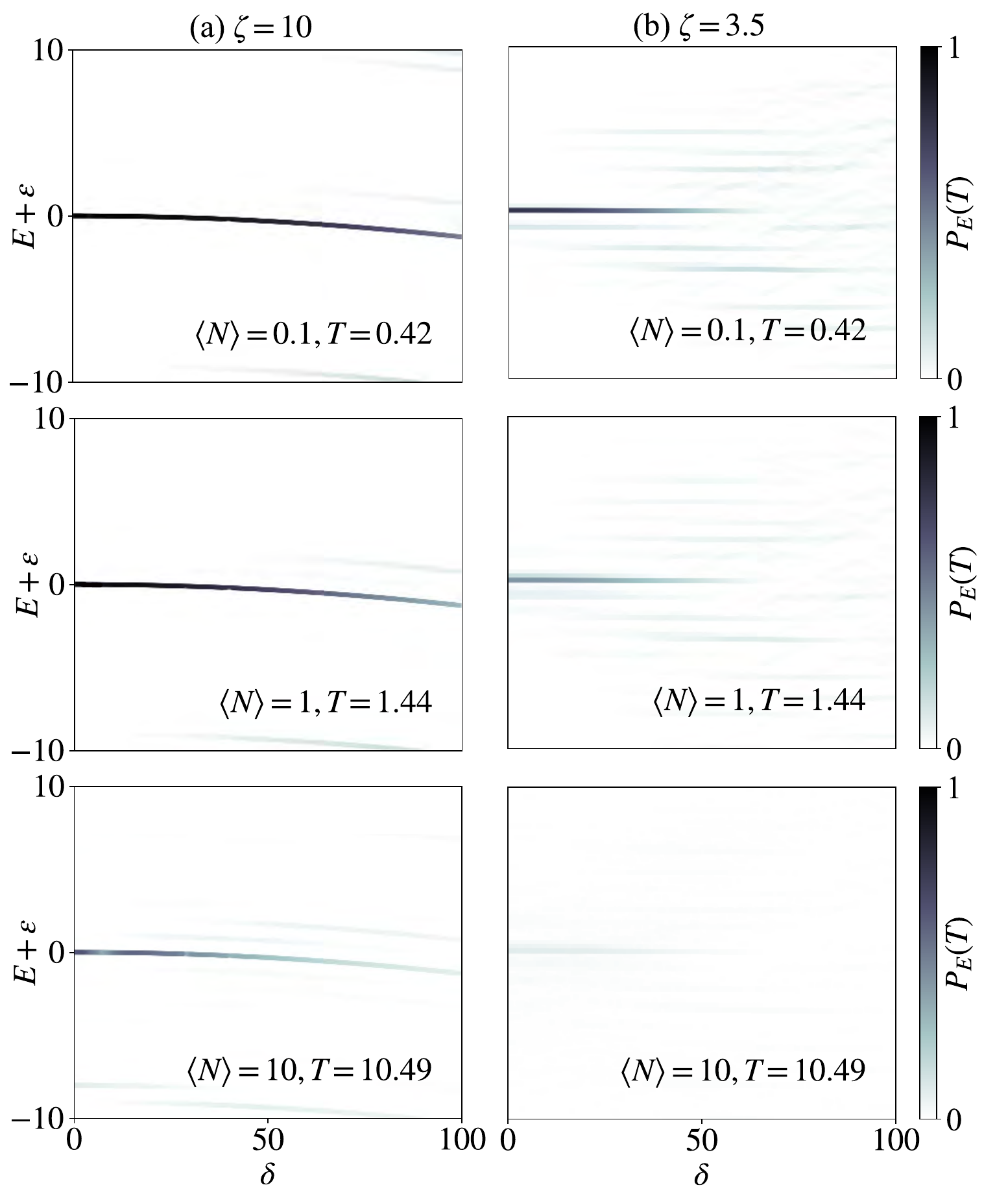}
    \caption{
        \textbf{Relative transition probability from the initial state $\rho_{\mathrm{th}}$ to the quasienergy eigenstates $\kket{E}$ as a function of the field displacement $\delta$.}
        Relative transition probability at finite temperature for two values of the oscillating field radio frequency: (a) $\zeta = 10$ and (b) $\zeta = 3.5$.
        The data is calculated following Eq.~\eqref{eq:relative-transition_probability-th}, which assumes that the occupation of the vibrational states is initially described by a thermal distribution with average phonon occupation number $\langle N \rangle$.
        As $\langle N \rangle$ increases, the relative transition probability spreads across a broader range of states with different $N$.
        The numerical methods for obtaining the relative transition probabilities and quasienergies shown in these plots are detailed in App.~\ref{app:thermal-plots}.}
    
    \label{fig:quasienergy-th-delta}
\end{figure}

\begin{figure}[t]
    \centering
    \includegraphics[scale=0.38]{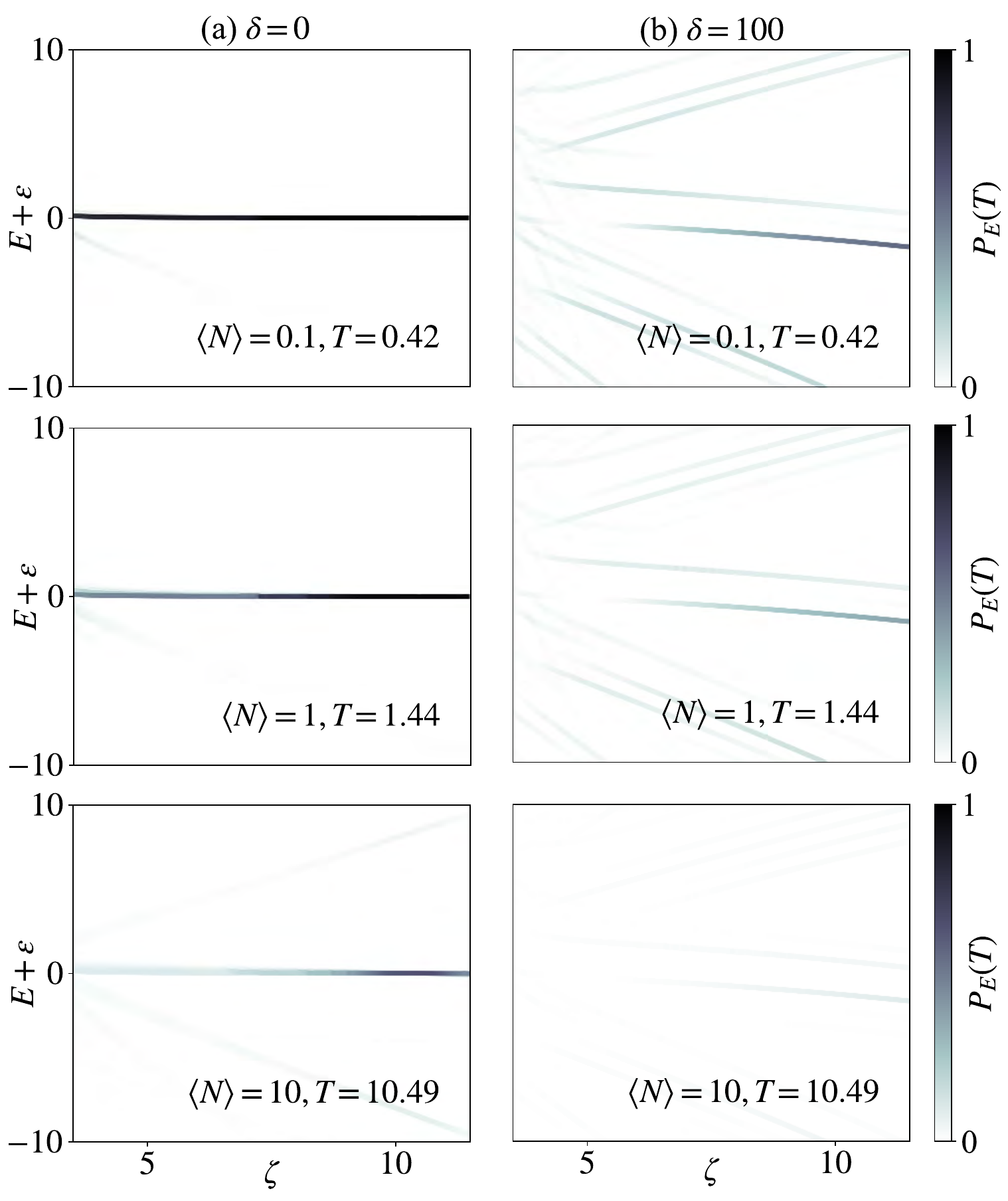}
    \caption{\textbf{Relative transition probability from the initial state $\rho_{\mathrm{th}}$ to the quasienergy eigenstates $\kket{E}$ as a function of the frequency of the oscillating field $\zeta$.}
        Relative transition probability at finite temperature for two relative field displacements: (a) $\delta = 0$ and (b) $\delta = 100$.
        The data is calculated following Eq.~\eqref{eq:relative-transition_probability-th}, which assumes that the occupation of the vibrational states is initially described by a thermal distribution  with average phonon occupation number $\langle N \rangle$.
        The phenomenology follows that presented in Fig.~\ref{fig:quasienergy-th-delta}; even for $\delta = 0$, the well resolved spectral line blurs as $\langle N \rangle$ increases.
        The numerical methods for obtaining the relative transition probabilities and associated quasienergies shown in these plots are detailed in App.~\ref{app:thermal-plots}.
    } 
    \label{fig:quasienergy-th-zeta}
\end{figure}

In Fig.~\ref{fig:quasienergy-spectrum}, we show quasienergy spectra, where each state is shaded according to its associated relative transition probability, with white and black corresponding to values of zero and one, respectively.
In order to obtain this data, we construct the quasienergy operator with the truncation parameters $m_{\mathrm{tr}} = 35$ and $N_{\mathrm{tr}} = 40$, which is sufficient to reach convergence.
For sufficiently large $\zeta$ and sufficiently small $\delta$ values, the relative transition probability is concentrated on one state, which has by far the largest overlap with the Rydberg $S$-state and zero phonon state.
Here, the quasienergy spectrum of the Rydberg ion, at least for the considered $S$-state, does not significantly deviate from the field-free case:
As can be seen in panel (a), the relative trap displacement $\delta$ merely introduces a second order Stark shift, but a clearly dominant spectral line persists in this parameter regime.
As $\zeta$ is decreased and $\delta$ increased, however, so-called sidebands emerge.
The reason is that the electronic and vibrational degrees of freedom become coupled, which allows transitions among different vibrational states to take place (i.e., phonon number changing transitions).
This goes to the point where the relative transition probabilities ``distribute'' over many states in the considered energy window and an isolated spectral line is no longer present.
In this situation, which is visible in panels (b), (c), (e), and (f), the coherent excitation of the ion to a well-defined state, i.e. spectrally isolated state, is no longer possible.
Figs.~\ref{fig:quasienergy-th-delta} and \ref{fig:quasienergy-th-zeta} illustrate the effects of finite temperature, where the average phonon occupation number is given by $\langle N \rangle = \tr \big(b^{\dagger} b \rho_{\mathrm{th}}\big)$.
As the temperature increases, spectral features become progressively smeared out.
Additionally, higher temperatures lead to the emergence of sidebands associated with phonon number changing transitions.
Experimentally, vibrational states can be laser-cooled to temperatures where 
$\langle N \rangle \ll 1$ \cite{wineland1998}, in which case finite-temperature effects on the transition probabilities remain negligible.

\subsection{Perturbation theory}\label{sec:perturbation-theory}

\begin{figure}[t]
    \centering
    \includegraphics[scale=0.38]{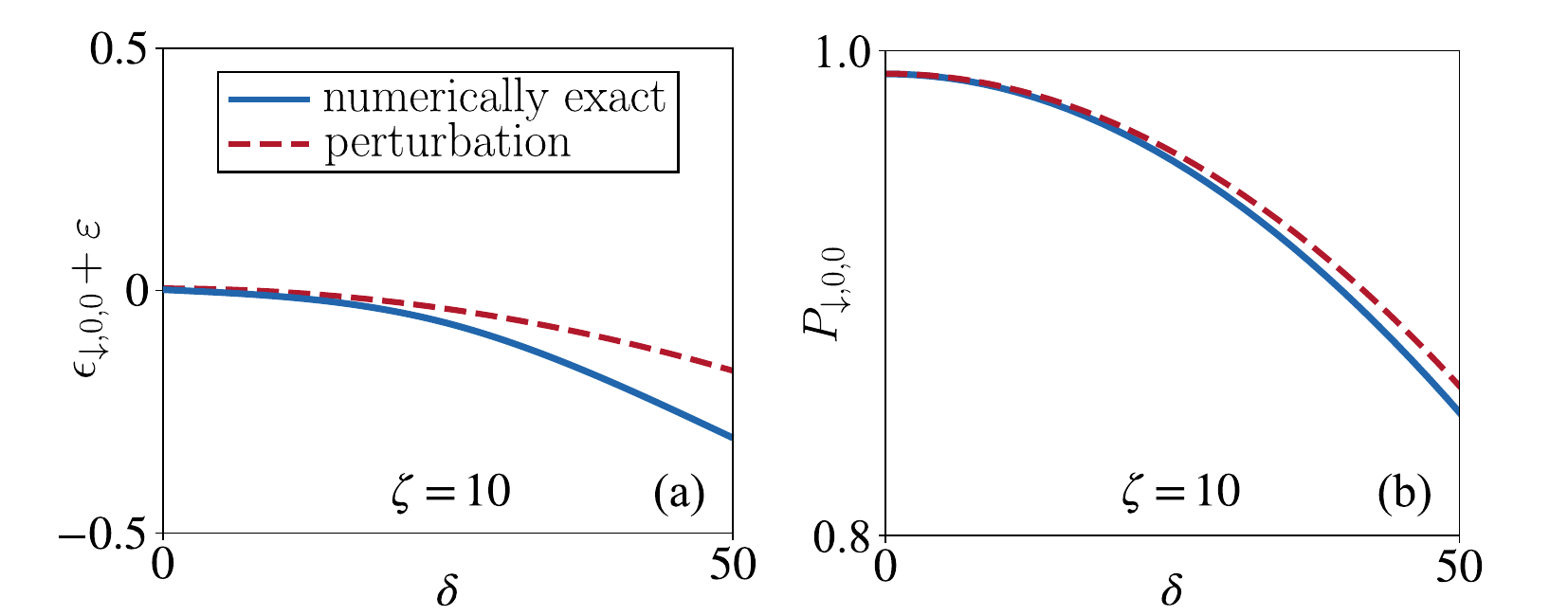}
    \caption{
        \textbf{Perturbatively calculated quasienergy and relative transition probability to the Rydberg $S$-state.}
        Quasienergy $\epsilon_{\downarrow, 0, 0}$ [panel (a)] and relative transition probability $P_{\downarrow, 0, 0}$ [panel (b)] to the state $\kket{u_{\downarrow, 0, 0}}$ obtained using numerical diagonalization (solid blue) and second order perturbation theory (dashed red).
        In the parameter regime considered, namely, large oscillating field frequency $\zeta = 10$ and field displacements $\delta \leq 50$, the two results are in good agreement.
        For larger $\delta$, contributions from couplings to higher order Floquet blocks become important, leading to nonnegligible perturbative corrections beyond second order.
    }
    \label{fig:perturbation-results}
\end{figure}

To gain some quantitative understanding of the impact of the micromotion on the quasienergy spectrum, we employ time-independent perturbation theory.
Accordingly, we consider the perturbative regime, in which the quasienergy operator is written in the form $\mathcal{H} = \mathcal{H}^{(0)} + \mathcal{V}$.
For the unperturbed quasienergy operator $\mathcal{H}^{(0)}$, we choose the terms in Eq.~\eqref{eq:spin-phonon-coupled-hamiltonian-terms-transformed} that are associated to the field-free external and internal dynamics. These are $H^0_\mathrm{ex}=b^{\dagger} b$ and the term $\varepsilon \sigma_{z}$ of $H^0_\mathrm{in}$. This yields the unperturbed quasienergy operator 
\begin{equation}
    \mathcal{H}^{(0)} = \sum_{\mathclap{m = - \infty}}^{\infty} [ \varepsilon \sigma_{z} + b^{\dagger} b + m \zeta ] \! \kket{m} \! \bbra{m} \!.
\end{equation}
The perturbation $\mathcal{V}$ then contains all remaining terms induced by the trap, see Eq.~\eqref{eq:spin-phonon-coupled-hamiltonian-terms-transformed}.
This includes those solely due to the static and oscillating electric field (the $\delta$-independent terms in $H_{\smash{\mathrm{co}}}^{0}$ and $H_{\smash{\mathrm{ex}}}^{1}$, $H_{\smash{\mathrm{co}}}^{1}$, and $H_{\smash{\mathrm{ex}}}^{2}$, respectively, the latter of which give rise to the intrinsic micromotion) and those caused by the misalignment of the two fields (the $\delta$-dependent terms in $H_{\smash{\mathrm{in}}}^{0}$, $H_{\smash{\mathrm{ex}}}^{1}$, $H_{\smash{\mathrm{in}}}^{1}$ and $H_{\smash{\mathrm{ex}}}^{2}$ that lead to the extrinsic micromotion).
Hence, in the Fourier mode basis,
\begin{equation}
\begin{aligned}
    \mathcal{V} & = \sum_{\mathclap{m = - \infty}}^{\infty} \ \bigg[ \! \bigg[ \frac{1}{4} [b^{\dagger} + b] \sigma_{x} - \frac{9 \delta}{32} \sigma_{x} \bigg] \! \kket{m} \! \bbra{m} - \bigg[ \frac{3 \zeta}{2} [b^{\dagger} + b] \sigma_{x} - \frac{3 \zeta \delta}{16} \sigma_{x} \bigg] [\kket{m + 1} \! \bbra{m} + \mathrm{H.c.}] \\
    & \qquad - \bigg[ \frac{3}{8} [b^{\dagger} + b] [b^{\dagger} - b] - \frac{3 \delta}{64} [b^{\dagger} - b] \bigg] [\kket{m + 1} \! \bbra{m} - \mathrm{H.c.}] \\
    & \qquad - \bigg[ \frac{9}{64} [b^{\dagger} + b] [b^{\dagger} + b] - \frac{9 \delta}{256} [b^{\dagger} + b] \bigg] [\kket{m + 2} \! \bbra{m} + \mathrm{H.c.}] \bigg].
\end{aligned}
\end{equation}
The eigenenergies and eigenstates of $\mathcal{H}^{(0)}$ are, respectively,
\begin{equation}\label{eq:unperturbed-values-states}
    \epsilon_{\sigma, N, m}^{(0)} = \pm \varepsilon + N + m \zeta, \qquad
    \kket{u_{\sigma, N, m}^{(0)}} = \kket{\sigma, N, m},
\end{equation}
where $\epsilon_{\sigma, N, m}^{(p)}$ and $\kket{u_{\sigma, N, m}^{(p)}}$ for $p > 0$ denote the higher order corrections to the unperturbed eigenenergies and eigenstates.
Since the perturbation $\mathcal{V}$ only possesses off-diagonal matrix elements [see Eq.~\eqref{eq:spin-phonon-coupled-hamiltonian-terms-transformed}],~the first order corrections to the eigenenergies are zero, i.e., $\epsilon_{\sigma, N, m}^{(1)} = 0$.
Moreover, due to the orthonormality of the unperturbed quasienergy eigenstates, $\ipe{u_{\smash{\sigma'\!, N'\!, m'}}^{\smash{(0)}}}{u_{\smash{\sigma, N, m}}^{\smash{(0)}}} = \ipe{\sigma'\!, N'\!, m'}{\sigma, N, m} = \delta_{\sigma, \sigma'} \delta_{N, N'} \delta_{m, m'}$, the first order corrections to the relative transition probabilities vanish.
Therefore, in order to analyze the effects of the micromotion on the Rydberg excitation spectrum shown in Fig.~\ref{fig:quasienergy-spectrum} analytically, we must necessarily consider second order corrections.

Accordingly, we define the following second order perturbed quasienergies, 
\begin{equation}
    \epsilon_{\sigma, N, m} = \epsilon_{\sigma, N, m}^{(0)} - \sum_{\mathclap{\sigma'\!, N'\!, m'}} \frac{\ab{\bbra{u_{\sigma'\!, N'\!, m'}^{(0)}} \! \mathcal{V} \! \kket{u_{\sigma, N, m}^{(0)}}}^{2}}{\epsilon_{\sigma'\!, N'\!, m'}^{(0)} - \epsilon_{\sigma, N, m}^{(0)}},
\end{equation}
together with their associated relative transition probabilities,
\begin{equation}
    \label{eq:prob}
    P_{\sigma, N, m} =
    \begin{cases}
        \displaystyle{1 - \sum_{\mathclap{\sigma'\!, N'\!, m'}} \frac{\ab{\bbra{u_{\sigma'\!, N'\!, m'}^{(0)}} \! \mathcal{V} \! \kket{u_{\downarrow, 0, 0}^{(0)}}}^{2}}{[\epsilon_{\sigma'\!, N'\!, m'}^{(0)} - \epsilon_{\downarrow, 0, 0}^{(0)}]^{2}}}, & \text{if $(\sigma, N, m) = (\downarrow, 0, 0)$}, \\
        \displaystyle{\frac{\ab{\bbra{u_{\sigma, N, m}^{(0)}} \! \mathcal{V} \! \kket{u_{\downarrow, 0, 0}^{(0)}}}^{2}}{[\epsilon_{\sigma, N, m}^{(0)} - \epsilon_{\downarrow, 0, 0}^{(0)}]^{2}}}, & \text{if $(\sigma, N, m) \neq (\downarrow, 0, 0)$}, \\
    \end{cases}
\end{equation}
where the summations are implicitly over $(\sigma'\!, N'\!, m') \neq (\sigma, N, m)$.
In the following we consider the state $\kket{\downarrow, 0, 0}$, i.e., the electron being in the Rydberg $S$-state and the ion being in the ground state of $H^{0}_{\mathrm{ex}}$.
We find that this state is perturbatively coupled to $18$ other states with $N' = 0, 1$ and $m' = 0, \pm 1, \pm 2$.
Computing its quasienergy and the relative transition probability yields
\begin{equation}\label{eq:second-order-perturbed-results}
\begin{aligned}
    \epsilon_{\downarrow, 0, 0} & = - \varepsilon - \frac{9 \zeta^{2} \delta^{2}}{256 \varepsilon} \bigg[ \! \bigg( \! \bigg[ \frac{4 \varepsilon^{2}}{4 \varepsilon^{2} - \zeta^{2}} + \frac{64}{\delta^{2}} \frac{2 \varepsilon [2 \varepsilon + 1]}{[2 \varepsilon + 1]^{2} - \zeta^{2}} \bigg] + \frac{9}{8 \zeta^{2}} \bigg[ 1 + \frac{64}{81 \delta^{2}} \frac{2 \varepsilon}{2 \varepsilon + 1} \bigg] \! \bigg) \\
    & \qquad - \frac{\varepsilon}{8 \zeta^{4}} \bigg( \! \bigg[ \frac{\zeta^{2}}{\zeta^{2} - 1} + \frac{9}{64} \frac{4 \zeta^{2}}{4 \zeta^{2} - 1} \bigg] + \frac{256}{\delta^{2}} \bigg[ \frac{\zeta^{2}}{\zeta^{2} - 4} + \frac{9}{256} \frac{\zeta^{2}}{\zeta^{2} - 1} \bigg] \! \bigg) \! \bigg], \\
    P_{\downarrow, 0, 0} & = 1 - \frac{9 \delta^{2}}{2048 \zeta^{2}} \bigg( \! \bigg[ \frac{\zeta^{2} [\zeta^{2} + 1]}{[\zeta^{2} - 1]^{2}} + \frac{9}{64} \frac{4 \zeta^{2} [4 \zeta^{2} + 1]}{[4 \zeta^{2} - 1]^{2}} \bigg] + \frac{128}{\delta^{2}} \bigg[ \frac{\zeta^{2} [\zeta^{2} + 4]}{[\zeta^{2} - 4]^{2}} + \frac{539}{1024} + \frac{9}{256} \frac{\zeta^{2} [\zeta^{2} + 1]}{[\zeta^{2} - 1]^{2}} \bigg] \! \bigg) \\
    & \qquad - \frac{72 \zeta^{4}}{64 \varepsilon^{2}} \bigg[ \frac{4 \varepsilon^{2} [[2 \varepsilon + 1]^{2} + \zeta^{2}]}{[[2 \varepsilon + 1]^{2} - \zeta^{2}]^{2}} + \frac{\delta^{2}}{64 \zeta^{2}} \bigg( \frac{4 \varepsilon^{2} [4 \varepsilon^{2} + \zeta^{2}]}{[4 \varepsilon^{2} - \zeta^{2}]^{2}} + \frac{9}{8 \zeta^{2}} \bigg[ 1 + \frac{64}{81 \delta^{2}} \frac{4 \varepsilon^{2}}{[2 \varepsilon + 1]^{2}} \bigg] \! \bigg) \! \bigg],
\end{aligned}
\end{equation}
where for clarity, we have grouped terms to make clearer the dependence on the relative field displacement $\delta$ and oscillating field frequency $\zeta$.
In Fig.~\ref{fig:perturbation-results}, we plot these expressions together with the values computed by numerical diagonalizing the truncated quasienergy operator, in a regime where perturbation theory is expected to be valid, i.e., for sufficiently large $\zeta$ and sufficiently small $\delta$.

These expressions can be further simplified by exploiting that for typical parameters we have that $\varepsilon \gg \zeta$ and $\zeta \gg 1$.
This yields
\begin{equation}\label{eq:approx}
\begin{aligned}
    \epsilon_{\downarrow, 0, 0} & \approx - \varepsilon - \frac{9}{4} \frac{\zeta^{2}}{\varepsilon} + \frac{2385}{2048} \frac{1}{\zeta^{2}} + \delta^{2} \bigg[ \frac{657}{131072} \frac{1}{\zeta^{2}} - \frac{9}{256} \frac{\zeta^{2}}{\varepsilon} \bigg], \\
    P_{\downarrow, 0, 0} & \approx 1 - \frac{14391}{16384} \frac{1}{\zeta^{2}} - \frac{657}{131072} \frac{\delta^{2}}{\zeta^{2}}.
\end{aligned}
\end{equation}
From these expressions, it becomes apparent that the shift of the quasienergy stems from both the coupling between the Rydberg $S$- and $P$-state (with energy separation $2 \varepsilon$) as well as from couplings to nearby Floquet blocks (separated by the radio frequency $\zeta$).
The relative transition probability $P_{\downarrow, 0, 0}$, on the other hand, is dominated by couplings to nearby Floquet blocks, as the approximate expression does not depend of the energy $\varepsilon$.


\section{Conclusions}\label{sec:conclusions}

In this work, we investigated the effects of micromotion on the Rydberg excitation spectrum of an ion in a linear Paul trap.
In particular, we considered extrinsic micromotion resulting from the relative displacement of the symmetry axes of the confining electric fields.
Such displacement can arise due to stray electric fields from imperfections in trap fabrication or calibration, but is inevitable in 2D and 3D trapped ion crystals.
To do so, we derived a minimal model Hamiltonian for the trapped Rydberg ion, which includes a Rydberg $S$- and $P$-state as well as one vibrational mode.
The corresponding time-periodic Hamiltonian was solved numerically using Floquet theory and analytically via perturbation theory.
This allowed us to gain a qualitative assessment of the impact of micromotion on the relative transition probabilities and their associated quasienergies.
We showed that when the electric fields are misaligned, the Rydberg states remain spectrally isolated, provided that the radio frequency of the oscillating electric field is sufficiently large.
These effects remain unchanged for temperatures relevant to experimental conditions.
For this regime we calculated perturbative expressions for the quasienergy of the Rydberg $S$-state and the relative transition probability for a laser excitation from the electronic ground state.

Building on our results, a number of future research directions may be explored.
These concern, for example, the inclusion of more than two electronic Rydberg states (e.g., within the same Zeeman manifold) and more trapped ions (e.g., in 2D and 3D configurations).
The former would yield a more detailed understanding of how the bound electronic quantum numbers and the couplings between electronic and vibrational dynamics are influenced by asymmetric electric potentials (e.g., due to stray electric fields or off-axis ion positions).
In contrast, the latter would allow one to study the effects of the micromotion on the interactions between the trapped Rydberg ions (e.g., due to the electric dipole-dipole and van der Waals forces).
Exploring these questions necessitates the augmentation of our model to include higher dimensional electronic and vibrational state spaces. 
The resulting generalized spin-boson models represent intriguing systems in their own right and demonstrate the capability of trapped Rydberg ions to serve as quantum simulators for complex many-body dynamics.

\textbf{Data access:} The codes used to produce the data supporting the findings within this manuscript are available on Zenodo~\cite{martins2024}.


\section*{Acknowledgements}

We acknowledge funding from the Deutsche Forschungsgemeinschaft (DFG, German Research Foundation) under Project Nos. 428276754 and 435696605 and through the Research Unit FOR 5413/1, Grant No. 465199066. We also acknowledge funding from EPSRC (Grant No. EP/W015641/1).
This project also received funding from the European Union’s Horizon Europe research and innovation program under Grant Agreement No. 101046968 (BRISQ).
This work was additionally supported by the University of Nottingham and the University of T\"{u}bingen’s funding as part of the Excellence Strategy of the German Federal and State Governments.
JW was supported by the University of T\"{u}bingen through a Research@Tübingen fellowship.
MH was supported by the Knut \& Alice Wallenberg Foundation through the Wallenberg Centre for Quantum Technology [WACQT] and by the Swedish Research Council (grant 2021-05811).

\renewcommand{\thesubsection}{\Alph{subsection}}
\renewcommand{\subsectionmark}[1]{\markboth{\MakeUppercase{\thesubsection\quad#1}}{}}

\numberwithin{equation}{subsection}
\numberwithin{figure}{subsection}

\appendix

\addcontentsline{toc}{section}{Appendices}

\section*{Appendices}

\subsection{Derivation of the trapped Rydberg ion Hamiltonian}\label{app:derivation}

In this appendix, we detail the derivation of the trapped Rydberg ion Hamiltonian from which the~spin-phonon coupled model Hamiltonian in Eq.~\eqref{eq:spin-phonon-coupled-hamiltonian} is obtained.
The starting point is the oscillating frame Hamiltonian represented in terms of the center of mass and relative coordinates given in Eq.~\eqref{eq:unitary-transformation},
\begin{equation}\label{eq:app-full-hamiltonian}
    H(t) = H_{\mathrm{ex}}(t) + H_{\mathrm{in}}(t) + H_{\mathrm{co}}(t).
\end{equation}
Expressing the oscillating electric field trigonometric functions as exponentials and separating the terms, the time-dependent Hamiltonian describing the external dynamics is rewritten as
\begin{equation}
    H_{\mathrm{ex}}(t) = H_{\mathrm{ex}}^{0} + [\e^{\i \nu t} - \e^{-\i \nu t}] H_{\mathrm{ex}}^{1} + [\e^{\i 2 \nu t} + \e^{-\i 2 \nu t}] H_{\mathrm{ex}}^{2},
\end{equation}
where the time-independent terms,
\begin{equation}
\begin{gathered}
    H_{\mathrm{ex}}^{0} = \frac{1}{2 M} \sum_{u} [P_{u}^{2} + M^{2} \omega_{u}^{2} R_{u}^{2}] + 2 e B [[1 + \epsilon] x_{\mathrm{dc}} R_{x} + [1 - \epsilon] y_{\mathrm{dc}} R_{y}], \\
    H_{\mathrm{ex}}^{1} = \frac{\i e A}{M \nu} [R_{x} P_{x} - R_{y} P_{y}], \qquad
    H_{\mathrm{ex}}^{2} = - \frac{e^{2} A^{2}}{2 M \nu^{2}} [R_{x}^{2} + R_{y}^{2}].
\end{gathered}
\end{equation}
Similarly, for the terms governing the time-dependent internal dynamics we have
\begin{equation}
    H_{\mathrm{in}}(t) = H_{\mathrm{in}}^{0} + [\e^{\i \nu t} + \e^{-\i \nu t}] H_{\mathrm{in}}^{1},
\end{equation}
where
\begin{equation}
\begin{gathered}
    H_{\mathrm{in}}^{0} = \frac{\b{p}^{2}}{2 m} + V(\ab{\b{r}}) + e B [r_{x}^{2} + r_{y}^{2} - 2 r_{z}^{2}] + \epsilon e B [r_{x}^{2} - r_{y}^{2}] - 2 e B [[1 + \epsilon] x_{\mathrm{dc}} r_{x} + [1 - \epsilon] y_{\mathrm{dc}} r_{y}], \\
    H_{\mathrm{in}}^{1} = - \frac{e A}{2} [r_{x}^{2} - r_{y}^{2}].
\end{gathered}
\end{equation}
Finally, for the terms accounting for the time-dependent coupled dynamics one finds
\begin{equation}
    H_{\mathrm{co}}(t) = H_{\mathrm{co}}^{0} + [\e^{\i \nu t} + \e^{-\i \nu t}] H_{\mathrm{co}}^{1},
\end{equation}
with
\begin{equation}
    H_{\mathrm{co}}^{0} = 2 e B [[1 + \epsilon] R_{x} r_{x} + [1 - \epsilon] R_{y} r_{y} - 2 R_{z} r_{z}], \qquad
    H_{\mathrm{co}}^{1} = - e A [R_{x} r_{x} - R_{y} r_{y}].
\end{equation}

In order to simplify these expressions, we parametrize the oscillating and static electric field gradients in terms of the natural units,
\begin{equation}
    A = \frac{\kappa \hbar \nu}{4 e \ell_{z}^{2}}, \qquad
    B = \frac{\hbar \omega_{z}}{8 e \ell_{z}^{2}},
\end{equation}
with $\kappa \equiv \sqrt{\smash{\gamma}_{\smash{x}}^{2} + \smash{\gamma}_{\smash{y}}^{2} + 1}$ and where the lengths $\ell_{u} = \sqrt{\hbar / 2 M \omega_{u}}$ and the strengths $\gamma_{u} \equiv \omega_{u} / \omega_{z} = \ell_{z}^{2} / \ell_{u}^{2}$.
Representing the internal and external degrees of freedom respectively in terms of the energy eigenbases of the field-free electronic and vibrational Hamiltonians introduced in Eqs.~\eqref{eq:electronic-states} and~\eqref{eq:vibrational-modes} and expressing the Hamiltonian in units of the natural vibrational energy $\hbar \omega_{z}$, it follows that the terms describing the external vibrational dynamics can be rewritten as
\begin{equation}\label{eq:app-external-dynamics}
\begin{aligned}
    \frac{H_{\mathrm{ex}}^{0}}{\hbar \omega_{z}} & = \gamma_{x} a_{x}^{\dagger} a_{x} + \gamma_{y} a_{y}^{\dagger} a_{y} + a_{z}^{\dagger} a_{z} + \frac{1 + \epsilon}{4 \sqrt{\gamma_{\smash{x}}}} \frac{x_{\mathrm{dc}}}{\ell_{z}} [a_{x}^{\dagger} + a_{x}] + \frac{1 - \epsilon}{4 \sqrt{\gamma_{\smash{y}}}} \frac{y_{\mathrm{dc}}}{\ell_{z}} [a_{y}^{\dagger} + a_{y}], \\
    \frac{H_{\mathrm{ex}}^{1}}{\hbar \omega_{z}} & = - \frac{\kappa}{4} [a_{x}^{\dagger} + a_{x}] [a_{x}^{\dagger} - a_{x}] + \frac{\kappa}{4} [a_{y}^{\dagger} + a_{y}] [a_{y}^{\dagger} - a_{y}], \\
    \frac{H_{\mathrm{ex}}^{2}}{\hbar \omega_{z}} & = - \frac{\kappa^{2}}{16 \gamma_{x}} [a_{x}^{\dagger} + a_{x}] [a_{x}^{\dagger} + a_{x}] - \frac{\kappa^{2}}{16 \gamma_{y}} [a_{y}^{\dagger} + a_{y}] [a_{y}^{\dagger} + a_{y}].
\end{aligned}
\end{equation}
Here, $a_{u}^{\dagger}$ and $a_{u}$ are the creation and annihilation operators introduced in Eq.~\eqref{eq:bosonic-operators}.
In order to ascertain the expressions for the internal and coupled terms, we expand the relative coordinates (i.e., the relative position operators $r_{u}^{k}$ where $k = 1, 2$ denotes the power and $u = x, y, z$ the axis) in terms of the electronic basis states.
Exploiting the fact that the effective central potential $V(\ab{\b{r}})$ approximating the electrostatic interaction between the valence electron and ionic core conserves the total angular momentum, it follows that the electronic basis states can be factored into their radial and angular components~\cite{schmidtkaler2011},
\begin{equation}
    \ket{\b{n}} = \ket{n, l, s, j, m_{j}} = \ket{n, l, s, j} \! \ket{l, s, j, m_{j}} \!.
\end{equation}
The angular components can then be factored further into their orbital and spin parts,
\begin{equation}
    \ket{l, s, j, m_{j}} = \sum_{\mathclap{m_{s} = -s}}^{s} \! \ip{l, m_{j} - m_{s}, s, m_{s}}{j, m_{j}} \! \ket{l, m_{j} - m_{s}} \! \ket{s, m_{s}} \!,
\end{equation}
where $\ip{l, m_{l}, s, m_{s}}{j, m_{j}}$ are Clebsch-Gordon coefficients with $m_{l}$ and $m_{s}$ the orbital and spin magnetic quantum numbers, respectively~\cite{friedrich2017} (see also Ref.~\cite{wilkinson2024b} for more details relevant to trapped Rydberg ions).
We then represent the relative position operators in spherical polar coordinates and subsequently employ the separation of radial and angular variables (i.e., $r$ and $\theta, \varphi$), with the latter expressed in terms of the spherical harmonics $Y_{\smash{k}}^{\smash{m_{k}}} \equiv Y_{\smash{k}}^{\smash{m_{k}}}(\theta, \varphi)$.
The terms governing the internal electronic dynamics are then
\begin{equation}\label{eq:app-internal-dynamics}
\begin{aligned}
    \frac{H_{\smash{\mathrm{in}}}^{0}}{\hbar \omega_{z}} & = \sum_{\mathclap{\b{n}}} \frac{E_{\b{n}}}{\hbar \omega_{z}} \! \op{\b{n}}{\b{n}} - \frac{\sqrt{\pi}}{2 \sqrt{5}} \sum_{\mathclap{\b{n}, \b{m}}} \frac{\me{\b{m}}{r^{2}}{\b{n}}}{\ell_{z}^{2}} \! \me{\b{m}}{Y_{2}^{0}}{\b{n}} \! \op{\b{m}}{\b{n}} \\
    & \qquad + \frac{\sqrt{\pi} \epsilon}{2 \sqrt{30}} \sum_{\mathclap{\b{n}, \b{m}}} \frac{\me{\b{m}}{r^{2}}{\b{n}}}{\ell_{z}^{2}} [\me{\b{m}}{Y_{2}^{-2}}{\b{n}} + \me{\b{m}}{Y_{2}^{2}}{\b{n}}] \! \op{\b{m}}{\b{n}} \\
    & \qquad - \frac{\sqrt{\pi} [1 + \epsilon]}{2 \sqrt{6}} \frac{x_{\mathrm{dc}}}{\ell_{z}} \sum_{\mathclap{\b{n}, \b{m}}} \frac{\me{\b{n}}{r}{\b{m}}}{\ell_{z}} [\me{\b{m}}{Y_{1}^{-1}}{\b{n}} - \me{\b{m}}{Y_{1}^{1}}{\b{n}}] \! \op{\b{m}}{\b{n}} \\
    & \qquad - \frac{\i \sqrt{\pi} [1 - \epsilon]}{2 \sqrt{6}} \frac{y_{\mathrm{dc}}}{\ell_{z}} \sum_{\mathclap{\b{n}, \b{m}}} \frac{\me{\b{n}}{r}{\b{m}}}{\ell_{z}} [\me{\b{m}}{Y_{1}^{-1}}{\b{n}} + \me{\b{m}}{Y_{1}^{1}}{\b{n}}] \! \op{\b{m}}{\b{n}} \!, \\
    \frac{H_{\smash{\mathrm{in}}}^{1}}{\hbar \omega_{z}} & = - \frac{\sqrt{\pi} \kappa}{2 \sqrt{30}} \frac{\nu}{\omega_{z}} \sum_{\mathclap{\b{n}, \b{m}}} \frac{\me{\b{m}}{r^{2}}{\b{n}}}{\ell_{z}^{2}} [\me{\b{m}}{Y_{2}^{-2}}{\b{n}} + \me{\b{m}}{Y_{2}^{-2}}{\b{n}}] \! \op{\b{m}}{\b{n}} \!,
\end{aligned}
\end{equation}
while those accounting for the coupled vibronic dynamics follow as
\begin{equation}\label{eq:app-coupled-dynamics}
\begin{aligned}
    \frac{H_{\mathrm{co}}^{0}}{\hbar \omega_{z}} & = \frac{\sqrt{\pi} [1 + \epsilon]}{2 \sqrt{6 \gamma_{\smash{x}}}} [a_{x}^{\dagger} + a_{x}] \sum_{\mathclap{\b{n}, \b{m}}} \frac{\me{\b{m}}{r}{\b{n}}}{\ell_{z}} [\me{\b{m}}{Y_{1}^{-1}}{\b{n}} - \me{\b{m}}{Y_{1}^{1}}{\b{n}}] \! \op{\b{m}}{\b{n}} \\
    & \qquad + \frac{\i \sqrt{\pi} [1 - \epsilon]}{2 \sqrt{6 \gamma_{\smash{y}}}} [a_{y}^{\dagger} + a_{y}] \sum_{\mathclap{\b{n}, \b{m}}} \frac{\me{\b{m}}{r}{\b{n}}}{\ell_{z}} [\me{\b{m}}{Y_{1}^{-1}}{\b{n}} + \me{\b{m}}{Y_{1}^{1}}{\b{n}}] \! \op{\b{m}}{\b{n}} \\
    & \qquad - \frac{\sqrt{\pi}}{\sqrt{3}} [a_{z}^{\dagger} + a_{z}] \sum_{\mathclap{\b{n}, \b{m}}} \frac{\me{\b{m}}{r}{\b{n}}}{\ell_{z}} \! \me{\b{m}}{Y_{1}^{0}}{\b{n}} \! \op{\b{m}}{\b{n}} \!, \\
    \frac{H_{\mathrm{co}}^{1}}{\hbar \omega_{z}} & = - \frac{\sqrt{\pi} \kappa}{2 \sqrt{6 \gamma_{\smash{x}}}} \frac{\nu}{\omega_{z}} [a_{x}^{\dagger} + a_{x}] \sum_{\mathclap{\b{n}, \b{m}}} \frac{\me{\b{m}}{r}{\b{n}}}{\ell_{z}} [\me{\b{m}}{Y_{1}^{-1}}{\b{n}} - \me{\b{m}}{Y_{1}^{1}}{\b{n}}] \! \op{\b{m}}{\b{n}} \\
    & \qquad + \frac{\i \sqrt{\pi} \kappa}{2 \sqrt{6 \gamma_{\smash{y}}}} \frac{\nu}{\omega_{z}} [a_{y}^{\dagger} + a_{y}] \sum_{\mathclap{\b{n}, \b{m}}} \frac{\me{\b{m}}{r}{\b{n}}}{\ell_{z}} [\me{\b{m}}{Y_{1}^{-1}}{\b{n}} + \me{\b{m}}{Y_{1}^{1}}{\b{n}}] \! \op{\b{m}}{\b{n}} \!.
\end{aligned}
\end{equation}

\subsection{Derivation of the spin-phonon coupled model Hamiltonian}\label{app:model}

In this appendix we develop a simplified model that captures the essential physics of the trapped Rydberg ion problem.

For the electronic state space, we restrict ourselves to the pair of low angular momentum energetically isolated Rydberg states given in Eq.~\eqref{eq:electronic-spin-states}.
Remarking that the total angular momentum quantum number $j = 1/2$ for these states, it follows that the angular matrix elements corresponding to electric transition quadrupole moments are zero~\cite{wilkinson2024b}.
Indeed, the only nonzero angular matrix elements are
\begin{equation}
    \me{\uparrow}{Y_{1}^{1}}{\downarrow} = - \! \me{\downarrow}{Y_{1}^{-1}}{\uparrow} = \frac{1}{\sqrt{6 \pi}},
\end{equation}
for which the corresponding radial matrix element $\me{\uparrow}{r}{\downarrow} = \me{\downarrow}{r}{\uparrow} = -\ab{\!\me{\uparrow}{r}{\downarrow}\!}$.
The Hamiltonian terms governing the internal dynamics are then
\begin{equation}\label{eq:app-internal-dynamics-restricted}
    \frac{H_{\smash{\mathrm{in}}}^{0}}{\hbar \omega_{z}} = \frac{E_{\uparrow} - E_{\downarrow}}{2 \hbar \omega_{z}} \sigma_{z} - \frac{1 + \epsilon}{4 \gamma_{\smash{x}}} \lambda_{x} \delta_{x} \sigma_{x} + \frac{1 - \epsilon}{4 \gamma_{\smash{y}}} \lambda_{y} \delta_{y} \sigma_{y},
\end{equation}
with $H_{\smash{\mathrm{in}}}^{0} = 0$, where we have introduced the electronic state (i.e., spin-$1/2$) Pauli operators,
\begin{equation}
    \sigma_{x} = \op{\uparrow}{\downarrow} + \op{\downarrow}{\uparrow} \!, \qquad
    \sigma_{y} = - \i \! \op{\uparrow}{\downarrow} + \i \! \op{\downarrow}{\uparrow} \!, \qquad
    \sigma_{z} = \op{\uparrow}{\uparrow} - \op{\downarrow}{\downarrow} \!,
\end{equation}
and for convenience shifted the electronic zero-point energy.
Furthermore, we have defined the following dimensionless (i.e., relative) quantities associated to the relative displacement of the static and oscillating electric fields and the electric dipole matrix elements between the Rydberg states, respectively,
\begin{equation}
    \delta_{x} \equiv \frac{\sqrt{\gamma_{x}} x_{\mathrm{dc}}}{\ell_{z}}, \qquad
    \delta_{y} \equiv \frac{\sqrt{\gamma_{y}} y_{\mathrm{dc}}}{\ell_{z}}, \qquad
    \lambda_{x} \equiv \frac{\sqrt{\gamma_{x}} \ab{\!\me{\uparrow}{r}{\downarrow}\!}}{3 \ell_{z}}, \qquad
    \lambda_{y} \equiv \frac{\sqrt{\gamma_{y}} \ab{\!\me{\uparrow}{r}{\downarrow}\!}}{3 \ell_{z}}.
\end{equation}
Similarly, the Hamiltonian terms describing the coupled dynamics simplify to,
\begin{equation}\label{eq:app-coupled-dynamics-restricted}
\begin{aligned}
    \frac{H_{\mathrm{co}}^{0}}{\hbar \omega_{z}} & = \frac{1 + \epsilon}{4 \gamma_{x}} \lambda_{x} [a_{x}^{\dagger} + a_{x}] \sigma_{x} + \frac{1 - \epsilon}{4 \gamma_{y}} \lambda_{y} [a_{y}^{\dagger} + a_{y}] \sigma_{y}, \\
    \frac{H_{\mathrm{co}}^{1}}{\hbar \omega_{z}} & = - \frac{\kappa}{4} \lambda_{x} \zeta_{x} [a_{x}^{\dagger} + a_{x}] \sigma_{x} + \frac{\kappa}{4} \lambda_{y} \zeta_{y} [a_{y}^{\dagger} + a_{y}] \sigma_{y},
\end{aligned}
\end{equation}
where we have further introduced the relative quantities relating the frequency of the oscillating electric quadrupole potential to that of the static harmonic trapping potential,
\begin{equation}
    \zeta_{x} \equiv \frac{\nu}{\omega_{x}}, \qquad
    \zeta_{y} \equiv \frac{\nu}{\omega_{y}}.
\end{equation}
Note that due to the restriction of the electronic states, the vibrational motion along the $z$-axis decouples from the dynamics, hence, it can be neglected allowing the Hamiltonian terms accounting for the external vibrational dynamics to be reduced to
\begin{equation}\label{eq:app-external-dynamics-restricted}
\begin{aligned}
    \frac{H_{\mathrm{ex}}^{0}}{\hbar \omega_{z}} & = \gamma_{x} a_{x}^{\dagger} a_{x} + \gamma_{y} a_{y}^{\dagger} a_{y} + \frac{1 + \epsilon}{4 \gamma_{x}} \delta_{x} [a_{x}^{\dagger} + a_{x}] + \frac{1 - \epsilon}{4 \gamma_{y}} \delta_{y} [a_{y}^{\dagger} + a_{y}], \\
    \frac{H_{\mathrm{ex}}^{1}}{\hbar \omega_{z}} & = - \frac{\kappa}{4} [a_{x}^{\dagger} + a_{x}] [a_{x}^{\dagger} - a_{x}] + \frac{\kappa}{4} [a_{y}^{\dagger} + a_{y}] [a_{y}^{\dagger} - a_{y}], \\
    \frac{H_{\mathrm{ex}}^{2}}{\hbar \omega_{z}} & = - \frac{\kappa^{2}}{16 \gamma_{x}} [a_{x}^{\dagger} + a_{x}] [a_{x}^{\dagger} + a_{x}] - \frac{\kappa^{2}}{16 \gamma_{y}} [a_{y}^{\dagger} + a_{y}] [a_{y}^{\dagger} + a_{y}].
\end{aligned}
\end{equation}

For the vibrational mode spaces, we limit our analysis to one spatial direction (i.e., the $x$-axis).
Since we are now neglecting motion along the $y$-axis, it is beneficial to define the terms in units of the natural vibrational energy $\hbar \omega_{x}$ rather than $\hbar \omega_{z}$.
Similarly, it is helpful to assume that the strength of the trapping along the $x$- and $y$-axes is identical (i.e., $\epsilon = 0$), which implies $\gamma \equiv \gamma_{x} = \gamma_{y}$.
This subsequently suggests that we can omit the subscript $x$ on the relative lengths $\delta \equiv \delta_{x}$ and $\lambda \equiv \lambda_{x}$ and frequency $\zeta \equiv \zeta_{x}$.
The latter then indicates that it is sensical to further define the relative time and energy,
\begin{equation}
    \tau \equiv \omega_{x} t, \qquad
    \varepsilon \equiv \frac{E_{\uparrow} - E_{\downarrow}}{2 \hbar \omega_{x}}.
\end{equation}
Taking this all together, the \textit{dimensionless} model Hamiltonian can be decomposed akin to Eq.~\eqref{eq:app-full-hamiltonian} as
\begin{equation}\label{eq:app-model-hamiltonian}
    H(\tau) = H_{\mathrm{ex}}(\tau) + H_{\mathrm{in}} + H_{\mathrm{co}}(\tau),
\end{equation}
where the time-dependent external, internal, and coupled Hamiltonians read
\begin{equation}\label{eq:app-model-hamiltonian-parts}
\begin{gathered}
    H_{\mathrm{ex}}(\tau) = H_{\mathrm{ex}}^{0} + [\e^{\i \zeta \tau} - \e^{- \i \zeta \tau}] H_{\mathrm{ex}}^{1} + [\e^{\i 2 \zeta \tau} + \e^{- \i 2 \zeta \tau}] H_{\mathrm{ex}}^{2}, \\
    H_{\mathrm{in}} = H_{\smash{\mathrm{in}}}^{0}, \qquad
    H_{\mathrm{co}}(\tau) = H_{\mathrm{co}}^{0} + [\e^{\i \zeta \tau} + \e^{- \i \zeta \tau}] H_{\mathrm{co}}^{1},
\end{gathered}
\end{equation}
whilst their respective time-independent components are given by
\begin{equation}\label{eq:app-model-hamiltonian-terms}
\begin{gathered}
    H_{\mathrm{ex}}^{0} = a^{\dagger} a + \frac{\delta}{4 \gamma^{2}} [a^{\dagger} + a], \qquad
    H_{\smash{\mathrm{in}}}^{0} = \varepsilon \sigma_{z} - \frac{\lambda \delta}{4 \gamma^{2}} \sigma_{x}, \qquad
    H_{\mathrm{co}}^{0} = \frac{\lambda}{4 \gamma^{2}} [a^{\dagger} + a] \sigma_{x}, \\
    H_{\mathrm{ex}}^{1} = - \frac{\kappa}{4 \gamma} [a^{\dagger} + a] [a^{\dagger} - a], \qquad
    H_{\mathrm{co}}^{1} = - \frac{\kappa \lambda \zeta}{4 \gamma} [a^{\dagger} + a] \sigma_{x}, \\
    H_{\mathrm{ex}}^{2} = - \frac{\kappa^{2}}{16 \gamma^{2}} [a^{\dagger} + a] [a^{\dagger} + a],
\end{gathered}
\end{equation}
as in Eq.~\eqref{eq:spin-phonon-coupled-hamiltonian-terms}, where for readability we have omitted the explicit natural vibrational energy units of $\hbar \omega_{x}$.
We can, however, simplify this expression further by partially diagonalizing the external dynamics.
This can be done by introducing the displaced creation and annihilation operators,
\begin{equation}
    b^{\dagger} = a^{\dagger} + \frac{\delta}{4 \gamma^{2}}, \qquad
    b = a + \frac{\delta}{4 \gamma^{2}}.
\end{equation}
Doing so yields the following terms for the external, internal and coupled dynamics,
\begin{equation}
\begin{gathered}
    H_{\mathrm{ex}}^{0} = b^{\dagger} b - \frac{\delta^{2}}{16 \gamma^{4}}, \qquad
    H_{\smash{\mathrm{in}}}^{0} = \varepsilon \sigma_{z} - \frac{\kappa^{2} \lambda \delta}{8 \gamma^{4}} \sigma_{x}, \qquad
    H_{\mathrm{co}}^{0} = \frac{\lambda}{4 \gamma^{2}} [b^{\dagger} + b] \sigma_{x}, \\
    H_{\mathrm{ex}}^{1} = - \frac{\kappa}{4 \gamma} [b^{\dagger} + b] [b^{\dagger} - b] + \frac{\kappa \delta}{8 \gamma^{3}} [b^{\dagger} - b], \qquad
    H_{\smash{\mathrm{in}}}^{1} = \frac{\kappa \lambda \zeta \delta}{8 \gamma^{3}} \sigma_{x}, \qquad
    H_{\mathrm{co}}^{1} = - \frac{\kappa \lambda \zeta}{4 \gamma} [b^{\dagger} + b] \sigma_{x}, \\
    H_{\mathrm{ex}}^{2} = - \frac{\kappa^{2}}{16 \gamma^{2}} [b^{\dagger} + b] [b^{\dagger} + b]  + \frac{\kappa^{2} \delta}{16 \gamma^{4}} [b^{\dagger} + b] - \frac{\kappa^{2} \delta^{2}}{64 \gamma^{6}}.
\end{gathered}
\end{equation}
Furthermore, we can eliminate the operator-independent terms (i.e., terms proportional to the identity) via the unitary transformation,
\begin{equation}
    \h{H} \mapsto \h{U} \h{H} \h{U}^{\dagger} + \i \hbar \frac{\partial \h{U}}{\partial \tau} \h{U}^{\dagger}, \qquad
    \h{U} = \exp \! \bigg( \frac{1}{\i \hbar} \frac{\delta^{2}}{16 \gamma^{4}} \tau \! \bigg) \! \exp \! \bigg( \frac{1}{\i \hbar} \frac{\kappa^{2} \delta^{2}}{64 \gamma^{6} \zeta} \sin(2 \zeta \tau) \! \bigg),
\end{equation}
to return the one-dimensional, time-independent, spin-phonon coupled model Hamiltonian terms,
\begin{equation}
\begin{gathered}
    H_{\mathrm{ex}}^{0} = b^{\dagger} b, \qquad
    H_{\smash{\mathrm{in}}}^{0} = \varepsilon \sigma_{z} - \frac{\kappa^{2} \lambda \delta}{8 \gamma^{4}} \sigma_{x}, \qquad
    H_{\mathrm{co}}^{0} = \frac{\lambda}{4 \gamma^{2}} [b^{\dagger} + b] \sigma_{x}, \\
    H_{\mathrm{ex}}^{1} = - \frac{\kappa}{4 \gamma} [b^{\dagger} + b] [b^{\dagger} - b] + \frac{\kappa \delta}{8 \gamma^{3}} [b^{\dagger} - b], \qquad
    H_{\smash{\mathrm{in}}}^{1} = \frac{\kappa \lambda \zeta \delta}{8 \gamma^{3}} \sigma_{x}, \qquad
    H_{\mathrm{co}}^{1} = - \frac{\kappa \lambda \zeta}{4 \gamma} [b^{\dagger} + b] \sigma_{x}, \\
    H_{\mathrm{ex}}^{2} = - \frac{\kappa^{2}}{16 \gamma^{2}} [b^{\dagger} + b] [b^{\dagger} + b]  + \frac{\kappa^{2} \delta}{16 \gamma^{4}} [b^{\dagger} + b].
\end{gathered}
\end{equation}

In order for the spin-phonon coupled model Hamiltonian to effectively model the trapped Rydberg~ion Hamiltonian, it is important the remaining dimensionless quantities (i.e., the relative energy~$\varepsilon$, lengths~$\lambda$ and~$\delta$, frequency~$\zeta$, and strength~$\gamma$) take physically meaningful values.
Since we are interested in understanding the impact of the intrinsic and extrinsic micromotion on the Rydberg excitation spectrum, which are encoded in $\zeta$ and $\delta$, respectively, we must necessarily fix the values of $\varepsilon$, $\lambda$, and $\gamma$.
Under typical experimental conditions (see, e.g., the recent review in Ref.~\cite{mokhberi2020}), the oscillating and static electric field gradients $A \sim 1 \, \unit{\giga\volt\per\meter\squared}$ and $B \sim 10 \, \unit{\mega\volt\per\meter\squared}$ whilst the frequency $\nu \sim 2 \pi \times 10 \, \unit{\mega\hertz}$.
For calcium $^{40}\mathrm{Ca}^{+}$ or strontium $^{88}\mathrm{Sr}^{+}$ ions, which are currently employed in trapped Rydberg ion experiments (see, e.g., Ref.~\cite{higgins2017a} and Ref.~\cite{feldker2015}), we obtain harmonic trap frequencies $\omega_{x} = \omega_{y} \approx 2 \pi \times 2 \, \unit{\mega\hertz}$ and $\omega_{z} \approx 2 \pi \times 1 \, \unit{\mega\hertz}$.
For the case of strontium $^{88}\mathrm{Sr}^{+}$ ions, this returns oscillator lengths $\ell_{x} = \ell_{y} \approx 5.36 \, \unit{\nano\meter}$ and $\ell_{z} \approx 7.58 \, \unit{\nano\meter}$, whilst for calcium $^{40}\mathrm{Ca}^{+}$ ions we get $\ell_{x} = \ell_{y} \approx 7.95 \, \unit{\nano\meter}$ and $\ell_{z} \approx 11.2 \, \unit{\nano\meter}$.
Accordingly, we choose to set $\omega_{z} = 2 \pi \times 1 \, \unit{\mega\hertz}$ and $\gamma = 2$ for the external vibrational parameters.

In current trapped Rydberg ion experiments, the choice of atomic quantum numbers varies (see, e.g., the recent review in Ref.~\cite{mokhberi2020} and references therein).
Indeed, principal quantum numbers from $n > 20$ to $n < 60$ have been considered for Rydberg states of both calcium $^{40}\mathrm{Ca}^{+}$ and strontium $^{88}\mathrm{Sr}^{+}$ ions.
Here, for specificity, we follow Refs.~\cite{zhang2020, wilkinson2024a} and consider the $n = 46$ states of a strontium $^{88}\mathrm{Sr}^{+}$ ion.
Utilizing an effective central potential to approximate the interaction between the valence electron and the ionic core (see the review in Ref.~\cite{aymar1996} and Ref.~\cite{pawlak2020} for further details), we obtain numerically computed theoretical values for the energy separation $E_{\smash{\uparrow}} - E_{\smash{\downarrow}} \equiv E_{\smash{n \mathrm{P}_{\smash{1/2}}}} - E_{\smash{n \mathrm{S}_{\smash{1/2}}}} = 121 \, \unit{\giga\hertz}$ and radial matrix element $\ab{\!\me{\uparrow}{r}{\downarrow}\!} \equiv \ab{\!\me{n \mathrm{P}_{\smash{1/2}}}{r}{n \mathrm{S}_{\smash{1/2}}}\!} = 63.7 \, \unit{\nano\meter}$, respectively.
This gives a relative energy splitting $\varepsilon \equiv [E_{\smash{\uparrow}} - E_{\smash{\downarrow}}] / 2 \hbar \omega_{x} = 30400$ and relative transition moment $\lambda \equiv \sqrt{\gamma} \ab{\!\me{\uparrow}{r}{\downarrow}\!} / 3 \ell_{z} = 3.96$.
For simplicity, we choose to fix $\varepsilon = 3 \times 10^{4}$ and $\lambda = 4$ for the internal electronic parameters.

To illustrate the effect of varying the principal quantum number $n$, we note that the relative energy splitting scales as $\varepsilon = 9.52 \times 10^{9}\, n^{-3.31}$, while the coupling strength follows $\lambda = 1.17 \times 10^{-3}\, n^{2.12}$, for strontium $^{88}\mathrm{Sr}^{+}$ ions. 
These scaling behaviors were determined using numerically obtained radial wavefunctions \cite{schmidtkaler2011, mokhberi2020, wilkinson2024b} and are consistent with values reported in Ref.~\cite{nist2024}.
For example, at $n = 20$, we find $\varepsilon = 467000$ and $\lambda = 0.65$, whereas for $n = 60$, these values change to $\varepsilon = 13200$ and $\lambda = 6.92$. 
This demonstrates the strong dependence of both energy splitting and coupling on the principal quantum number, highlighting the significant enhancement of interactions at higher $n$.
By analogy with the equations obtained in Eq.~\eqref{eq:approx}, the simplified expressions for the energy correction and relative transition probability, for $\varepsilon \gg \zeta$ and $\zeta \gg 1$ and incorporating the parameter $\lambda$, are given by
\begin{equation}
\begin{aligned}
    \epsilon_{\downarrow, 0, 0} & \approx - \varepsilon - \frac{\lambda^{2}}{512 \varepsilon} \bigg[1 + \frac{9 \delta^{2} \zeta^{2}}{8}  +  \frac{81}{64} \frac{\delta^{2}}{\lambda^{2}} + 72 \zeta^{2}\bigg] + \frac{2385}{2048} \frac{1}{\zeta^{2}} + \frac{657}{131072} \frac{\delta^{2}}{\zeta^{2}}, \\
    P_{\downarrow, 0, 0} & \approx 1 - \frac{\lambda^{2}}{64 \varepsilon^{2}} \bigg[1 + \frac{81 \delta^{2}}{256} + \frac{9\zeta^{2} \delta^{2}}{128} + \frac{9 \zeta^{2}}{2}\bigg] - \frac{14391}{16384} \frac{1}{\zeta^{2}} - \frac{657}{131072} \frac{\delta^{2}}{\zeta^{2}}.
\end{aligned}
\end{equation}
Comparing these expressions with Eq.~\eqref{eq:approx}, it is possible to see that arbitrary couplings add a $\lambda^{2}$ dependence in terms coupling $S$- and $P$-states (terms with $\epsilon$), which are sub-dominant in the approximated relative transition probability.
The energy correction and the relative transition probability explicitly show dependencies $\lambda^{2}/\varepsilon = 1.45 \times 10^{-16} \, n^{7.55}$ and $\lambda^{2}/\varepsilon^{2} = 1.51 \times 10^{-26} \, n^{10.86}$, where the scaling suggests decrease in the energy corrections and transition probabilities as $n$ increases.
However, the parameters reach order $1$, i.e., $\lambda^{2}/\varepsilon \sim 1$ and $\lambda^{2}/\varepsilon^{2} \sim 1$, only for $n = 126$ and $n = 239$, respectively.
  
\subsection{Dressed ground state of the vibrational motion}\label{app:ground-state}

In Sec.~\ref{sec:transition-probability}, we investigate the relative transition probabilities for the laser excitation of the trapped ion from the electronic ground state to the Rydberg $S$-state.
We choose $\kket{g, 0, 0}$ as the initial state, which assumes that the ion initially has no vibrational excitations.
This means that the vibrational state is described by the Fock state with zero phonons $N = 0$, which represents the ground state of the bare Hamiltonian $H^{0}_{\mathrm{ex}}$.
However, this is an approximation, as contributions due to the intrinsic micromotion, which add time-periodic terms to the Hamiltonian, are neglected.
Rather, the full Hamiltonian a trapped ion in the ground state is exposed to (within the approximations we have been making throughout) reads
\begin{equation}\label{eq:ground_state_hamiltonian}
    H_{g} = [\varepsilon_{g} + H_{\smash{\mathrm{ex}}}^{0} + [\e^{\i \zeta \tau} - \e^{- \i \zeta \tau}] H_{\smash{\mathrm{ex}}}^{1} + [\e^{\i 2 \zeta \tau} + \e^{- \i 2 \zeta \tau}] H_{\smash{\mathrm{ex}}}^{2}] \! \op{g}{g} \!,
\end{equation}
where $\varepsilon_{g}$ is the bare electronic ground state energy.
The additional terms, $H^{1}_{\mathrm{ex}}$ and $H^{2}_{\mathrm{ex}}$, ``dress'' the ground state of $H^{0}_{\mathrm{ex}}$.
In the Floquet theory treatment, this means that $\kket{g, 0, 0}$ is no longer an eigenstate of the quasienergy operator associated to the instantaneous Hamiltonian of Eq.~\eqref{eq:ground_state_hamiltonian}: $\mathcal{H}_{g}(\tau) = H_{\smash{g}}(\tau)- \i \frac{\d}{\d \tau}$.

\begin{figure}[t]
    \centering
    \includegraphics[scale=0.38]{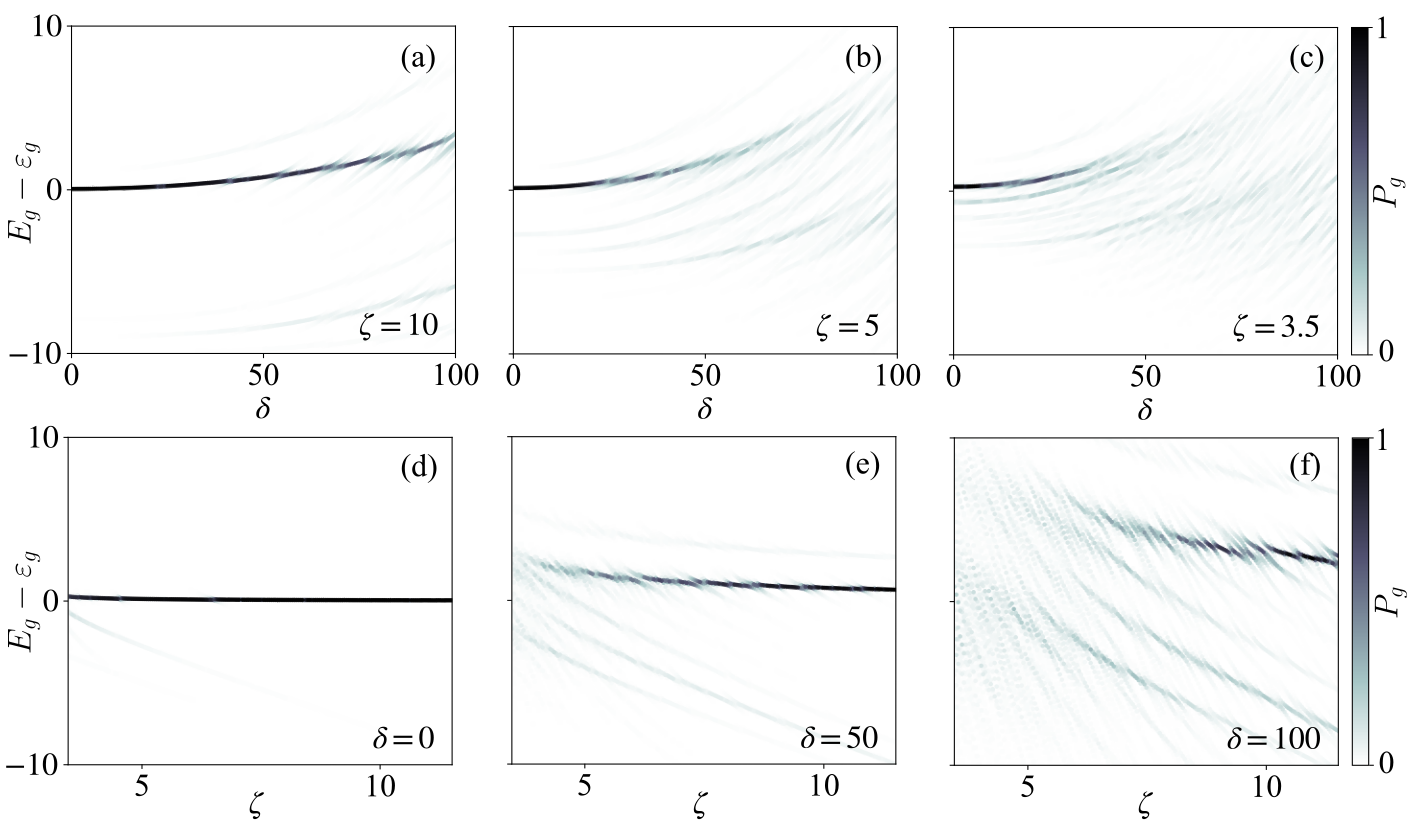}
    \caption{
        \textbf{Overlap of the quasienergy eigenstates with the state $\kket{g, 0, 0}$.}
        We show the eigenenergies of the quasienergy operator $\mathcal{H}_{g}$ [Eq.~\eqref{eq:H_ground_state}] as a function of the relative field displacement $\delta$ and field radio frequency $\zeta$.
        The shading is given by the overlap probability in Eq.~\eqref{eq:ground_state_overlap} between the quasienergy eigenstate $\kket{E_{g}}$ and the basis state $\kket{g, 0, 0}$, which is the ground state of the Hamiltonian $H^{0}_{\mathrm{ex}}$ [see Eq.~\eqref{eq:spin-phonon-coupled-hamiltonian-terms-transformed}].
        The results are obtained for the same parameter set of Fig.~\ref{fig:quasienergy-spectrum}.
    }
    \label{fig:gs-spectrum}
\end{figure}

In the following, we quantify the strength of the dressing by calculating the overlap,
\begin{equation}\label{eq:ground_state_overlap}
    P_{g} = \ab{\ipe{g, 0, 0}{E_{g}}}^{2},
\end{equation}
between the state $\kket{g, 0, 0}$ and the quasienergy eigenstates $\kket{E_{g}}$ obeying $\mathcal{H}_{g} \! \kket{E_{g}} = E_{g} \! \kket{E_{g}}$.
The corresponding data is shown in Fig.~\ref{fig:gs-spectrum}, where the truncated parameters are $N_{\mathrm{tr}} = 40$ and $m_{\mathrm{tr}} = 35$.
One can see that for sufficiently large $\zeta$ and sufficiently small $\delta$ there is one quasienergy eigenstate which is mostly composed of the state $\kket{g, 0, 0}$.
Here, the dressing of $\kket{g, 0, 0}$ is weak and it is justified to calculate the relative transition probabilities to the Rydberg states using Eq.~\eqref{eq:state-transition-probabilities} in Sec.~\ref{sec:transition-probability}.
For small $\zeta$ and large $\delta$, on the other hand, the time-dependent contributions of the electric field lead to strong mixing of the state $\kket{g, 0, 0}$ with other states.
Here, the use of Eq.~\eqref{eq:state-transition-probabilities} is no longer justified.
This is, however, anyway the regime in which the Rydberg $S$-state also gets distributed over a broad range of quasienergies (see Fig. \ref{fig:quasienergy-spectrum}) and no well-defined Rydberg lines exist. 

To obtain some quantitative insights into the quasienergy of the dressed vibrational ground state and its overlap with $\kket{g, 0, 0}$, we use perturbation theory, following Sec.~\ref{sec:perturbation-theory}.
The unperturbed quasienergy operator is
\begin{equation}\label{eq:H_ground_state}
    \mathcal{H}^{(0)}_{g} = \sum_{\mathclap{m = - \infty}}^{\infty} [ [\varepsilon_{g} + b^{\dagger} b  + m \zeta] \! \op{g}{g}] \! \kket{m} \! \bbra{m} \!,
\end{equation}
with eigenenergies and eigenstates $\epsilon^{(0)}_{g, N, m} = \varepsilon_{g} + N + m\zeta$ and $\kket{u^{(0)}_{g, N, m}} = \kket{g, N, m}$.
The perturbation $\mathcal{V}_{g}$ contains terms associated with the oscillating electric field and reads
\begin{equation}
\begin{aligned}
    \mathcal{V}_{g} & = - \sum_{\mathclap{m = - \infty}}^{\infty} \ \bigg[ \! \bigg[ \frac{3}{8} [b^{\dagger} + b] [b^{\dagger} - b] - \frac{3 \delta}{64} [b^{\dagger} - b] \bigg] \! \op{g}{g} \! [\kket{m + 1} \! \bbra{m} - \mathrm{H.c.}] \\
    & \qquad + \bigg[ \frac{9}{64} [b^{\dagger} + b] [b^{\dagger} + b] - \frac{9 \delta}{256} [b^{\dagger} + b] \bigg] \! \op{g}{g} \! [\kket{m + 2} \! \bbra{m} + \mathrm{H.c.}] \bigg].
\end{aligned}
\end{equation}
Using Eq.~\eqref{eq:prob}, the perturbed quasienergy and the overlap of the dressed ground state with $\kket{g, 0, 0}$ read
\begin{equation}
    \begin{aligned}
    \epsilon_{g, 0,0} & = \varepsilon_{g} + \frac{9 \delta^{2}}{2048 \zeta^{2}} \bigg[ \! \bigg[ \frac{\zeta^{2}}{\zeta^{2} - 1} + \frac{9}{128} \frac{4 \zeta^{2}}{4 \zeta^{2} - 1} \bigg] + \frac{256}{\delta^{2}} \bigg[ \frac{\zeta^{2}}{\zeta^{2} - 4} + \frac{1}{256} \frac{\zeta^{2}}{\zeta^{2} - 1} \bigg] \! \bigg], \\
    P_{g, 0, 0} & = 1 - \frac{9 \delta^{2}}{2048 \zeta^{2}} \bigg[ \! \bigg[ \frac{\zeta^{2} [\zeta^{2} + 1]}{[\zeta^{2} - 1]^{2}} + \frac{9}{64} \frac{4 \zeta^{2} [4 \zeta^{2} + 1]}{[4 \zeta^{2} - 1]^{2}} \bigg] + \frac{128}{\delta^{2}} \bigg[ \frac{\zeta^{2} [\zeta^{2} + 4]}{[\zeta^{2} - 4]^{2}} + \frac{539}{1024} + \frac{9}{256} \frac{\zeta^{2} [\zeta^{2} + 1]}{[\zeta^{2} - 1]^{2}} \bigg] \! \bigg].
    \end{aligned}
\end{equation} 
Using the approximations $\varepsilon \gg \zeta$ and $\zeta \gg 1$, we obtain 
\begin{equation}
\begin{aligned}
    \epsilon_{g, 0,0} & \approx \varepsilon_{g} + \frac{2313}{2048} \frac{1}{\zeta^{2}} + \frac{1233}{262144} \frac{\delta^{2}}{\zeta^{2}}, \\
    P_{g, 0, 0} & \approx 1 - \frac{14391}{16385} \frac{1}{\zeta^{2}} - \frac{657}{131072} \frac{\delta^{2}}{\zeta^{2}}.
\end{aligned}
\end{equation}
In these expressions, we can see contributions from intrinsic micromotion, which depend only on $\zeta$, and extrinsic micromotion, which depend on $\delta$. 
The approximated expression for the overlap $P_{g, 0, 0}$ matches the relative transition probability $P_{\downarrow, 0, 0}$ from Eq.~\eqref{eq:approx}. 
This suggests that in the regime where the coupling between the Rydberg $S$- and $P$-state is negligible, the vibrational state of the Rydberg ion is altered in the same way as the vibrational state of the ground state ion. 
Here, one would expect that the relative transition probability into the perturbed Rydberg $S$-state is actually even higher than the result obtained from Eq.~\eqref{eq:state-transition-probabilities}.

\begin{figure}[t]
    \centering
    \includegraphics[scale=0.38]{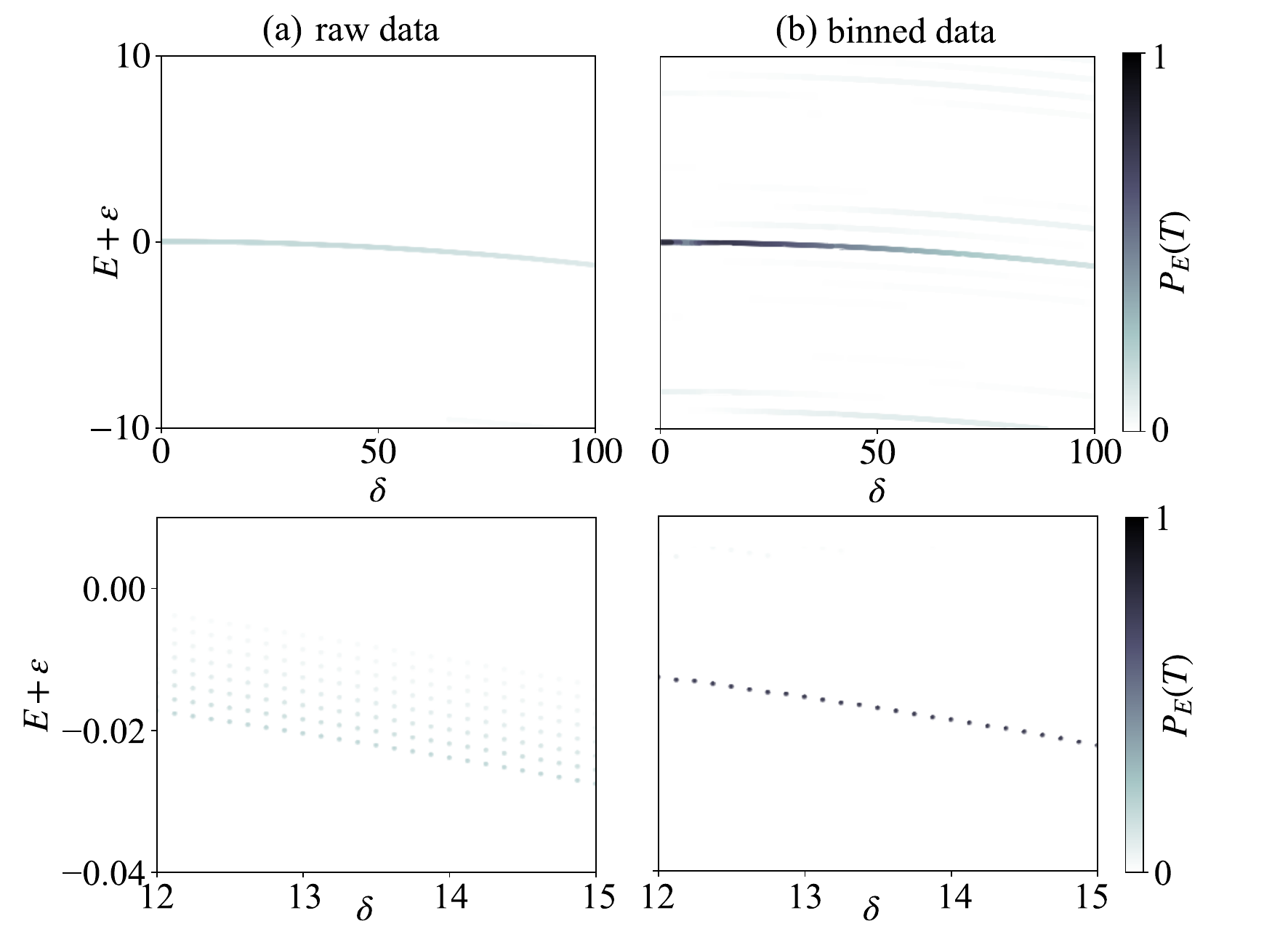}
    \caption{
        \textbf{Difference between raw and binned data for relative transition probability plots.}
        (a) The left panel shows the excitation spectrum without binning, revealing line broadening due to excitations involving different phonon numbers $N$. 
        The top curve is magnified to highlight the broadening in the lower plot.
        (b) The right panel presents the binned data, with a finite resolution $\Delta E = 0.1$. 
        The resulting binned points are shown in the lower plot.
        Here, we use $T = 5$ and $\zeta = 10$.}
    \label{fig:binning}
\end{figure}

\subsection{Relative transition probabilities for finite temperature initial states\label{app:thermal-plots}}
To obtain the relative transition probabilities in Eq.~\eqref{eq:state-transition-probabilities} and Eq.~\eqref{eq:relative-transition_probability-th}, we diagonalize the truncated quasienergy operator from Eq.~\eqref{eq:quasi_op}. 
The resulting quasienergies satisfy the eigenvalue equation 
$\mathcal{H}\kket{E} = E\kket{E}$.
At zero temperature $(T = 0)$, the computed eigenenergies $E$ and eigenstates $\kket{E}$ directly determine the relative transition probabilities shown in Fig.~\ref{fig:quasienergy-spectrum}. 
For finite temperatures ($T > 0$), these probabilities are obtained by summing contributions from different phonon numbers, indexed by $N$, as expressed in Eq.~\eqref{eq:relative-transition_probability-th}. 
The energy shifts associated with different phonon numbers $N$ lead to spectral broadening, which is observed in the magnified plot shown in Fig.~\ref{fig:binning} (a).
To produce the plots in the main text we assume a finite energy resolution and introduce quasienergy bins of width $\Delta E$. 
Each bin is assigned an energy value given by the weighted average of the quasienergies within each $\Delta E$, while the corresponding transition probability is computed by summing the relative transition probabilities of those quasienergies.
The resulting plot obtained from binned energies and relative transition probabilities is shown in Fig.~\ref{fig:binning} (b). 
In Figs.~\ref{fig:quasienergy-th-delta} and \ref{fig:quasienergy-th-zeta} in the main text we set $\Delta E = 0.1$.


\addcontentsline{toc}{section}{References}

\bibliographystyle{apsrev4-2}

\bibliography{main_v2.bib}

\begin{thebibliography}{77}%
\makeatletter
\providecommand \@ifxundefined [1]{%
 \@ifx{#1\undefined}
}%
\providecommand \@ifnum [1]{%
 \ifnum #1\expandafter \@firstoftwo
 \else \expandafter \@secondoftwo
 \fi
}%
\providecommand \@ifx [1]{%
 \ifx #1\expandafter \@firstoftwo
 \else \expandafter \@secondoftwo
 \fi
}%
\providecommand \natexlab [1]{#1}%
\providecommand \enquote  [1]{``#1''}%
\providecommand \bibnamefont  [1]{#1}%
\providecommand \bibfnamefont [1]{#1}%
\providecommand \citenamefont [1]{#1}%
\providecommand \href@noop [0]{\@secondoftwo}%
\providecommand \href [0]{\begingroup \@sanitize@url \@href}%
\providecommand \@href[1]{\@@startlink{#1}\@@href}%
\providecommand \@@href[1]{\endgroup#1\@@endlink}%
\providecommand \@sanitize@url [0]{\catcode `\\12\catcode `\$12\catcode
  `\&12\catcode `\#12\catcode `\^12\catcode `\_12\catcode `\%12\relax}%
\providecommand \@@startlink[1]{}%
\providecommand \@@endlink[0]{}%
\providecommand \url  [0]{\begingroup\@sanitize@url \@url }%
\providecommand \@url [1]{\endgroup\@href {#1}{\urlprefix }}%
\providecommand \urlprefix  [0]{URL }%
\providecommand \Eprint [0]{\href }%
\providecommand \doibase [0]{https://doi.org/}%
\providecommand \selectlanguage [0]{\@gobble}%
\providecommand \bibinfo  [0]{\@secondoftwo}%
\providecommand \bibfield  [0]{\@secondoftwo}%
\providecommand \translation [1]{[#1]}%
\providecommand \BibitemOpen [0]{}%
\providecommand \bibitemStop [0]{}%
\providecommand \bibitemNoStop [0]{.\EOS\space}%
\providecommand \EOS [0]{\spacefactor3000\relax}%
\providecommand \BibitemShut  [1]{\csname bibitem#1\endcsname}%
\let\auto@bib@innerbib\@empty
\bibitem [{\citenamefont {Cirac}\ and\ \citenamefont
  {Zoller}(1995)}]{cirac1995}%
  \BibitemOpen
  \bibfield  {author} {\bibinfo {author} {\bibfnamefont {J.~I.}\ \bibnamefont
  {Cirac}}\ and\ \bibinfo {author} {\bibfnamefont {P.}~\bibnamefont {Zoller}},\
  }\href {https://doi.org/10.1103/PhysRevLett.74.4091} {\bibfield  {journal}
  {\bibinfo  {journal} {Phys. Rev. Lett.}\ }\textbf {\bibinfo {volume} {74}},\
  \bibinfo {pages} {4091} (\bibinfo {year} {1995})}\BibitemShut {NoStop}%
\bibitem [{\citenamefont {Cirac}\ and\ \citenamefont
  {Zoller}(2000)}]{cirac2000}%
  \BibitemOpen
  \bibfield  {author} {\bibinfo {author} {\bibfnamefont {J.~I.}\ \bibnamefont
  {Cirac}}\ and\ \bibinfo {author} {\bibfnamefont {P.}~\bibnamefont {Zoller}},\
  }\href {https://doi.org/10.1038/35007021} {\bibfield  {journal} {\bibinfo
  {journal} {Nature}\ }\textbf {\bibinfo {volume} {404}},\ \bibinfo {pages}
  {579} (\bibinfo {year} {2000})}\BibitemShut {NoStop}%
\bibitem [{\citenamefont {Blatt}\ and\ \citenamefont
  {Wineland}(2008)}]{blatt2008}%
  \BibitemOpen
  \bibfield  {author} {\bibinfo {author} {\bibfnamefont {R.}~\bibnamefont
  {Blatt}}\ and\ \bibinfo {author} {\bibfnamefont {D.}~\bibnamefont
  {Wineland}},\ }\href {https://doi.org/10.1038/nature07125} {\bibfield
  {journal} {\bibinfo  {journal} {Nature}\ }\textbf {\bibinfo {volume} {453}},\
  \bibinfo {pages} {1008} (\bibinfo {year} {2008})}\BibitemShut {NoStop}%
\bibitem [{\citenamefont {H{\" a}ffner}\ \emph {et~al.}(2008)\citenamefont
  {H{\" a}ffner}, \citenamefont {Roos},\ and\ \citenamefont
  {Blatt}}]{haffner2008}%
  \BibitemOpen
  \bibfield  {author} {\bibinfo {author} {\bibfnamefont {H.}~\bibnamefont {H{\"
  a}ffner}}, \bibinfo {author} {\bibfnamefont {C.~F.}\ \bibnamefont {Roos}},\
  and\ \bibinfo {author} {\bibfnamefont {R.}~\bibnamefont {Blatt}},\ }\href
  {https://doi.org/10.1016/j.physrep.2008.09.003} {\bibfield  {journal}
  {\bibinfo  {journal} {Phys. Rep.}\ }\textbf {\bibinfo {volume} {469}},\
  \bibinfo {pages} {155} (\bibinfo {year} {2008})}\BibitemShut {NoStop}%
\bibitem [{\citenamefont {Schindler}\ \emph {et~al.}(2013)\citenamefont
  {Schindler}, \citenamefont {Nigg}, \citenamefont {Monz}, \citenamefont
  {Barreiro}, \citenamefont {Martinez}, \citenamefont {Wang}, \citenamefont
  {Quint}, \citenamefont {Brandl}, \citenamefont {Nebendahl}, \citenamefont
  {Roos}, \citenamefont {Chwalla}, \citenamefont {Hennrich},\ and\
  \citenamefont {Blatt}}]{schindler2013}%
  \BibitemOpen
  \bibfield  {author} {\bibinfo {author} {\bibfnamefont {P.}~\bibnamefont
  {Schindler}}, \bibinfo {author} {\bibfnamefont {D.}~\bibnamefont {Nigg}},
  \bibinfo {author} {\bibfnamefont {T.}~\bibnamefont {Monz}}, \bibinfo {author}
  {\bibfnamefont {J.~T.}\ \bibnamefont {Barreiro}}, \bibinfo {author}
  {\bibfnamefont {E.}~\bibnamefont {Martinez}}, \bibinfo {author}
  {\bibfnamefont {S.~X.}\ \bibnamefont {Wang}}, \bibinfo {author}
  {\bibfnamefont {S.}~\bibnamefont {Quint}}, \bibinfo {author} {\bibfnamefont
  {M.~F.}\ \bibnamefont {Brandl}}, \bibinfo {author} {\bibfnamefont
  {V.}~\bibnamefont {Nebendahl}}, \bibinfo {author} {\bibfnamefont {C.~F.}\
  \bibnamefont {Roos}}, \bibinfo {author} {\bibfnamefont {M.}~\bibnamefont
  {Chwalla}}, \bibinfo {author} {\bibfnamefont {M.}~\bibnamefont {Hennrich}},\
  and\ \bibinfo {author} {\bibfnamefont {R.}~\bibnamefont {Blatt}},\ }\href
  {https://doi.org/10.1088/1367-2630/15/12/123012} {\bibfield  {journal}
  {\bibinfo  {journal} {New J. Phys.}\ }\textbf {\bibinfo {volume} {15}},\
  \bibinfo {pages} {123012} (\bibinfo {year} {2013})}\BibitemShut {NoStop}%
\bibitem [{\citenamefont {Duan}\ and\ \citenamefont {Monroe}(2010)}]{duan2010}%
  \BibitemOpen
  \bibfield  {author} {\bibinfo {author} {\bibfnamefont {L.-M.}\ \bibnamefont
  {Duan}}\ and\ \bibinfo {author} {\bibfnamefont {C.}~\bibnamefont {Monroe}},\
  }\href {https://doi.org/10.1103/RevModPhys.82.1209} {\bibfield  {journal}
  {\bibinfo  {journal} {Rev. Mod. Phys.}\ }\textbf {\bibinfo {volume} {82}},\
  \bibinfo {pages} {1209} (\bibinfo {year} {2010})}\BibitemShut {NoStop}%
\bibitem [{\citenamefont {Bermudez}\ \emph {et~al.}(2017)\citenamefont
  {Bermudez}, \citenamefont {Xu}, \citenamefont {Nigmatullin}, \citenamefont
  {O'Gorman}, \citenamefont {Negnevitsky}, \citenamefont {Schindler},
  \citenamefont {Monz}, \citenamefont {Poschinger}, \citenamefont {Hempel},
  \citenamefont {Home}, \citenamefont {Schmidt-Kaler}, \citenamefont {Biercuk},
  \citenamefont {Blatt}, \citenamefont {Benjamin},\ and\ \citenamefont
  {M\"{u}ller}}]{bermudez2017}%
  \BibitemOpen
  \bibfield  {author} {\bibinfo {author} {\bibfnamefont {A.}~\bibnamefont
  {Bermudez}}, \bibinfo {author} {\bibfnamefont {X.}~\bibnamefont {Xu}},
  \bibinfo {author} {\bibfnamefont {R.}~\bibnamefont {Nigmatullin}}, \bibinfo
  {author} {\bibfnamefont {J.}~\bibnamefont {O'Gorman}}, \bibinfo {author}
  {\bibfnamefont {V.}~\bibnamefont {Negnevitsky}}, \bibinfo {author}
  {\bibfnamefont {P.}~\bibnamefont {Schindler}}, \bibinfo {author}
  {\bibfnamefont {T.}~\bibnamefont {Monz}}, \bibinfo {author} {\bibfnamefont
  {U.~G.}\ \bibnamefont {Poschinger}}, \bibinfo {author} {\bibfnamefont
  {C.}~\bibnamefont {Hempel}}, \bibinfo {author} {\bibfnamefont
  {J.}~\bibnamefont {Home}}, \bibinfo {author} {\bibfnamefont {F.}~\bibnamefont
  {Schmidt-Kaler}}, \bibinfo {author} {\bibfnamefont {M.}~\bibnamefont
  {Biercuk}}, \bibinfo {author} {\bibfnamefont {R.}~\bibnamefont {Blatt}},
  \bibinfo {author} {\bibfnamefont {S.}~\bibnamefont {Benjamin}},\ and\
  \bibinfo {author} {\bibfnamefont {M.}~\bibnamefont {M\"{u}ller}},\ }\href
  {https://doi.org/10.1103/PhysRevX.7.041061} {\bibfield  {journal} {\bibinfo
  {journal} {Phys. Rev. X}\ }\textbf {\bibinfo {volume} {7}},\ \bibinfo {pages}
  {041061} (\bibinfo {year} {2017})}\BibitemShut {NoStop}%
\bibitem [{\citenamefont {Bruzewicz}\ \emph {et~al.}(2019)\citenamefont
  {Bruzewicz}, \citenamefont {Chiaverini}, \citenamefont {McConnell},\ and\
  \citenamefont {Sage}}]{bruzewicz2019}%
  \BibitemOpen
  \bibfield  {author} {\bibinfo {author} {\bibfnamefont {C.~D.}\ \bibnamefont
  {Bruzewicz}}, \bibinfo {author} {\bibfnamefont {J.}~\bibnamefont
  {Chiaverini}}, \bibinfo {author} {\bibfnamefont {R.}~\bibnamefont
  {McConnell}},\ and\ \bibinfo {author} {\bibfnamefont {J.~M.}\ \bibnamefont
  {Sage}},\ }\href {https://doi.org/10.1063/1.5088164} {\bibfield  {journal}
  {\bibinfo  {journal} {Appl. Phys. Rev.}\ }\textbf {\bibinfo {volume} {6}},\
  \bibinfo {pages} {021314} (\bibinfo {year} {2019})}\BibitemShut {NoStop}%
\bibitem [{\citenamefont {Porras}\ and\ \citenamefont
  {Cirac}(2004)}]{porras2004}%
  \BibitemOpen
  \bibfield  {author} {\bibinfo {author} {\bibfnamefont {D.}~\bibnamefont
  {Porras}}\ and\ \bibinfo {author} {\bibfnamefont {J.~I.}\ \bibnamefont
  {Cirac}},\ }\href {https://doi.org/10.1103/PhysRevLett.92.207901} {\bibfield
  {journal} {\bibinfo  {journal} {Phys. Rev. Lett.}\ }\textbf {\bibinfo
  {volume} {92}},\ \bibinfo {pages} {207901} (\bibinfo {year}
  {2004})}\BibitemShut {NoStop}%
\bibitem [{\citenamefont {Friedenauer}\ \emph {et~al.}(2008)\citenamefont
  {Friedenauer}, \citenamefont {Schmitz}, \citenamefont {Glueckert},
  \citenamefont {Porras},\ and\ \citenamefont {Schaetz}}]{friedenauer2008}%
  \BibitemOpen
  \bibfield  {author} {\bibinfo {author} {\bibfnamefont {A.}~\bibnamefont
  {Friedenauer}}, \bibinfo {author} {\bibfnamefont {H.}~\bibnamefont
  {Schmitz}}, \bibinfo {author} {\bibfnamefont {J.~T.}\ \bibnamefont
  {Glueckert}}, \bibinfo {author} {\bibfnamefont {D.}~\bibnamefont {Porras}},\
  and\ \bibinfo {author} {\bibfnamefont {T.}~\bibnamefont {Schaetz}},\ }\href
  {https://doi.org/10.1038/nphys1032} {\bibfield  {journal} {\bibinfo
  {journal} {Nature Phys.}\ }\textbf {\bibinfo {volume} {4}},\ \bibinfo {pages}
  {757} (\bibinfo {year} {2008})}\BibitemShut {NoStop}%
\bibitem [{\citenamefont {Kim}\ \emph {et~al.}(2010)\citenamefont {Kim},
  \citenamefont {Chang}, \citenamefont {Korenblit}, \citenamefont {Islam},
  \citenamefont {Edwards}, \citenamefont {Freericks}, \citenamefont {Lin},
  \citenamefont {Duan},\ and\ \citenamefont {Monroe}}]{kim2010}%
  \BibitemOpen
  \bibfield  {author} {\bibinfo {author} {\bibfnamefont {K.}~\bibnamefont
  {Kim}}, \bibinfo {author} {\bibfnamefont {M.-S.}\ \bibnamefont {Chang}},
  \bibinfo {author} {\bibfnamefont {S.}~\bibnamefont {Korenblit}}, \bibinfo
  {author} {\bibfnamefont {R.}~\bibnamefont {Islam}}, \bibinfo {author}
  {\bibfnamefont {E.~E.}\ \bibnamefont {Edwards}}, \bibinfo {author}
  {\bibfnamefont {J.~K.}\ \bibnamefont {Freericks}}, \bibinfo {author}
  {\bibfnamefont {G.-D.}\ \bibnamefont {Lin}}, \bibinfo {author} {\bibfnamefont
  {L.-M.}\ \bibnamefont {Duan}},\ and\ \bibinfo {author} {\bibfnamefont
  {C.}~\bibnamefont {Monroe}},\ }\href {https://doi.org/10.1038/nature09071}
  {\bibfield  {journal} {\bibinfo  {journal} {Nature}\ }\textbf {\bibinfo
  {volume} {465}},\ \bibinfo {pages} {590} (\bibinfo {year}
  {2010})}\BibitemShut {NoStop}%
\bibitem [{\citenamefont {Barreiro}\ \emph {et~al.}(2011)\citenamefont
  {Barreiro}, \citenamefont {M{\" u}ller}, \citenamefont {Schindler},
  \citenamefont {Nigg}, \citenamefont {Monz}, \citenamefont {Chwalla},
  \citenamefont {Hennrich}, \citenamefont {Roos}, \citenamefont {Zoller},\ and\
  \citenamefont {Blatt}}]{barreiro2011}%
  \BibitemOpen
  \bibfield  {author} {\bibinfo {author} {\bibfnamefont {J.~T.}\ \bibnamefont
  {Barreiro}}, \bibinfo {author} {\bibfnamefont {M.}~\bibnamefont {M{\"
  u}ller}}, \bibinfo {author} {\bibfnamefont {P.}~\bibnamefont {Schindler}},
  \bibinfo {author} {\bibfnamefont {D.}~\bibnamefont {Nigg}}, \bibinfo {author}
  {\bibfnamefont {T.}~\bibnamefont {Monz}}, \bibinfo {author} {\bibfnamefont
  {M.}~\bibnamefont {Chwalla}}, \bibinfo {author} {\bibfnamefont
  {M.}~\bibnamefont {Hennrich}}, \bibinfo {author} {\bibfnamefont {C.~F.}\
  \bibnamefont {Roos}}, \bibinfo {author} {\bibfnamefont {P.}~\bibnamefont
  {Zoller}},\ and\ \bibinfo {author} {\bibfnamefont {R.}~\bibnamefont
  {Blatt}},\ }\href {https://doi.org/10.1038/nature09801} {\bibfield  {journal}
  {\bibinfo  {journal} {Nature}\ }\textbf {\bibinfo {volume} {470}},\ \bibinfo
  {pages} {486} (\bibinfo {year} {2011})}\BibitemShut {NoStop}%
\bibitem [{\citenamefont {Blatt}\ and\ \citenamefont {Roos}(2012)}]{blatt2012}%
  \BibitemOpen
  \bibfield  {author} {\bibinfo {author} {\bibfnamefont {R.}~\bibnamefont
  {Blatt}}\ and\ \bibinfo {author} {\bibfnamefont {C.~F.}\ \bibnamefont
  {Roos}},\ }\href {https://doi.org/10.1038/nphys2252} {\bibfield  {journal}
  {\bibinfo  {journal} {Nat. Phys.}\ }\textbf {\bibinfo {volume} {8}},\
  \bibinfo {pages} {277} (\bibinfo {year} {2012})}\BibitemShut {NoStop}%
\bibitem [{\citenamefont {Georgescu}\ \emph {et~al.}(2014)\citenamefont
  {Georgescu}, \citenamefont {Ashhab},\ and\ \citenamefont
  {Nori}}]{georgescu2014}%
  \BibitemOpen
  \bibfield  {author} {\bibinfo {author} {\bibfnamefont {I.~M.}\ \bibnamefont
  {Georgescu}}, \bibinfo {author} {\bibfnamefont {S.}~\bibnamefont {Ashhab}},\
  and\ \bibinfo {author} {\bibfnamefont {F.}~\bibnamefont {Nori}},\ }\href
  {https://doi.org/10.1103/RevModPhys.86.153} {\bibfield  {journal} {\bibinfo
  {journal} {Rev. Mod. Phys.}\ }\textbf {\bibinfo {volume} {86}},\ \bibinfo
  {pages} {153} (\bibinfo {year} {2014})}\BibitemShut {NoStop}%
\bibitem [{\citenamefont {Monroe}\ \emph {et~al.}(2021)\citenamefont {Monroe},
  \citenamefont {Campbell}, \citenamefont {Duan}, \citenamefont {Gong},
  \citenamefont {Gorshkov}, \citenamefont {Hess}, \citenamefont {Islam},
  \citenamefont {Kim}, \citenamefont {Linke}, \citenamefont {Pagano},
  \citenamefont {Richerme}, \citenamefont {Senko},\ and\ \citenamefont
  {Yao}}]{monroe2021}%
  \BibitemOpen
  \bibfield  {author} {\bibinfo {author} {\bibfnamefont {C.}~\bibnamefont
  {Monroe}}, \bibinfo {author} {\bibfnamefont {W.~C.}\ \bibnamefont
  {Campbell}}, \bibinfo {author} {\bibfnamefont {L.-M.}\ \bibnamefont {Duan}},
  \bibinfo {author} {\bibfnamefont {Z.-X.}\ \bibnamefont {Gong}}, \bibinfo
  {author} {\bibfnamefont {A.~V.}\ \bibnamefont {Gorshkov}}, \bibinfo {author}
  {\bibfnamefont {P.~W.}\ \bibnamefont {Hess}}, \bibinfo {author}
  {\bibfnamefont {R.}~\bibnamefont {Islam}}, \bibinfo {author} {\bibfnamefont
  {K.}~\bibnamefont {Kim}}, \bibinfo {author} {\bibfnamefont {N.~M.}\
  \bibnamefont {Linke}}, \bibinfo {author} {\bibfnamefont {G.}~\bibnamefont
  {Pagano}}, \bibinfo {author} {\bibfnamefont {P.}~\bibnamefont {Richerme}},
  \bibinfo {author} {\bibfnamefont {C.}~\bibnamefont {Senko}},\ and\ \bibinfo
  {author} {\bibfnamefont {N.~Y.}\ \bibnamefont {Yao}},\ }\href
  {https://doi.org/10.1103/RevModPhys.93.025001} {\bibfield  {journal}
  {\bibinfo  {journal} {Rev. Mod. Phys.}\ }\textbf {\bibinfo {volume} {93}},\
  \bibinfo {pages} {025001} (\bibinfo {year} {2021})}\BibitemShut {NoStop}%
\bibitem [{\citenamefont {Kotler}\ \emph {et~al.}(2011)\citenamefont {Kotler},
  \citenamefont {Akerman}, \citenamefont {Glickman}, \citenamefont {Keselman},\
  and\ \citenamefont {Ozeri}}]{kotler2011}%
  \BibitemOpen
  \bibfield  {author} {\bibinfo {author} {\bibfnamefont {S.}~\bibnamefont
  {Kotler}}, \bibinfo {author} {\bibfnamefont {N.}~\bibnamefont {Akerman}},
  \bibinfo {author} {\bibfnamefont {Y.}~\bibnamefont {Glickman}}, \bibinfo
  {author} {\bibfnamefont {A.}~\bibnamefont {Keselman}},\ and\ \bibinfo
  {author} {\bibfnamefont {R.}~\bibnamefont {Ozeri}},\ }\href
  {https://doi.org/10.1038/nature10010} {\bibfield  {journal} {\bibinfo
  {journal} {Nature}\ }\textbf {\bibinfo {volume} {473}},\ \bibinfo {pages}
  {61} (\bibinfo {year} {2011})}\BibitemShut {NoStop}%
\bibitem [{\citenamefont {Hempel}\ \emph {et~al.}(2013)\citenamefont {Hempel},
  \citenamefont {Lanyon}, \citenamefont {Jurcevic}, \citenamefont {Gerritsma},
  \citenamefont {Blatt},\ and\ \citenamefont {Roos}}]{hempel2013}%
  \BibitemOpen
  \bibfield  {author} {\bibinfo {author} {\bibfnamefont {C.}~\bibnamefont
  {Hempel}}, \bibinfo {author} {\bibfnamefont {B.~P.}\ \bibnamefont {Lanyon}},
  \bibinfo {author} {\bibfnamefont {P.}~\bibnamefont {Jurcevic}}, \bibinfo
  {author} {\bibfnamefont {R.}~\bibnamefont {Gerritsma}}, \bibinfo {author}
  {\bibfnamefont {R.}~\bibnamefont {Blatt}},\ and\ \bibinfo {author}
  {\bibfnamefont {C.~F.}\ \bibnamefont {Roos}},\ }\href
  {https://doi.org/10.1038/nphoton.2013.172} {\bibfield  {journal} {\bibinfo
  {journal} {Nat. Photonics}\ }\textbf {\bibinfo {volume} {7}},\ \bibinfo
  {pages} {630} (\bibinfo {year} {2013})}\BibitemShut {NoStop}%
\bibitem [{\citenamefont {Degen}\ \emph {et~al.}(2017)\citenamefont {Degen},
  \citenamefont {Reinhard},\ and\ \citenamefont {Cappellaro}}]{degen2017}%
  \BibitemOpen
  \bibfield  {author} {\bibinfo {author} {\bibfnamefont {C.~L.}\ \bibnamefont
  {Degen}}, \bibinfo {author} {\bibfnamefont {F.}~\bibnamefont {Reinhard}},\
  and\ \bibinfo {author} {\bibfnamefont {P.}~\bibnamefont {Cappellaro}},\
  }\href {https://doi.org/10.1103/RevModPhys.89.035002} {\bibfield  {journal}
  {\bibinfo  {journal} {Rev. Mod. Phys.}\ }\textbf {\bibinfo {volume} {89}},\
  \bibinfo {pages} {035002} (\bibinfo {year} {2017})}\BibitemShut {NoStop}%
\bibitem [{\citenamefont {Campbell}\ and\ \citenamefont
  {Hamilton}(2017)}]{campbell2017}%
  \BibitemOpen
  \bibfield  {author} {\bibinfo {author} {\bibfnamefont {W.~C.}\ \bibnamefont
  {Campbell}}\ and\ \bibinfo {author} {\bibfnamefont {P.}~\bibnamefont
  {Hamilton}},\ }\href {https://doi.org/10.1088/1361-6455/aa5a8f} {\bibfield
  {journal} {\bibinfo  {journal} {J. Phys. B: Atom. Mol. Opt. Phys.}\ }\textbf
  {\bibinfo {volume} {50}},\ \bibinfo {pages} {064002} (\bibinfo {year}
  {2017})}\BibitemShut {NoStop}%
\bibitem [{\citenamefont {Gilmore}\ \emph {et~al.}(2021)\citenamefont
  {Gilmore}, \citenamefont {Affolter}, \citenamefont {Lewis-Swan},
  \citenamefont {Barberena}, \citenamefont {Jordan}, \citenamefont {Rey},\ and\
  \citenamefont {Bollinger}}]{gilmore2021}%
  \BibitemOpen
  \bibfield  {author} {\bibinfo {author} {\bibfnamefont {K.~A.}\ \bibnamefont
  {Gilmore}}, \bibinfo {author} {\bibfnamefont {M.}~\bibnamefont {Affolter}},
  \bibinfo {author} {\bibfnamefont {R.~J.}\ \bibnamefont {Lewis-Swan}},
  \bibinfo {author} {\bibfnamefont {D.}~\bibnamefont {Barberena}}, \bibinfo
  {author} {\bibfnamefont {E.}~\bibnamefont {Jordan}}, \bibinfo {author}
  {\bibfnamefont {A.~M.}\ \bibnamefont {Rey}},\ and\ \bibinfo {author}
  {\bibfnamefont {J.~J.}\ \bibnamefont {Bollinger}},\ }\href
  {https://doi.org/10.1126/science.abi5226} {\bibfield  {journal} {\bibinfo
  {journal} {Science}\ }\textbf {\bibinfo {volume} {373}},\ \bibinfo {pages}
  {673} (\bibinfo {year} {2021})}\BibitemShut {NoStop}%
\bibitem [{\citenamefont {Kim}\ \emph {et~al.}(2009)\citenamefont {Kim},
  \citenamefont {Chang}, \citenamefont {Islam}, \citenamefont {Korenblit},
  \citenamefont {Duan},\ and\ \citenamefont {Monroe}}]{kim2009}%
  \BibitemOpen
  \bibfield  {author} {\bibinfo {author} {\bibfnamefont {K.}~\bibnamefont
  {Kim}}, \bibinfo {author} {\bibfnamefont {M.-S.}\ \bibnamefont {Chang}},
  \bibinfo {author} {\bibfnamefont {R.}~\bibnamefont {Islam}}, \bibinfo
  {author} {\bibfnamefont {S.}~\bibnamefont {Korenblit}}, \bibinfo {author}
  {\bibfnamefont {L.-M.}\ \bibnamefont {Duan}},\ and\ \bibinfo {author}
  {\bibfnamefont {C.}~\bibnamefont {Monroe}},\ }\href
  {https://doi.org/10.1103/PhysRevLett.103.120502} {\bibfield  {journal}
  {\bibinfo  {journal} {Phys. Rev. Lett.}\ }\textbf {\bibinfo {volume} {103}},\
  \bibinfo {pages} {120502} (\bibinfo {year} {2009})}\BibitemShut {NoStop}%
\bibitem [{\citenamefont {M{\"u}ller}\ \emph {et~al.}(2011)\citenamefont
  {M{\"u}ller}, \citenamefont {Hammerer}, \citenamefont {Zhou}, \citenamefont
  {Roos},\ and\ \citenamefont {Zoller}}]{muller2011}%
  \BibitemOpen
  \bibfield  {author} {\bibinfo {author} {\bibfnamefont {M.}~\bibnamefont
  {M{\"u}ller}}, \bibinfo {author} {\bibfnamefont {K.}~\bibnamefont
  {Hammerer}}, \bibinfo {author} {\bibfnamefont {Y.~L.}\ \bibnamefont {Zhou}},
  \bibinfo {author} {\bibfnamefont {C.~F.}\ \bibnamefont {Roos}},\ and\
  \bibinfo {author} {\bibfnamefont {P.}~\bibnamefont {Zoller}},\ }\href
  {https://doi.org/10.1088/1367-2630/13/8/085007} {\bibfield  {journal}
  {\bibinfo  {journal} {New J. Phys.}\ }\textbf {\bibinfo {volume} {13}},\
  \bibinfo {pages} {085007} (\bibinfo {year} {2011})}\BibitemShut {NoStop}%
\bibitem [{\citenamefont {Schneider}\ \emph {et~al.}(2012)\citenamefont
  {Schneider}, \citenamefont {Porras},\ and\ \citenamefont
  {Schaetz}}]{schneider2012}%
  \BibitemOpen
  \bibfield  {author} {\bibinfo {author} {\bibfnamefont {C.}~\bibnamefont
  {Schneider}}, \bibinfo {author} {\bibfnamefont {D.}~\bibnamefont {Porras}},\
  and\ \bibinfo {author} {\bibfnamefont {T.}~\bibnamefont {Schaetz}},\ }\href
  {https://doi.org/10.1088/0034-4885/75/2/024401} {\bibfield  {journal}
  {\bibinfo  {journal} {Rep. Prog. Phys.}\ }\textbf {\bibinfo {volume} {75}},\
  \bibinfo {pages} {024401} (\bibinfo {year} {2012})}\BibitemShut {NoStop}%
\bibitem [{\citenamefont {Leibfried}\ \emph {et~al.}(2003)\citenamefont
  {Leibfried}, \citenamefont {Blatt}, \citenamefont {Monroe},\ and\
  \citenamefont {Wineland}}]{leibfried2003}%
  \BibitemOpen
  \bibfield  {author} {\bibinfo {author} {\bibfnamefont {D.}~\bibnamefont
  {Leibfried}}, \bibinfo {author} {\bibfnamefont {R.}~\bibnamefont {Blatt}},
  \bibinfo {author} {\bibfnamefont {C.}~\bibnamefont {Monroe}},\ and\ \bibinfo
  {author} {\bibfnamefont {D.~J.}\ \bibnamefont {Wineland}},\ }\href
  {https://doi.org/10.1103/RevModPhys.75.281} {\bibfield  {journal} {\bibinfo
  {journal} {Rev. Mod. Phys.}\ }\textbf {\bibinfo {volume} {75}},\ \bibinfo
  {pages} {281} (\bibinfo {year} {2003})}\BibitemShut {NoStop}%
\bibitem [{\citenamefont {Haljan}\ \emph {et~al.}(2005)\citenamefont {Haljan},
  \citenamefont {Brickman}, \citenamefont {Deslauriers}, \citenamefont {Lee},\
  and\ \citenamefont {Monroe}}]{haljan2005}%
  \BibitemOpen
  \bibfield  {author} {\bibinfo {author} {\bibfnamefont {P.~C.}\ \bibnamefont
  {Haljan}}, \bibinfo {author} {\bibfnamefont {K.-A.}\ \bibnamefont
  {Brickman}}, \bibinfo {author} {\bibfnamefont {L.}~\bibnamefont
  {Deslauriers}}, \bibinfo {author} {\bibfnamefont {P.~J.}\ \bibnamefont
  {Lee}},\ and\ \bibinfo {author} {\bibfnamefont {C.}~\bibnamefont {Monroe}},\
  }\href {https://doi.org/10.1103/PhysRevLett.94.153602} {\bibfield  {journal}
  {\bibinfo  {journal} {Phys. Rev. Lett.}\ }\textbf {\bibinfo {volume} {94}},\
  \bibinfo {pages} {153602} (\bibinfo {year} {2005})}\BibitemShut {NoStop}%
\bibitem [{\citenamefont {Behrle}\ \emph {et~al.}(2023)\citenamefont {Behrle},
  \citenamefont {Nguyen}, \citenamefont {Reiter}, \citenamefont {Baur},
  \citenamefont {de~Neeve}, \citenamefont {Stadler}, \citenamefont {Marinelli},
  \citenamefont {Lancellotti}, \citenamefont {Yelin},\ and\ \citenamefont
  {Home}}]{behrle2023}%
  \BibitemOpen
  \bibfield  {author} {\bibinfo {author} {\bibfnamefont {T.}~\bibnamefont
  {Behrle}}, \bibinfo {author} {\bibfnamefont {T.~L.}\ \bibnamefont {Nguyen}},
  \bibinfo {author} {\bibfnamefont {F.}~\bibnamefont {Reiter}}, \bibinfo
  {author} {\bibfnamefont {D.}~\bibnamefont {Baur}}, \bibinfo {author}
  {\bibfnamefont {B.}~\bibnamefont {de~Neeve}}, \bibinfo {author}
  {\bibfnamefont {M.}~\bibnamefont {Stadler}}, \bibinfo {author} {\bibfnamefont
  {M.}~\bibnamefont {Marinelli}}, \bibinfo {author} {\bibfnamefont
  {F.}~\bibnamefont {Lancellotti}}, \bibinfo {author} {\bibfnamefont {S.~F.}\
  \bibnamefont {Yelin}},\ and\ \bibinfo {author} {\bibfnamefont {J.~P.}\
  \bibnamefont {Home}},\ }\href
  {https://doi.org/10.1103/PhysRevLett.131.043605} {\bibfield  {journal}
  {\bibinfo  {journal} {Phys. Rev. Lett.}\ }\textbf {\bibinfo {volume} {131}},\
  \bibinfo {pages} {043605} (\bibinfo {year} {2023})}\BibitemShut {NoStop}%
\bibitem [{\citenamefont {M{\"u}ller}\ \emph {et~al.}(2008)\citenamefont
  {M{\"u}ller}, \citenamefont {Liang}, \citenamefont {Lesanovsky},\ and\
  \citenamefont {Zoller}}]{muller2008}%
  \BibitemOpen
  \bibfield  {author} {\bibinfo {author} {\bibfnamefont {M.}~\bibnamefont
  {M{\"u}ller}}, \bibinfo {author} {\bibfnamefont {L.}~\bibnamefont {Liang}},
  \bibinfo {author} {\bibfnamefont {I.}~\bibnamefont {Lesanovsky}},\ and\
  \bibinfo {author} {\bibfnamefont {P.}~\bibnamefont {Zoller}},\ }\href
  {https://doi.org/10.1088/1367-2630/10/9/093009} {\bibfield  {journal}
  {\bibinfo  {journal} {New J. Phys.}\ }\textbf {\bibinfo {volume} {10}},\
  \bibinfo {pages} {093009} (\bibinfo {year} {2008})}\BibitemShut {NoStop}%
\bibitem [{\citenamefont {Schmidt-Kaler}\ \emph {et~al.}(2011)\citenamefont
  {Schmidt-Kaler}, \citenamefont {Feldker}, \citenamefont {Kolbe},
  \citenamefont {Walz}, \citenamefont {M\"{u}ller}, \citenamefont {Zoller},
  \citenamefont {Li},\ and\ \citenamefont {Lesanovsky}}]{schmidtkaler2011}%
  \BibitemOpen
  \bibfield  {author} {\bibinfo {author} {\bibfnamefont {F.}~\bibnamefont
  {Schmidt-Kaler}}, \bibinfo {author} {\bibfnamefont {T.}~\bibnamefont
  {Feldker}}, \bibinfo {author} {\bibfnamefont {D.}~\bibnamefont {Kolbe}},
  \bibinfo {author} {\bibfnamefont {J.}~\bibnamefont {Walz}}, \bibinfo {author}
  {\bibfnamefont {M.}~\bibnamefont {M\"{u}ller}}, \bibinfo {author}
  {\bibfnamefont {P.}~\bibnamefont {Zoller}}, \bibinfo {author} {\bibfnamefont
  {W.}~\bibnamefont {Li}},\ and\ \bibinfo {author} {\bibfnamefont
  {I.}~\bibnamefont {Lesanovsky}},\ }\href
  {https://doi.org/10.1088/1367-2630/13/7/075014} {\bibfield  {journal}
  {\bibinfo  {journal} {New J. Phys.}\ }\textbf {\bibinfo {volume} {13}},\
  \bibinfo {pages} {075014} (\bibinfo {year} {2011})}\BibitemShut {NoStop}%
\bibitem [{\citenamefont {Feldker}\ \emph {et~al.}(2015)\citenamefont
  {Feldker}, \citenamefont {Bachor}, \citenamefont {Stappel}, \citenamefont
  {Kolbe}, \citenamefont {Gerritsma}, \citenamefont {Walz},\ and\ \citenamefont
  {Schmidt-Kaler}}]{feldker2015}%
  \BibitemOpen
  \bibfield  {author} {\bibinfo {author} {\bibfnamefont {T.}~\bibnamefont
  {Feldker}}, \bibinfo {author} {\bibfnamefont {P.}~\bibnamefont {Bachor}},
  \bibinfo {author} {\bibfnamefont {M.}~\bibnamefont {Stappel}}, \bibinfo
  {author} {\bibfnamefont {D.}~\bibnamefont {Kolbe}}, \bibinfo {author}
  {\bibfnamefont {R.}~\bibnamefont {Gerritsma}}, \bibinfo {author}
  {\bibfnamefont {J.}~\bibnamefont {Walz}},\ and\ \bibinfo {author}
  {\bibfnamefont {F.}~\bibnamefont {Schmidt-Kaler}},\ }\href
  {https://doi.org/10.1103/PhysRevLett.115.173001} {\bibfield  {journal}
  {\bibinfo  {journal} {Phys. Rev. Lett.}\ }\textbf {\bibinfo {volume} {115}},\
  \bibinfo {pages} {173001} (\bibinfo {year} {2015})}\BibitemShut {NoStop}%
\bibitem [{\citenamefont {Bachor}\ \emph {et~al.}(2016)\citenamefont {Bachor},
  \citenamefont {Feldker}, \citenamefont {Walz},\ and\ \citenamefont
  {Schmidt-Kaler}}]{bachor2016}%
  \BibitemOpen
  \bibfield  {author} {\bibinfo {author} {\bibfnamefont {P.}~\bibnamefont
  {Bachor}}, \bibinfo {author} {\bibfnamefont {T.}~\bibnamefont {Feldker}},
  \bibinfo {author} {\bibfnamefont {J.}~\bibnamefont {Walz}},\ and\ \bibinfo
  {author} {\bibfnamefont {F.}~\bibnamefont {Schmidt-Kaler}},\ }\href
  {https://doi.org/10.1088/0953-4075/49/15/154004} {\bibfield  {journal}
  {\bibinfo  {journal} {J. Phys. B: At. Mol. Opt. Phys.}\ }\textbf {\bibinfo
  {volume} {49}},\ \bibinfo {pages} {154004} (\bibinfo {year}
  {2016})}\BibitemShut {NoStop}%
\bibitem [{\citenamefont {Higgins}\ \emph
  {et~al.}(2017{\natexlab{a}})\citenamefont {Higgins}, \citenamefont {Li},
  \citenamefont {Pokorny}, \citenamefont {Zhang}, \citenamefont {Kress},
  \citenamefont {Maier}, \citenamefont {Haag}, \citenamefont {Bodart},
  \citenamefont {Lesanovsky},\ and\ \citenamefont {Hennrich}}]{higgins2017a}%
  \BibitemOpen
  \bibfield  {author} {\bibinfo {author} {\bibfnamefont {G.}~\bibnamefont
  {Higgins}}, \bibinfo {author} {\bibfnamefont {W.}~\bibnamefont {Li}},
  \bibinfo {author} {\bibfnamefont {F.}~\bibnamefont {Pokorny}}, \bibinfo
  {author} {\bibfnamefont {C.}~\bibnamefont {Zhang}}, \bibinfo {author}
  {\bibfnamefont {F.}~\bibnamefont {Kress}}, \bibinfo {author} {\bibfnamefont
  {C.}~\bibnamefont {Maier}}, \bibinfo {author} {\bibfnamefont
  {J.}~\bibnamefont {Haag}}, \bibinfo {author} {\bibfnamefont {Q.}~\bibnamefont
  {Bodart}}, \bibinfo {author} {\bibfnamefont {I.}~\bibnamefont {Lesanovsky}},\
  and\ \bibinfo {author} {\bibfnamefont {M.}~\bibnamefont {Hennrich}},\ }\href
  {https://doi.org/10.1103/PhysRevX.7.021038} {\bibfield  {journal} {\bibinfo
  {journal} {Phys. Rev. X}\ }\textbf {\bibinfo {volume} {7}},\ \bibinfo {pages}
  {021038} (\bibinfo {year} {2017}{\natexlab{a}})}\BibitemShut {NoStop}%
\bibitem [{\citenamefont {Higgins}\ \emph
  {et~al.}(2017{\natexlab{b}})\citenamefont {Higgins}, \citenamefont {Pokorny},
  \citenamefont {Zhang}, \citenamefont {Bodart},\ and\ \citenamefont
  {Hennrich}}]{higgins2017b}%
  \BibitemOpen
  \bibfield  {author} {\bibinfo {author} {\bibfnamefont {G.}~\bibnamefont
  {Higgins}}, \bibinfo {author} {\bibfnamefont {F.}~\bibnamefont {Pokorny}},
  \bibinfo {author} {\bibfnamefont {C.}~\bibnamefont {Zhang}}, \bibinfo
  {author} {\bibfnamefont {Q.}~\bibnamefont {Bodart}},\ and\ \bibinfo {author}
  {\bibfnamefont {M.}~\bibnamefont {Hennrich}},\ }\href
  {https://doi.org/10.1103/PhysRevLett.119.220501} {\bibfield  {journal}
  {\bibinfo  {journal} {Phys. Rev. Lett.}\ }\textbf {\bibinfo {volume} {119}},\
  \bibinfo {pages} {220501} (\bibinfo {year} {2017}{\natexlab{b}})}\BibitemShut
  {NoStop}%
\bibitem [{\citenamefont {Higgins}(2018)}]{higgins2018}%
  \BibitemOpen
  \bibfield  {author} {\bibinfo {author} {\bibfnamefont {G.}~\bibnamefont
  {Higgins}},\ }\emph {\bibinfo {title} {{A single trapped Rydberg ion}}},\
  \href {https://doi.org/10.1007/978-3-030-33770-4} {Ph.D. thesis},\ \bibinfo
  {school} {Stockholm University} (\bibinfo {year} {2018})\BibitemShut
  {NoStop}%
\bibitem [{\citenamefont {Higgins}\ \emph {et~al.}(2019)\citenamefont
  {Higgins}, \citenamefont {Pokorny}, \citenamefont {Zhang},\ and\
  \citenamefont {Hennrich}}]{higgins2019}%
  \BibitemOpen
  \bibfield  {author} {\bibinfo {author} {\bibfnamefont {G.}~\bibnamefont
  {Higgins}}, \bibinfo {author} {\bibfnamefont {F.}~\bibnamefont {Pokorny}},
  \bibinfo {author} {\bibfnamefont {C.}~\bibnamefont {Zhang}},\ and\ \bibinfo
  {author} {\bibfnamefont {M.}~\bibnamefont {Hennrich}},\ }\href
  {https://doi.org/10.1103/PhysRevLett.123.153602} {\bibfield  {journal}
  {\bibinfo  {journal} {Phys. Rev. Lett.}\ }\textbf {\bibinfo {volume} {123}},\
  \bibinfo {pages} {153602} (\bibinfo {year} {2019})}\BibitemShut {NoStop}%
\bibitem [{\citenamefont {Mokhberi}\ \emph {et~al.}(2019)\citenamefont
  {Mokhberi}, \citenamefont {Vogel}, \citenamefont {Andrijauskas},
  \citenamefont {Bachor}, \citenamefont {Walz},\ and\ \citenamefont
  {Schmidt-Kaler}}]{mokhberi2019}%
  \BibitemOpen
  \bibfield  {author} {\bibinfo {author} {\bibfnamefont {A.}~\bibnamefont
  {Mokhberi}}, \bibinfo {author} {\bibfnamefont {J.}~\bibnamefont {Vogel}},
  \bibinfo {author} {\bibfnamefont {J.}~\bibnamefont {Andrijauskas}}, \bibinfo
  {author} {\bibfnamefont {P.}~\bibnamefont {Bachor}}, \bibinfo {author}
  {\bibfnamefont {J.}~\bibnamefont {Walz}},\ and\ \bibinfo {author}
  {\bibfnamefont {F.}~\bibnamefont {Schmidt-Kaler}},\ }\href
  {https://doi.org/10.1088/1361-6455/ab402b} {\bibfield  {journal} {\bibinfo
  {journal} {J. Phys. B: Atom. Mol. Opt. Phys.}\ }\textbf {\bibinfo {volume}
  {52}},\ \bibinfo {pages} {214001} (\bibinfo {year} {2019})}\BibitemShut
  {NoStop}%
\bibitem [{\citenamefont {Vogel}\ \emph {et~al.}(2019)\citenamefont {Vogel},
  \citenamefont {Li}, \citenamefont {Mokhberi}, \citenamefont {Lesanovsky},\
  and\ \citenamefont {Schmidt-Kaler}}]{vogel2019}%
  \BibitemOpen
  \bibfield  {author} {\bibinfo {author} {\bibfnamefont {J.}~\bibnamefont
  {Vogel}}, \bibinfo {author} {\bibfnamefont {W.}~\bibnamefont {Li}}, \bibinfo
  {author} {\bibfnamefont {A.}~\bibnamefont {Mokhberi}}, \bibinfo {author}
  {\bibfnamefont {I.}~\bibnamefont {Lesanovsky}},\ and\ \bibinfo {author}
  {\bibfnamefont {F.}~\bibnamefont {Schmidt-Kaler}},\ }\href
  {https://doi.org/10.1103/PhysRevLett.123.153603} {\bibfield  {journal}
  {\bibinfo  {journal} {Phys. Rev. Lett.}\ }\textbf {\bibinfo {volume} {123}},\
  \bibinfo {pages} {153603} (\bibinfo {year} {2019})}\BibitemShut {NoStop}%
\bibitem [{\citenamefont {Andrijauskas}\ \emph {et~al.}(2021)\citenamefont
  {Andrijauskas}, \citenamefont {Vogel}, \citenamefont {Mokhberi},\ and\
  \citenamefont {Schmidt-Kaler}}]{andrijauskas2021}%
  \BibitemOpen
  \bibfield  {author} {\bibinfo {author} {\bibfnamefont {J.}~\bibnamefont
  {Andrijauskas}}, \bibinfo {author} {\bibfnamefont {J.}~\bibnamefont {Vogel}},
  \bibinfo {author} {\bibfnamefont {A.}~\bibnamefont {Mokhberi}},\ and\
  \bibinfo {author} {\bibfnamefont {F.}~\bibnamefont {Schmidt-Kaler}},\ }\href
  {https://doi.org/10.1103/PhysRevLett.127.203001} {\bibfield  {journal}
  {\bibinfo  {journal} {Phys. Rev. Lett.}\ }\textbf {\bibinfo {volume} {127}},\
  \bibinfo {pages} {203001} (\bibinfo {year} {2021})}\BibitemShut {NoStop}%
\bibitem [{\citenamefont {Mokhberi}\ \emph {et~al.}(2020)\citenamefont
  {Mokhberi}, \citenamefont {Hennrich},\ and\ \citenamefont
  {Schmidt-Kaler}}]{mokhberi2020}%
  \BibitemOpen
  \bibfield  {author} {\bibinfo {author} {\bibfnamefont {A.}~\bibnamefont
  {Mokhberi}}, \bibinfo {author} {\bibfnamefont {M.}~\bibnamefont {Hennrich}},\
  and\ \bibinfo {author} {\bibfnamefont {F.}~\bibnamefont {Schmidt-Kaler}}\
  }(\bibinfo  {publisher} {Academic Press},\ \bibinfo {year} {2020})\
  Chap.~\bibinfo {chapter} {4}, pp.\ \bibinfo {pages} {233--306}\BibitemShut
  {NoStop}%
\bibitem [{\citenamefont {Weber}\ \emph {et~al.}(2017)\citenamefont {Weber},
  \citenamefont {Tresp}, \citenamefont {Menke}, \citenamefont {Urvoy},
  \citenamefont {Firstenberg}, \citenamefont {B{\"u}chler},\ and\ \citenamefont
  {Hofferberth}}]{weber2017}%
  \BibitemOpen
  \bibfield  {author} {\bibinfo {author} {\bibfnamefont {S.}~\bibnamefont
  {Weber}}, \bibinfo {author} {\bibfnamefont {C.}~\bibnamefont {Tresp}},
  \bibinfo {author} {\bibfnamefont {H.}~\bibnamefont {Menke}}, \bibinfo
  {author} {\bibfnamefont {A.}~\bibnamefont {Urvoy}}, \bibinfo {author}
  {\bibfnamefont {O.}~\bibnamefont {Firstenberg}}, \bibinfo {author}
  {\bibfnamefont {H.~P.}\ \bibnamefont {B{\"u}chler}},\ and\ \bibinfo {author}
  {\bibfnamefont {S.}~\bibnamefont {Hofferberth}},\ }\href
  {https://doi.org/10.1088/1361-6455/aa743a} {\bibfield  {journal} {\bibinfo
  {journal} {J. Phys. B}\ }\textbf {\bibinfo {volume} {50}},\ \bibinfo {pages}
  {133001} (\bibinfo {year} {2017})}\BibitemShut {NoStop}%
\bibitem [{\citenamefont {Zhang}\ \emph {et~al.}(2020)\citenamefont {Zhang},
  \citenamefont {Pokorny}, \citenamefont {Li}, \citenamefont {Higgins},
  \citenamefont {P{\" o}schl}, \citenamefont {Lesanovsky},\ and\ \citenamefont
  {Hennrich}}]{zhang2020}%
  \BibitemOpen
  \bibfield  {author} {\bibinfo {author} {\bibfnamefont {C.}~\bibnamefont
  {Zhang}}, \bibinfo {author} {\bibfnamefont {F.}~\bibnamefont {Pokorny}},
  \bibinfo {author} {\bibfnamefont {W.}~\bibnamefont {Li}}, \bibinfo {author}
  {\bibfnamefont {G.}~\bibnamefont {Higgins}}, \bibinfo {author} {\bibfnamefont
  {A.}~\bibnamefont {P{\" o}schl}}, \bibinfo {author} {\bibfnamefont
  {I.}~\bibnamefont {Lesanovsky}},\ and\ \bibinfo {author} {\bibfnamefont
  {M.}~\bibnamefont {Hennrich}},\ }\href
  {https://doi.org/10.1038/s41586-020-2152-9} {\bibfield  {journal} {\bibinfo
  {journal} {Nature}\ }\textbf {\bibinfo {volume} {580}},\ \bibinfo {pages}
  {345} (\bibinfo {year} {2020})}\BibitemShut {NoStop}%
\bibitem [{\citenamefont {Paul}(1990)}]{paul1990}%
  \BibitemOpen
  \bibfield  {author} {\bibinfo {author} {\bibfnamefont {W.}~\bibnamefont
  {Paul}},\ }\href {https://doi.org/10.1103/RevModPhys.62.531} {\bibfield
  {journal} {\bibinfo  {journal} {Rev. Mod. Phys.}\ }\textbf {\bibinfo {volume}
  {62}},\ \bibinfo {pages} {531} (\bibinfo {year} {1990})}\BibitemShut
  {NoStop}%
\bibitem [{\citenamefont {Wineland}\ \emph {et~al.}(1998)\citenamefont
  {Wineland}, \citenamefont {Monroe}, \citenamefont {Itano}, \citenamefont
  {Leibfried}, \citenamefont {King},\ and\ \citenamefont
  {Meekhof}}]{wineland1998}%
  \BibitemOpen
  \bibfield  {author} {\bibinfo {author} {\bibfnamefont {D.~J.}\ \bibnamefont
  {Wineland}}, \bibinfo {author} {\bibfnamefont {C.}~\bibnamefont {Monroe}},
  \bibinfo {author} {\bibfnamefont {W.~M.}\ \bibnamefont {Itano}}, \bibinfo
  {author} {\bibfnamefont {D.}~\bibnamefont {Leibfried}}, \bibinfo {author}
  {\bibfnamefont {B.~E.}\ \bibnamefont {King}},\ and\ \bibinfo {author}
  {\bibfnamefont {D.~M.}\ \bibnamefont {Meekhof}},\ }\href
  {https://doi.org/10.6028/jres.103.019} {\bibfield  {journal} {\bibinfo
  {journal} {J. Res. Natl. Inst. Stand. Technol.}\ }\textbf {\bibinfo {volume}
  {103}},\ \bibinfo {pages} {259} (\bibinfo {year} {1998})}\BibitemShut
  {NoStop}%
\bibitem [{\citenamefont {Major}\ \emph {et~al.}(2005)\citenamefont {Major},
  \citenamefont {Gheorghe},\ and\ \citenamefont {Werth}}]{major2005}%
  \BibitemOpen
  \bibfield  {author} {\bibinfo {author} {\bibfnamefont {F.~G.}\ \bibnamefont
  {Major}}, \bibinfo {author} {\bibfnamefont {V.~N.}\ \bibnamefont
  {Gheorghe}},\ and\ \bibinfo {author} {\bibfnamefont {G.}~\bibnamefont
  {Werth}},\ }\href {https://doi.org/10.1007/b137836} {\emph {\bibinfo {title}
  {{Charged Particle Traps}}}},\ \bibinfo {edition} {1st}\ ed.\ (\bibinfo
  {publisher} {Springer},\ \bibinfo {year} {2005})\BibitemShut {NoStop}%
\bibitem [{\citenamefont {Brown}(1991)}]{brown1991}%
  \BibitemOpen
  \bibfield  {author} {\bibinfo {author} {\bibfnamefont {L.~S.}\ \bibnamefont
  {Brown}},\ }\href {https://doi.org/10.1103/PhysRevLett.66.527} {\bibfield
  {journal} {\bibinfo  {journal} {Phys. Rev. Lett.}\ }\textbf {\bibinfo
  {volume} {66}},\ \bibinfo {pages} {527} (\bibinfo {year} {1991})}\BibitemShut
  {NoStop}%
\bibitem [{\citenamefont {Raizen}\ \emph {et~al.}(1992)\citenamefont {Raizen},
  \citenamefont {Gilligan}, \citenamefont {Bergquist}, \citenamefont {Itano},\
  and\ \citenamefont {Wineland}}]{raizen1992}%
  \BibitemOpen
  \bibfield  {author} {\bibinfo {author} {\bibfnamefont {M.~G.}\ \bibnamefont
  {Raizen}}, \bibinfo {author} {\bibfnamefont {J.~M.}\ \bibnamefont
  {Gilligan}}, \bibinfo {author} {\bibfnamefont {J.~C.}\ \bibnamefont
  {Bergquist}}, \bibinfo {author} {\bibfnamefont {W.~M.}\ \bibnamefont
  {Itano}},\ and\ \bibinfo {author} {\bibfnamefont {D.~J.}\ \bibnamefont
  {Wineland}},\ }\href {https://doi.org/10.1103/PhysRevA.45.6493} {\bibfield
  {journal} {\bibinfo  {journal} {Phys. Rev. A}\ }\textbf {\bibinfo {volume}
  {45}},\ \bibinfo {pages} {6493} (\bibinfo {year} {1992})}\BibitemShut
  {NoStop}%
\bibitem [{\citenamefont {Drewsen}\ and\ \citenamefont
  {Br\o{}ner}(2000)}]{drewsen2000}%
  \BibitemOpen
  \bibfield  {author} {\bibinfo {author} {\bibfnamefont {M.}~\bibnamefont
  {Drewsen}}\ and\ \bibinfo {author} {\bibfnamefont {A.}~\bibnamefont
  {Br\o{}ner}},\ }\href {https://doi.org/10.1103/PhysRevA.62.045401} {\bibfield
   {journal} {\bibinfo  {journal} {Phys. Rev. A}\ }\textbf {\bibinfo {volume}
  {62}},\ \bibinfo {pages} {045401} (\bibinfo {year} {2000})}\BibitemShut
  {NoStop}%
\bibitem [{\citenamefont {Cook}\ \emph {et~al.}(1985)\citenamefont {Cook},
  \citenamefont {Shankland},\ and\ \citenamefont {Wells}}]{cook1985}%
  \BibitemOpen
  \bibfield  {author} {\bibinfo {author} {\bibfnamefont {R.~J.}\ \bibnamefont
  {Cook}}, \bibinfo {author} {\bibfnamefont {D.~G.}\ \bibnamefont
  {Shankland}},\ and\ \bibinfo {author} {\bibfnamefont {A.~L.}\ \bibnamefont
  {Wells}},\ }\href {https://doi.org/10.1103/PhysRevA.31.564} {\bibfield
  {journal} {\bibinfo  {journal} {Phys. Rev. A}\ }\textbf {\bibinfo {volume}
  {31}},\ \bibinfo {pages} {564} (\bibinfo {year} {1985})}\BibitemShut
  {NoStop}%
\bibitem [{\citenamefont {Berkeland}\ \emph {et~al.}(1998)\citenamefont
  {Berkeland}, \citenamefont {Miller}, \citenamefont {Bergquist}, \citenamefont
  {Itano},\ and\ \citenamefont {Wineland}}]{berkeland1998}%
  \BibitemOpen
  \bibfield  {author} {\bibinfo {author} {\bibfnamefont {D.~J.}\ \bibnamefont
  {Berkeland}}, \bibinfo {author} {\bibfnamefont {J.~D.}\ \bibnamefont
  {Miller}}, \bibinfo {author} {\bibfnamefont {J.~C.}\ \bibnamefont
  {Bergquist}}, \bibinfo {author} {\bibfnamefont {W.~M.}\ \bibnamefont
  {Itano}},\ and\ \bibinfo {author} {\bibfnamefont {D.~J.}\ \bibnamefont
  {Wineland}},\ }\href {https://doi.org/10.1063/1.367318} {\bibfield  {journal}
  {\bibinfo  {journal} {J. Appl. Phys.}\ }\textbf {\bibinfo {volume} {83}},\
  \bibinfo {pages} {5025} (\bibinfo {year} {1998})}\BibitemShut {NoStop}%
\bibitem [{\citenamefont {Braak}(2011)}]{braak2011}%
  \BibitemOpen
  \bibfield  {author} {\bibinfo {author} {\bibfnamefont {D.}~\bibnamefont
  {Braak}},\ }\href {https://doi.org/10.1103/PhysRevLett.107.100401} {\bibfield
   {journal} {\bibinfo  {journal} {Phys. Rev. Lett.}\ }\textbf {\bibinfo
  {volume} {107}},\ \bibinfo {pages} {100401} (\bibinfo {year}
  {2011})}\BibitemShut {NoStop}%
\bibitem [{\citenamefont {Eckle}\ and\ \citenamefont
  {Johannesson}(2017)}]{eckle2017}%
  \BibitemOpen
  \bibfield  {author} {\bibinfo {author} {\bibfnamefont {H.-P.}\ \bibnamefont
  {Eckle}}\ and\ \bibinfo {author} {\bibfnamefont {H.}~\bibnamefont
  {Johannesson}},\ }\href {https://doi.org/10.1088/1751-8121/aa785a} {\bibfield
   {journal} {\bibinfo  {journal} {J. Phys. A: Math. Theor.}\ }\textbf
  {\bibinfo {volume} {50}},\ \bibinfo {pages} {294004} (\bibinfo {year}
  {2017})}\BibitemShut {NoStop}%
\bibitem [{\citenamefont {Xie}\ \emph {et~al.}(2017)\citenamefont {Xie},
  \citenamefont {Zhong}, \citenamefont {Batchelor},\ and\ \citenamefont
  {Lee}}]{xie2017}%
  \BibitemOpen
  \bibfield  {author} {\bibinfo {author} {\bibfnamefont {Q.}~\bibnamefont
  {Xie}}, \bibinfo {author} {\bibfnamefont {H.}~\bibnamefont {Zhong}}, \bibinfo
  {author} {\bibfnamefont {M.~T.}\ \bibnamefont {Batchelor}},\ and\ \bibinfo
  {author} {\bibfnamefont {C.}~\bibnamefont {Lee}},\ }\href
  {https://doi.org/10.1088/1751-8121/aa5a65} {\bibfield  {journal} {\bibinfo
  {journal} {J. Phys. A}\ }\textbf {\bibinfo {volume} {50}},\ \bibinfo {pages}
  {113001} (\bibinfo {year} {2017})}\BibitemShut {NoStop}%
\bibitem [{\citenamefont {Sambe}(1973)}]{sambe1973}%
  \BibitemOpen
  \bibfield  {author} {\bibinfo {author} {\bibfnamefont {H.}~\bibnamefont
  {Sambe}},\ }\href {https://doi.org/10.1103/PhysRevA.7.2203} {\bibfield
  {journal} {\bibinfo  {journal} {Phys. Rev. A}\ }\textbf {\bibinfo {volume}
  {7}},\ \bibinfo {pages} {2203} (\bibinfo {year} {1973})}\BibitemShut
  {NoStop}%
\bibitem [{\citenamefont {Eckardt}\ and\ \citenamefont
  {Anisimovas}(2015)}]{eckardt2015}%
  \BibitemOpen
  \bibfield  {author} {\bibinfo {author} {\bibfnamefont {A.}~\bibnamefont
  {Eckardt}}\ and\ \bibinfo {author} {\bibfnamefont {E.}~\bibnamefont
  {Anisimovas}},\ }\href {https://doi.org/10.1088/1367-2630/17/9/093039}
  {\bibfield  {journal} {\bibinfo  {journal} {New J. Phys.}\ }\textbf {\bibinfo
  {volume} {17}},\ \bibinfo {pages} {093039} (\bibinfo {year}
  {2015})}\BibitemShut {NoStop}%
\bibitem [{\citenamefont {Novi{\v c}enko}\ \emph {et~al.}(2017)\citenamefont
  {Novi{\v c}enko}, \citenamefont {Anisimovas},\ and\ \citenamefont {Juzeli{\=
  u}nas}}]{novicenko2017}%
  \BibitemOpen
  \bibfield  {author} {\bibinfo {author} {\bibfnamefont {V.}~\bibnamefont
  {Novi{\v c}enko}}, \bibinfo {author} {\bibfnamefont {E.}~\bibnamefont
  {Anisimovas}},\ and\ \bibinfo {author} {\bibfnamefont {G.}~\bibnamefont
  {Juzeli{\= u}nas}},\ }\href {https://doi.org/10.1103/PhysRevA.95.023615}
  {\bibfield  {journal} {\bibinfo  {journal} {Phys. Rev. A}\ }\textbf {\bibinfo
  {volume} {95}},\ \bibinfo {pages} {023615} (\bibinfo {year}
  {2017})}\BibitemShut {NoStop}%
\bibitem [{\citenamefont {Rodriguez-Vega}\ \emph {et~al.}(2021)\citenamefont
  {Rodriguez-Vega}, \citenamefont {Vogl},\ and\ \citenamefont
  {Fiete}}]{rodriguez2021}%
  \BibitemOpen
  \bibfield  {author} {\bibinfo {author} {\bibfnamefont {M.}~\bibnamefont
  {Rodriguez-Vega}}, \bibinfo {author} {\bibfnamefont {M.}~\bibnamefont
  {Vogl}},\ and\ \bibinfo {author} {\bibfnamefont {G.~A.}\ \bibnamefont
  {Fiete}},\ }\href {https://doi.org/10.1016/j.aop.2021.168434} {\bibfield
  {journal} {\bibinfo  {journal} {Ann. Phys.}\ }\textbf {\bibinfo {volume}
  {435}},\ \bibinfo {pages} {168434} (\bibinfo {year} {2021})}\BibitemShut
  {NoStop}%
\bibitem [{\citenamefont {Djerad}(1991)}]{djerad1991}%
  \BibitemOpen
  \bibfield  {author} {\bibinfo {author} {\bibfnamefont {M.~T.}\ \bibnamefont
  {Djerad}},\ }\href {https://doi.org/10.1051/jp2:1991135} {\bibfield
  {journal} {\bibinfo  {journal} {J. Phys. II}\ }\textbf {\bibinfo {volume}
  {1}},\ \bibinfo {pages} {1} (\bibinfo {year} {1991})}\BibitemShut {NoStop}%
\bibitem [{\citenamefont {Gallagher}(2023)}]{gallagher2023}%
  \BibitemOpen
  \bibfield  {author} {\bibinfo {author} {\bibfnamefont {T.~F.}\ \bibnamefont
  {Gallagher}},\ }in\ \href {https://doi.org/10.1007/978-3-030-73893-8} {\emph
  {\bibinfo {booktitle} {Springer Handbook of Atomic, Molecular, and Optical
  Physics}}},\ \bibinfo {editor} {edited by\ \bibinfo {editor} {\bibfnamefont
  {G.~W.~F.}\ \bibnamefont {Drake}}}\ (\bibinfo  {publisher} {Springer},\
  \bibinfo {year} {2023})\ Chap.~\bibinfo {chapter} {15}, pp.\ \bibinfo {pages}
  {231--240}\BibitemShut {NoStop}%
\bibitem [{\citenamefont {Steck}(2024)}]{steck2024}%
  \BibitemOpen
  \bibfield  {author} {\bibinfo {author} {\bibfnamefont {D.~A.}\ \bibnamefont
  {Steck}},\ }\href
  {https://atomoptics.uoregon.edu/~dsteck/teaching/quantum-optics/} {\bibinfo
  {title} {{Quantum and Atom Optics}}} (\bibinfo {year} {2024})\BibitemShut
  {NoStop}%
\bibitem [{\citenamefont {Wilkinson}\ \emph
  {et~al.}(2024{\natexlab{a}})\citenamefont {Wilkinson}, \citenamefont
  {Bolsmann}, \citenamefont {Guedes}, \citenamefont {Müller},\ and\
  \citenamefont {Lesanovsky}}]{wilkinson2024b}%
  \BibitemOpen
  \bibfield  {author} {\bibinfo {author} {\bibfnamefont {J.~W.~P.}\
  \bibnamefont {Wilkinson}}, \bibinfo {author} {\bibfnamefont {K.}~\bibnamefont
  {Bolsmann}}, \bibinfo {author} {\bibfnamefont {T.~L.~M.}\ \bibnamefont
  {Guedes}}, \bibinfo {author} {\bibfnamefont {M.}~\bibnamefont {Müller}},\
  and\ \bibinfo {author} {\bibfnamefont {I.}~\bibnamefont {Lesanovsky}},\
  }\href {https://arxiv.org/abs/2412.13699} {\bibinfo {title} {Two-qubit gate
  protocols with microwave-dressed {R}ydberg ions in a linear {P}aul trap}}
  (\bibinfo {year} {2024}{\natexlab{a}}),\ \Eprint
  {https://arxiv.org/abs/2412.13699} {arXiv:2412.13699 [quant-ph]} \BibitemShut
  {NoStop}%
\bibitem [{\citenamefont {Wilkinson}\ \emph
  {et~al.}(2024{\natexlab{b}})\citenamefont {Wilkinson}, \citenamefont {Li},\
  and\ \citenamefont {Lesanovsky}}]{wilkinson2024a}%
  \BibitemOpen
  \bibfield  {author} {\bibinfo {author} {\bibfnamefont {J.~W.~P.}\
  \bibnamefont {Wilkinson}}, \bibinfo {author} {\bibfnamefont {W.}~\bibnamefont
  {Li}},\ and\ \bibinfo {author} {\bibfnamefont {I.}~\bibnamefont
  {Lesanovsky}},\ }\href {https://doi.org/10.1103/PhysRevLett.132.223401}
  {\bibfield  {journal} {\bibinfo  {journal} {Phys. Rev. Lett.}\ }\textbf
  {\bibinfo {volume} {132}},\ \bibinfo {pages} {223401} (\bibinfo {year}
  {2024}{\natexlab{b}})}\BibitemShut {NoStop}%
\bibitem [{\citenamefont {Shirley}(1965)}]{shirley1965}%
  \BibitemOpen
  \bibfield  {author} {\bibinfo {author} {\bibfnamefont {J.~H.}\ \bibnamefont
  {Shirley}},\ }\href {https://doi.org/10.1103/PhysRev.138.B979} {\bibfield
  {journal} {\bibinfo  {journal} {Phys. Rev.}\ }\textbf {\bibinfo {volume}
  {138}},\ \bibinfo {pages} {B979} (\bibinfo {year} {1965})}\BibitemShut
  {NoStop}%
\bibitem [{\citenamefont {Barone}\ \emph {et~al.}(1977)\citenamefont {Barone},
  \citenamefont {Narcowich},\ and\ \citenamefont {Narcowich}}]{barone1977}%
  \BibitemOpen
  \bibfield  {author} {\bibinfo {author} {\bibfnamefont {S.~R.}\ \bibnamefont
  {Barone}}, \bibinfo {author} {\bibfnamefont {M.~A.}\ \bibnamefont
  {Narcowich}},\ and\ \bibinfo {author} {\bibfnamefont {F.~J.}\ \bibnamefont
  {Narcowich}},\ }\href {https://doi.org/10.1103/PhysRevA.15.1109} {\bibfield
  {journal} {\bibinfo  {journal} {Phys. Rev. A}\ }\textbf {\bibinfo {volume}
  {15}},\ \bibinfo {pages} {1109} (\bibinfo {year} {1977})}\BibitemShut
  {NoStop}%
\bibitem [{\citenamefont {Ho}\ \emph {et~al.}(1983)\citenamefont {Ho},
  \citenamefont {Chu},\ and\ \citenamefont {Tietz}}]{ho1983}%
  \BibitemOpen
  \bibfield  {author} {\bibinfo {author} {\bibfnamefont {T.-S.}\ \bibnamefont
  {Ho}}, \bibinfo {author} {\bibfnamefont {S.-I.}\ \bibnamefont {Chu}},\ and\
  \bibinfo {author} {\bibfnamefont {J.~V.}\ \bibnamefont {Tietz}},\ }\href
  {https://doi.org/10.1016/0009-2614(83)80732-5} {\bibfield  {journal}
  {\bibinfo  {journal} {Chem. Phys. Lett.}\ }\textbf {\bibinfo {volume} {96}},\
  \bibinfo {pages} {464} (\bibinfo {year} {1983})}\BibitemShut {NoStop}%
\bibitem [{\citenamefont {Blanes}\ \emph {et~al.}(2009)\citenamefont {Blanes},
  \citenamefont {Casas}, \citenamefont {Oteo},\ and\ \citenamefont
  {Ros}}]{blanes2009}%
  \BibitemOpen
  \bibfield  {author} {\bibinfo {author} {\bibfnamefont {S.}~\bibnamefont
  {Blanes}}, \bibinfo {author} {\bibfnamefont {F.}~\bibnamefont {Casas}},
  \bibinfo {author} {\bibfnamefont {J.-A.}\ \bibnamefont {Oteo}},\ and\
  \bibinfo {author} {\bibfnamefont {J.}~\bibnamefont {Ros}},\ }\href
  {https://doi.org/10.1016/j.physrep.2008.11.001} {\bibfield  {journal}
  {\bibinfo  {journal} {Phys. Rep.}\ }\textbf {\bibinfo {volume} {470}},\
  \bibinfo {pages} {151} (\bibinfo {year} {2009})}\BibitemShut {NoStop}%
\bibitem [{\citenamefont {Mananga}\ and\ \citenamefont
  {Charpentier}(2011)}]{mananga2011}%
  \BibitemOpen
  \bibfield  {author} {\bibinfo {author} {\bibfnamefont {E.~S.}\ \bibnamefont
  {Mananga}}\ and\ \bibinfo {author} {\bibfnamefont {T.}~\bibnamefont
  {Charpentier}},\ }\href {https://doi.org/10.1063/1.3610943} {\bibfield
  {journal} {\bibinfo  {journal} {J. Chem. Phys.}\ }\textbf {\bibinfo {volume}
  {135}},\ \bibinfo {pages} {044109} (\bibinfo {year} {2011})}\BibitemShut
  {NoStop}%
\bibitem [{\citenamefont {Mananga}\ and\ \citenamefont
  {Charpentier}(2016)}]{mananga2016}%
  \BibitemOpen
  \bibfield  {author} {\bibinfo {author} {\bibfnamefont {E.~S.}\ \bibnamefont
  {Mananga}}\ and\ \bibinfo {author} {\bibfnamefont {T.}~\bibnamefont
  {Charpentier}},\ }\href {https://doi.org/10.1016/j.physrep.2015.10.005}
  {\bibfield  {journal} {\bibinfo  {journal} {Physics Reports}\ }\textbf
  {\bibinfo {volume} {609}},\ \bibinfo {pages} {1} (\bibinfo {year}
  {2016})}\BibitemShut {NoStop}%
\bibitem [{\citenamefont {Kuwahara}\ \emph {et~al.}(2016)\citenamefont
  {Kuwahara}, \citenamefont {Mori},\ and\ \citenamefont
  {Saito}}]{kuwahara2016}%
  \BibitemOpen
  \bibfield  {author} {\bibinfo {author} {\bibfnamefont {T.}~\bibnamefont
  {Kuwahara}}, \bibinfo {author} {\bibfnamefont {T.}~\bibnamefont {Mori}},\
  and\ \bibinfo {author} {\bibfnamefont {K.}~\bibnamefont {Saito}},\ }\href
  {https://doi.org/10.1016/j.aop.2016.01.012} {\bibfield  {journal} {\bibinfo
  {journal} {Ann. Phys.}\ }\textbf {\bibinfo {volume} {367}},\ \bibinfo {pages}
  {96} (\bibinfo {year} {2016})}\BibitemShut {NoStop}%
\bibitem [{\citenamefont {Shavitt}\ and\ \citenamefont
  {Redmon}(1980)}]{shavitt1980}%
  \BibitemOpen
  \bibfield  {author} {\bibinfo {author} {\bibfnamefont {I.}~\bibnamefont
  {Shavitt}}\ and\ \bibinfo {author} {\bibfnamefont {L.~T.}\ \bibnamefont
  {Redmon}},\ }\href {https://doi.org/10.1063/1.440050} {\bibfield  {journal}
  {\bibinfo  {journal} {J. Chem. Phys.}\ }\textbf {\bibinfo {volume} {73}},\
  \bibinfo {pages} {5711} (\bibinfo {year} {1980})}\BibitemShut {NoStop}%
\bibitem [{\citenamefont {Kirtman}(1981)}]{kirtman1981}%
  \BibitemOpen
  \bibfield  {author} {\bibinfo {author} {\bibfnamefont {B.}~\bibnamefont
  {Kirtman}},\ }\href {https://doi.org/10.1063/1.442123} {\bibfield  {journal}
  {\bibinfo  {journal} {J. Chem. Phys.}\ }\textbf {\bibinfo {volume} {75}},\
  \bibinfo {pages} {798} (\bibinfo {year} {1981})}\BibitemShut {NoStop}%
\bibitem [{\citenamefont {Sibert}(1988)}]{sibert1988}%
  \BibitemOpen
  \bibfield  {author} {\bibinfo {author} {\bibfnamefont {E.~L.}\ \bibnamefont
  {Sibert}},\ }\href {https://doi.org/10.1063/1.453797} {\bibfield  {journal}
  {\bibinfo  {journal} {J. Chem. Phys.}\ }\textbf {\bibinfo {volume} {88}},\
  \bibinfo {pages} {4378} (\bibinfo {year} {1988})}\BibitemShut {NoStop}%
\bibitem [{\citenamefont {De~Giovannini}\ and\ \citenamefont {H{\"
  u}bener}(2019)}]{giovannini2019}%
  \BibitemOpen
  \bibfield  {author} {\bibinfo {author} {\bibfnamefont {U.}~\bibnamefont
  {De~Giovannini}}\ and\ \bibinfo {author} {\bibfnamefont {H.}~\bibnamefont
  {H{\" u}bener}},\ }\href {https://doi.org/10.1088/2515-7639/ab387b}
  {\bibfield  {journal} {\bibinfo  {journal} {J. Phys. Mat.}\ }\textbf
  {\bibinfo {volume} {3}},\ \bibinfo {pages} {012001} (\bibinfo {year}
  {2019})}\BibitemShut {NoStop}%
\bibitem [{\citenamefont {Vogl}\ \emph {et~al.}(2020)\citenamefont {Vogl},
  \citenamefont {Rodriguez-Vega},\ and\ \citenamefont {Fiete}}]{vogl2020}%
  \BibitemOpen
  \bibfield  {author} {\bibinfo {author} {\bibfnamefont {M.}~\bibnamefont
  {Vogl}}, \bibinfo {author} {\bibfnamefont {M.}~\bibnamefont
  {Rodriguez-Vega}},\ and\ \bibinfo {author} {\bibfnamefont {G.~A.}\
  \bibnamefont {Fiete}},\ }\href {https://doi.org/10.1103/PhysRevB.101.024303}
  {\bibfield  {journal} {\bibinfo  {journal} {Phys. Rev. B}\ }\textbf {\bibinfo
  {volume} {101}},\ \bibinfo {pages} {024303} (\bibinfo {year}
  {2020})}\BibitemShut {NoStop}%
\bibitem [{\citenamefont {Martins}(2024)}]{martins2024}%
  \BibitemOpen
  \bibfield  {author} {\bibinfo {author} {\bibfnamefont {W.~S.}\ \bibnamefont
  {Martins}},\ }\href {https://doi.org/10.5281/zenodo.14988944} {\bibinfo
  {title} {{AG-Lesanovsky/2024mmotion: Impact of micromotion and field-axis
  misalignment on the excitation of Rydberg states of ions in a Paul trap,
  10.5281/zenodo.14988944}}} (\bibinfo {year} {2024})\BibitemShut {NoStop}%
\bibitem [{\citenamefont {Friedrich}(2017)}]{friedrich2017}%
  \BibitemOpen
  \bibfield  {author} {\bibinfo {author} {\bibfnamefont {H.}~\bibnamefont
  {Friedrich}},\ }\href {https://doi.org/10.1007/978-3-319-47769-5} {\emph
  {\bibinfo {title} {{Theoretical Atomic Physics}}}},\ \bibinfo {edition}
  {4th}\ ed.\ (\bibinfo  {publisher} {Springer},\ \bibinfo {year}
  {2017})\BibitemShut {NoStop}%
\bibitem [{\citenamefont {Aymar}\ \emph {et~al.}(1996)\citenamefont {Aymar},
  \citenamefont {Greene},\ and\ \citenamefont {Luc-Koenig}}]{aymar1996}%
  \BibitemOpen
  \bibfield  {author} {\bibinfo {author} {\bibfnamefont {M.}~\bibnamefont
  {Aymar}}, \bibinfo {author} {\bibfnamefont {C.~H.}\ \bibnamefont {Greene}},\
  and\ \bibinfo {author} {\bibfnamefont {E.}~\bibnamefont {Luc-Koenig}},\
  }\href {https://doi.org/10.1103/RevModPhys.68.1015} {\bibfield  {journal}
  {\bibinfo  {journal} {Rev. Mod. Phys.}\ }\textbf {\bibinfo {volume} {68}},\
  \bibinfo {pages} {1015} (\bibinfo {year} {1996})}\BibitemShut {NoStop}%
\bibitem [{\citenamefont {Pawlak}\ and\ \citenamefont
  {Sadeghpour}(2020)}]{pawlak2020}%
  \BibitemOpen
  \bibfield  {author} {\bibinfo {author} {\bibfnamefont {M.}~\bibnamefont
  {Pawlak}}\ and\ \bibinfo {author} {\bibfnamefont {H.~R.}\ \bibnamefont
  {Sadeghpour}},\ }\href {https://doi.org/10.1103/PhysRevA.101.052510}
  {\bibfield  {journal} {\bibinfo  {journal} {Phys. Rev. A}\ }\textbf {\bibinfo
  {volume} {101}},\ \bibinfo {pages} {052510} (\bibinfo {year}
  {2020})}\BibitemShut {NoStop}%
\bibitem [{\citenamefont {Kramida}\ \emph {et~al.}(2024)\citenamefont
  {Kramida}, \citenamefont {Ralchenko}, \citenamefont {Reader},\ and\
  \citenamefont {{{NIST} {ASD} Team}}}]{nist2024}%
  \BibitemOpen
  \bibfield  {author} {\bibinfo {author} {\bibfnamefont {A.}~\bibnamefont
  {Kramida}}, \bibinfo {author} {\bibfnamefont {Y.}~\bibnamefont {Ralchenko}},
  \bibinfo {author} {\bibfnamefont {J.}~\bibnamefont {Reader}},\ and\ \bibinfo
  {author} {\bibnamefont {{{NIST} {ASD} Team}}},\ }\href
  {https://physics.nist.gov/asd} {\bibinfo {title} {{NIST Atomic Spectra
  Database (v. 5.11)}}} (\bibinfo {year} {2024})\BibitemShut {NoStop}%
\end{thebibliography}%

\end{document}